\documentclass[a4paper,11pt]{article}

\usepackage{jheppub,bm} 

\usepackage{graphicx}
\usepackage{amsmath}
\usepackage{axodraw4j}

\newcommand{\nn}{\nonumber}
\newcommand{\bea}{\begin{eqnarray}}
\newcommand{\eea}{\end{eqnarray}}
\newcommand{\be}{\begin{equation}}
\newcommand{\ee}{\end{equation}}

\newcommand{\muf}{{\mu_{\scriptscriptstyle F} }}

\newcommand{\aas}{{\alpha\alpha_s }}

\newcommand{\smallw}{{\scriptscriptstyle W}}
\newcommand{\smallz}{{\scriptscriptstyle Z}}
\newcommand{\smallv}{{\scriptscriptstyle V}}

\newcommand{\oaas}{\mbox{${\cal O}(\alpha \alpha_s)$}\,}

\newcommand{\oa}{${\cal O}(\alpha)$}

\allowdisplaybreaks[4]

\newcommand{\pts}[1]{\left( #1 \right)}
\newcommand{\ptsq}[1]{\left[ #1 \right]}

\newcommand{\eps}{\varepsilon}

\newcommand{\rmd}{\mathrm{d}}

\newcommand{\Ncal}{\mathcal{N}}

\newcommand{\Ocal}{\mathcal{O}}
\newcommand{\zed}[1]{\zeta_{#1}}

\newcommand{\SWsq}{\mathrm{sin}^{2} \theta_{W} }
\newcommand{\CWsq}{\mathrm{cos}^{2} \theta_{W} }
\newcommand{\deltz}{\delta\left(1 - z \right)}

\newcommand{\Hpl}{\mathrm{H}}
\newcommand{\Log}[1]{\mathrm{ln}\left( #1 \right)}
\newcommand{\PLog}[2]{\mathrm{ln}^{#2} \left( #1 \right)}
\newcommand{\Li}[2]{\mathrm{Li}_{#1} \left( #2 \right)}
\newcommand{\PlusD}[1]{\left( #1 \right)_{+}}

\newcommand{\amplitudeone}{
	\scalebox{0.75}{
		\begin{picture}(77,64) (244,-82.5)
		
			 \SetWidth{1.0}
  			 \SetColor{Black}
    			 \Line[arrow,arrowpos=0.5,arrowlength=1.75,arrowwidth=0.7,arrowinset=0.2](269,-33)(269,-55)
    			 \Line[arrow,arrowpos=0.5,arrowlength=1.75,arrowwidth=0.7,arrowinset=0.2](269,-55)(269,-77)

   			 \SetWidth{1.3}
  		     \Photon(269,-33)(293,-30){2.7}{4.5}
   			 \SetWidth{0.8}
   			 \Photon(269,-55)(293,-55){2.3}{4.5}
   			 \Gluon(269,-77)(293,-80){2.3}{4.5}
   			 
   			 \SetWidth{1.0}
   			 \Line[arrow,arrowpos=0.5,arrowlength=1.75,arrowwidth=0.7,arrowinset=0.2](252,-30)(269,-33)
   			 \Line[arrow,arrowpos=0.5,arrowlength=1.75,arrowwidth=0.7,arrowinset=0.2](269,-77)(252,-80)
    
    \Text(242,-30)[lb]{\small{\Black{$p_{1}$}}}
    \Text(242,-82)[lb]{\small{\Black{$p_{2}$}}}  
    \Text(297,-31)[lb]{\small{\Black{$W,Z$}}}
    \Text(297,-83)[lb]{\small{\Black{$p_{4}$}}}
    \Text(297,-57)[lb]{\small{\Black{$p_{5}$}}}

		\end{picture}   }
}

\newcommand{\amplitudetwo}{
	\scalebox{0.75}{
		\begin{picture}(77,64) (244,-82.5)
		
			 \SetWidth{1.0}
  			 \SetColor{Black}
    			 \Line[arrow,arrowpos=0.5,arrowlength=1.75,arrowwidth=0.7,arrowinset=0.2](269,-33)(269,-55)
    			 \Line[arrow,arrowpos=0.5,arrowlength=1.75,arrowwidth=0.7,arrowinset=0.2](269,-55)(269,-77)

   			 \SetWidth{1.3}
  		     \Photon(269,-33)(293,-30){2.7}{4.5}
   			 \SetWidth{0.8}
   			 \Photon(269,-77)(293,-55){2.3}{4.5}
   			 \Gluon(269,-55)(293,-80){2.3}{4.5}
   			 
   			 \SetWidth{1.0}
   			 \Line[arrow,arrowpos=0.5,arrowlength=1.75,arrowwidth=0.7,arrowinset=0.2](252,-30)(269,-33)
   			 \Line[arrow,arrowpos=0.5,arrowlength=1.75,arrowwidth=0.7,arrowinset=0.2](269,-77)(252,-80)
    
    \Text(242,-30)[lb]{\small{\Black{$p_{1}$}}}
    \Text(242,-82)[lb]{\small{\Black{$p_{2}$}}}  
    \Text(297,-31)[lb]{\small{\Black{$W,Z$}}}
    \Text(297,-83)[lb]{\small{\Black{$p_{4}$}}}
    \Text(297,-57)[lb]{\small{\Black{$p_{5}$}}}

		\end{picture}   }
}

\newcommand{\amplitudethree}{
	\scalebox{0.75}{
		\begin{picture}(77,64) (244,-82.5)
		
			 \SetWidth{1.0}
  			 \SetColor{Black}
    			 \Line[arrow,arrowpos=0.5,arrowlength=1.75,arrowwidth=0.7,arrowinset=0.2](269,-33)(269,-55)
    			 \Line[arrow,arrowpos=0.5,arrowlength=1.75,arrowwidth=0.7,arrowinset=0.2](269,-55)(269,-77)

   			 \SetWidth{1.3}
  		     \Photon(269,-55)(293,-30){2.7}{4.5}
   			 \SetWidth{0.8}
   			 \Photon(269,-33)(293,-55){2.3}{4.5}
   			 \Gluon(269,-77)(293,-80){2.3}{4.5}
   			 
   			 \SetWidth{1.0}
   			 \Line[arrow,arrowpos=0.5,arrowlength=1.75,arrowwidth=0.7,arrowinset=0.2](252,-30)(269,-33)
   			 \Line[arrow,arrowpos=0.5,arrowlength=1.75,arrowwidth=0.7,arrowinset=0.2](269,-77)(252,-80)
    
    \Text(242,-30)[lb]{\small{\Black{$p_{1}$}}}
    \Text(242,-82)[lb]{\small{\Black{$p_{2}$}}}  
    \Text(297,-31)[lb]{\small{\Black{$W,Z$}}}
    \Text(297,-83)[lb]{\small{\Black{$p_{4}$}}}
    \Text(297,-57)[lb]{\small{\Black{$p_{5}$}}}

		\end{picture}   }
}

\newcommand{\amplitudefour}{
	\scalebox{0.75}{
		\begin{picture}(77,64) (244,-82.5)
		
			 \SetWidth{1.0}
  			 \SetColor{Black}
    			 \Line[arrow,arrowpos=0.5,arrowlength=1.75,arrowwidth=0.7,arrowinset=0.2](269,-33)(269,-55)
    			 \Line[arrow,arrowpos=0.5,arrowlength=1.75,arrowwidth=0.7,arrowinset=0.2](269,-55)(269,-77)

   			 \SetWidth{1.3}
  		     \Photon(269,-55)(293,-30){2.7}{4.5}
   			 \SetWidth{0.8}
   			 \Photon(269,-77)(293,-55){2.3}{4.5}
   			 \Gluon(269,-33)(293,-80){2.3}{4.5}
   			 
   			 \SetWidth{1.0}
   			 \Line[arrow,arrowpos=0.5,arrowlength=1.75,arrowwidth=0.7,arrowinset=0.2](252,-30)(269,-33)
   			 \Line[arrow,arrowpos=0.5,arrowlength=1.75,arrowwidth=0.7,arrowinset=0.2](269,-77)(252,-80)
    
    \Text(242,-30)[lb]{\small{\Black{$p_{1}$}}}
    \Text(242,-82)[lb]{\small{\Black{$p_{2}$}}}  
    \Text(297,-31)[lb]{\small{\Black{$W,Z$}}}
    \Text(297,-83)[lb]{\small{\Black{$p_{4}$}}}
    \Text(297,-57)[lb]{\small{\Black{$p_{5}$}}}

		\end{picture}   }
}

\newcommand{\amplitudefive}{
	\scalebox{0.75}{
		\begin{picture}(77,64) (244,-82.5)
		
			 \SetWidth{1.0}
  			 \SetColor{Black}
    			 \Line[arrow,arrowpos=0.5,arrowlength=1.75,arrowwidth=0.7,arrowinset=0.2](269,-33)(269,-55)
    			 \Line[arrow,arrowpos=0.5,arrowlength=1.75,arrowwidth=0.7,arrowinset=0.2](269,-55)(269,-77)

   			 \SetWidth{1.3}
  		     \Photon(269,-77)(293,-30){2.7}{4.5}
   			 \SetWidth{0.8}
   			 \Photon(269,-33)(293,-55){2.3}{4.5}
   			 \Gluon(269,-55)(293,-80){2.3}{4.5}
   			 
   			 \SetWidth{1.0}
   			 \Line[arrow,arrowpos=0.5,arrowlength=1.75,arrowwidth=0.7,arrowinset=0.2](252,-30)(269,-33)
   			 \Line[arrow,arrowpos=0.5,arrowlength=1.75,arrowwidth=0.7,arrowinset=0.2](269,-77)(252,-80)
    
    \Text(242,-30)[lb]{\small{\Black{$p_{1}$}}}
    \Text(242,-82)[lb]{\small{\Black{$p_{2}$}}}  
    \Text(297,-31)[lb]{\small{\Black{$W,Z$}}}
    \Text(297,-83)[lb]{\small{\Black{$p_{4}$}}}
    \Text(297,-57)[lb]{\small{\Black{$p_{5}$}}}

		\end{picture}   }
}

\newcommand{\amplitudesix}{
	\scalebox{0.75}{
		\begin{picture}(77,64) (244,-82.5)
		
			 \SetWidth{1.0}
  			 \SetColor{Black}
    			 \Line[arrow,arrowpos=0.5,arrowlength=1.75,arrowwidth=0.7,arrowinset=0.2](269,-33)(269,-55)
    			 \Line[arrow,arrowpos=0.5,arrowlength=1.75,arrowwidth=0.7,arrowinset=0.2](269,-55)(269,-77)

   			 \SetWidth{1.3}
  		     \Photon(269,-77)(293,-30){2.7}{4.5}
   			 \SetWidth{0.8}
   			 \Photon(269,-55)(293,-55){2.3}{4.5}
   			 \Gluon(269,-33)(293,-80){2.3}{4.5}
   			 
   			 \SetWidth{1.0}
   			 \Line[arrow,arrowpos=0.5,arrowlength=1.75,arrowwidth=0.7,arrowinset=0.2](252,-30)(269,-33)
   			 \Line[arrow,arrowpos=0.5,arrowlength=1.75,arrowwidth=0.7,arrowinset=0.2](269,-77)(252,-80)
    
    \Text(242,-30)[lb]{\small{\Black{$p_{1}$}}}
    \Text(242,-82)[lb]{\small{\Black{$p_{2}$}}}  
    \Text(297,-31)[lb]{\small{\Black{$W,Z$}}}
    \Text(297,-83)[lb]{\small{\Black{$p_{4}$}}}
    \Text(297,-57)[lb]{\small{\Black{$p_{5}$}}}

		\end{picture}   }
}

\newcommand{\amplitudeseven}{
	\scalebox{0.75}{
		\begin{picture}(77,64) (244,-82.5)
		
			 \SetWidth{1.0}
  			 \SetColor{Black}
    			 \Line[arrow,arrowpos=0.5,arrowlength=1.75,arrowwidth=0.7,arrowinset=0.2](269,-33)(269,-77)

   			 \SetWidth{1.3}
  		     \Photon(269,-33)(293,-30){2.7}{4.5}
   			 \SetWidth{0.8}
   			 \Photon(281,-31.5)(293,-55){2.3}{4.5}
   			 \Gluon(269,-77)(293,-80){2.3}{4.5}
   			 
   			 \SetWidth{1.0}
   			 \Line[arrow,arrowpos=0.5,arrowlength=1.75,arrowwidth=0.7,arrowinset=0.2](252,-30)(269,-33)
   			 \Line[arrow,arrowpos=0.5,arrowlength=1.75,arrowwidth=0.7,arrowinset=0.2](269,-77)(252,-80)
    
    \Text(242,-30)[lb]{\small{\Black{$p_{1}$}}}
    \Text(242,-82)[lb]{\small{\Black{$p_{2}$}}}  
    \Text(297,-31)[lb]{\small{\Black{$W$}}}
    \Text(297,-83)[lb]{\small{\Black{$p_{4}$}}}
    \Text(297,-57)[lb]{\small{\Black{$p_{5}$}}}

		\end{picture}   }
}

\newcommand{\amplitudeeight}{
	\scalebox{0.75}{
		\begin{picture}(77,64) (244,-82.5)
		
			 \SetWidth{1.0}
  			 \SetColor{Black}
    			 \Line[arrow,arrowpos=0.5,arrowlength=1.75,arrowwidth=0.7,arrowinset=0.2](269,-33)(269,-77)

   			 \SetWidth{1.3}
  		     \Photon(269,-77)(293,-80){2.7}{4.5}
   			 \SetWidth{0.8}
   			 \Photon(281,-78.5)(293,-55){2.3}{4.5}
   			 \Gluon(269,-33)(293,-30){2.3}{4.5}
   			 
   			 \SetWidth{1.0}
   			 \Line[arrow,arrowpos=0.5,arrowlength=1.75,arrowwidth=0.7,arrowinset=0.2](252,-30)(269,-33)
   			 \Line[arrow,arrowpos=0.5,arrowlength=1.75,arrowwidth=0.7,arrowinset=0.2](269,-77)(252,-80)
    
    \Text(242,-30)[lb]{\small{\Black{$p_{1}$}}}
    \Text(242,-82)[lb]{\small{\Black{$p_{2}$}}}  
    \Text(297,-83)[lb]{\small{\Black{$W$}}}
    \Text(297,-31)[lb]{\small{\Black{$p_{4}$}}}
    \Text(297,-57)[lb]{\small{\Black{$p_{5}$}}}

		\end{picture}   }
}

\newcommand{\amplitudescollection}{
\scalebox{1.0}{

	\begin{picture}(300,200)(0,0)
	
\put(0,140){\scalebox{1.3}{\amplitudeone}}
\put(100,140){\scalebox{1.3}{\amplitudetwo}}
\put(200,140){\scalebox{1.3}{\amplitudethree}}

\put(0,70){\scalebox{1.3}{\amplitudefour}}
\put(100,70){\scalebox{1.3}{\amplitudefive}}
\put(200,70){\scalebox{1.3}{\amplitudesix}}

\put(50,0){\scalebox{1.3}{\amplitudeseven}}
\put(150,00){\scalebox{1.3}{\amplitudeeight}}

\end{picture} } 
}

\newcommand{\masterone}{
	\scalebox{0.75}{
		\begin{picture}(63,55) (92,-22)
		
			\SetWidth{2.7}
    			\Arc(123,5)(22,0,180)
    			\SetWidth{1.0}
    			\SetColor{Black}
    			\Arc(123,5)(22,180,360)
   			\SetWidth{1.3}
   			\Line[dash,dashsize=5](123,30)(123,-20)
   			\SetWidth{1.0}
   			\Line(101,5)(145,5)
    			\Line[arrow,arrowpos=0.5,arrowlength=1.75,arrowwidth=0.7,arrowinset=0.2](88,5)(101,5)
   			\Line[arrow,arrowpos=0.5,arrowlength=1.75,arrowwidth=0.7,arrowinset=0.2](145,5)(158,5)
  		    \Text(90,8)[lb]{\small{\Black{$p$}}}
    			\Text(153,8)[lb]{\small{\Black{$p$}}}
		\end{picture}	}
}

\newcommand{\mastertwo}{
	\scalebox{0.75}{
		\begin{picture}(65,55) (215,-22)
		    \SetWidth{2.7}
    \Arc(246,5)(22,0,180)
    \SetWidth{1.0}
    \SetColor{Black}
    \Arc(246,5)(22,180,360)
    \SetWidth{1.3}
    \Line[dash,dashsize=5](246,30)(246,-20)
    \SetWidth{1.0}
    \Line(224,5)(268,5)
    \Line[arrow,arrowpos=0.5,arrowlength=1.75,arrowwidth=0.7,arrowinset=0.2](211,5)(224,5)
    \Line[arrow,arrowpos=0.5,arrowlength=1.75,arrowwidth=0.7,arrowinset=0.2](268,5)(281,5)
    \Text(213,8)[lb]{\small{\Black{$p$}}}
    \Text(276,8)[lb]{\small{\Black{$p$}}}
    \Text(264,20)[lb]{\small{\Black{$\left(p_{1} - p_{4} \right)^{2}$}}}
			
		\end{picture}	}
}

\newcommand{\masterthree}{
	\scalebox{0.75}{
		\begin{picture}(65,55) (344,-22)

\SetWidth{2.7}
    \SetColor{Black}
	\Line(353,27)(397,27)
	\SetWidth{1.3}
    \Line[dash,dashsize=5](375,30)(375,-20)
    \SetWidth{1.0}
    \Arc(375,27)(44,240,300)
    \Arc(375,-49.21)(44,60,120)
    \Line(353,27)(353,-11.105)
    \Line(397,27)(397,-11.105)
    \Line[arrow,arrowpos=0.5,arrowlength=1.75,arrowwidth=0.7,arrowinset=0.2](340,27)(353,27)
    \Line[arrow,arrowpos=0.5,arrowlength=1.75,arrowwidth=0.7,arrowinset=0.2](397,27)(410,27)
    \Line[arrow,arrowpos=0.5,arrowlength=1.75,arrowwidth=0.7,arrowinset=0.2](340,-11.105)(353,-11.105)
    \Line[arrow,arrowpos=0.5,arrowlength=1.75,arrowwidth=0.7,arrowinset=0.2](397,-11.105)(410,-11.105)
    
    \Text(342,30)[lb]{\small{\Black{$p_{1}$}}}
    \Text(403,30)[lb]{\small{\Black{$p_{2}$}}}
    \Text(342,-8)[lb]{\small{\Black{$p_{2}$}}}
    \Text(403,-8)[lb]{\small{\Black{$p_{1}$}}}

		\end{picture}	}
}

\newcommand{\masterfour}{
	\scalebox{0.75}{
		\begin{picture}(65,55) (91,-81)

\SetWidth{2.7}
    \SetColor{Black}
    \Line(101,-55)(145,-55)
    \SetWidth{1.3}
    \Line[dash,dashsize=5](123,-30)(123,-80)
    \SetWidth{1.0}
    \Line(101,-33)(145,-55)
    \Line(101,-77)(145,-55)
    \Line(101,-33)(101,-77)    
    \Line[arrow,arrowpos=0.5,arrowlength=1.75,arrowwidth=0.7,arrowinset=0.2](88,-33)(101,-33)
    \Line[arrow,arrowpos=0.5,arrowlength=1.75,arrowwidth=0.7,arrowinset=0.2](88,-77)(101,-77)
    \Line[arrow,arrowpos=0.5,arrowlength=1.75,arrowwidth=0.7,arrowinset=0.2](145,-55)(158,-55)
    
    \Text(90,-30)[lb]{\small{\Black{$p_{1}$}}}
    \Text(90,-74)[lb]{\small{\Black{$p_{2}$}}}
    \Text(153,-52)[lb]{\small{\Black{$p$}}}

		\end{picture} }
}

\newcommand{\masterfive}{
	\scalebox{0.75}{
		\begin{picture}(65,55) (175,-81)
		
			\SetWidth{2.7}
   		    \SetColor{Black}
    			\Line(185,-77)(229,-55)
   		    \SetWidth{1.3}
    			\Line[dash,dashsize=5](196,-30)(196,-80)
    			\SetWidth{1.0}
    \Line(185,-33)(207,-66)
    \Line(185,-33)(229,-55)
    \Line(185,-33)(185,-77)    
    \Line[arrow,arrowpos=0.5,arrowlength=1.75,arrowwidth=0.7,arrowinset=0.2](172,-33)(185,-33)
    \Line[arrow,arrowpos=0.5,arrowlength=1.75,arrowwidth=0.7,arrowinset=0.2](172,-77)(185,-77)
    \Line[arrow,arrowpos=0.5,arrowlength=1.75,arrowwidth=0.7,arrowinset=0.2](229,-55)(242,-55)
    
    \Text(174,-30)[lb]{\small{\Black{$p_{1}$}}}
    \Text(174,-74)[lb]{\small{\Black{$p_{2}$}}}  
    \Text(237,-52)[lb]{\small{\Black{$p$}}} 
      
		\end{picture} }
}

\newcommand{\mastersix}{
	\scalebox{0.75}{
		\begin{picture}(65,55) (259,-81)
		
			 \SetWidth{2.7}
  			 \SetColor{Black}
    			 \Line(269,-77)(313,-55)
    \SetWidth{1.3}
    \Line[dash,dashsize=5](280,-30)(280,-80)
    \SetWidth{1.0}
    \Line(269,-33)(291,-66)
    \Line(269,-33)(313,-55)
    \Line(269,-33)(269,-77)    
    \Line[arrow,arrowpos=0.5,arrowlength=1.75,arrowwidth=0.7,arrowinset=0.2](256,-33)(269,-33)
    \Line[arrow,arrowpos=0.5,arrowlength=1.75,arrowwidth=0.7,arrowinset=0.2](256,-77)(269,-77)
    \Line[arrow,arrowpos=0.5,arrowlength=1.75,arrowwidth=0.7,arrowinset=0.2](313,-55)(326,-55)
    \SetWidth{1.0}
    \Vertex(269,-55){2.3}
    
    \Text(258,-30)[lb]{\small{\Black{$p_{1}$}}}
    \Text(258,-74)[lb]{\small{\Black{$p_{2}$}}}  
    \Text(319,-52)[lb]{\small{\Black{$p$}}} 
		\end{picture} }
}

\newcommand{\masterseven}{
	\scalebox{0.75}{
		\begin{picture}(65,60) (343,-81)
     \SetWidth{2.7}
    \SetColor{Black}
    \Line(353,-77)(397,-55)
    \SetWidth{1.3}
    \Line[dash,dashsize=5](364,-30)(363,-80)
    \SetWidth{1.0}
    \Line(353,-33)(375,-66)
    \Line(353,-55)(397,-55)
    \Line(353,-33)(353,-77)    
    \Line[arrow,arrowpos=0.5,arrowlength=1.75,arrowwidth=0.7,arrowinset=0.2](340,-33)(353,-33)
    \Line[arrow,arrowpos=0.5,arrowlength=1.75,arrowwidth=0.7,arrowinset=0.2](340,-77)(353,-77)
    \Line[arrow,arrowpos=0.5,arrowlength=1.75,arrowwidth=0.7,arrowinset=0.2](393,-55)(410,-55)

    \Text(407,-52)[lb]{\small{\Black{$p$}}}
    \Text(342,-30)[lb]{\small{\Black{$p_{1}$}}}
    \Text(342,-74)[lb]{\small{\Black{$p_{2}$}}}

		\end{picture}  }
}

\newcommand{\mastereight}{
	\scalebox{0.75}{
		\begin{picture}(65,55) (91,-148)

    \SetWidth{2.7}
    \SetColor{Black}
    \Line(101,-142)(145,-142)
    \SetWidth{1.3}
    \Line[dash,dashsize=5](123,-90)(123,-150)
    \SetWidth{1.0}
    \Line(101,-98)(145,-98)
    \Line(101,-120)(145,-120)
    \Line(101,-98)(101,-142)    
    \Line(145,-98)(145,-142)    
    \Line[arrow,arrowpos=0.5,arrowlength=1.75,arrowwidth=0.7,arrowinset=0.2](88,-98)(101,-98)
    \Line[arrow,arrowpos=0.5,arrowlength=1.75,arrowwidth=0.7,arrowinset=0.2](145,-98)(158,-98)
    \Line[arrow,arrowpos=0.5,arrowlength=1.75,arrowwidth=0.7,arrowinset=0.2](88,-142)(101,-142)
    \Line[arrow,arrowpos=0.5,arrowlength=1.75,arrowwidth=0.7,arrowinset=0.2](145,-142)(158,-142)

    \Text(90,-95)[lb]{\small{\Black{$p_{1}$}}}
    \Text(151,-95)[lb]{\small{\Black{$p_{2}$}}}
    \Text(90,-139)[lb]{\small{\Black{$p_{2}$}}}
    \Text(151,-139)[lb]{\small{\Black{$p_{1}$}}}
    
    \end{picture} }
    
    }

\newcommand{\masternine}{
	\scalebox{0.75}{
		\begin{picture}(65,55) (175,-148)
    \SetWidth{2.7}
    \SetColor{Black}
    \Line(185,-120)(229,-120)
    \SetWidth{1.3}
    \Line[dash,dashsize=5](207,-90)(207,-150)
    \SetWidth{1.0}
    \Line(185,-98)(229,-98)
    \Line(185,-142)(229,-142)
    \Line(185,-98)(185,-142)    
    \Line(229,-98)(229,-142)    
    \Line[arrow,arrowpos=0.5,arrowlength=1.75,arrowwidth=0.7,arrowinset=0.2](172,-98)(185,-98)
    \Line[arrow,arrowpos=0.5,arrowlength=1.75,arrowwidth=0.7,arrowinset=0.2](229,-98)(242,-98)
    \Line[arrow,arrowpos=0.5,arrowlength=1.75,arrowwidth=0.7,arrowinset=0.2](172,-142)(185,-142)
    \Line[arrow,arrowpos=0.5,arrowlength=1.75,arrowwidth=0.7,arrowinset=0.2](229,-142)(242,-142)

    \Text(174,-95)[lb]{\small{\Black{$p_{1}$}}}
    \Text(235,-95)[lb]{\small{\Black{$p_{2}$}}}
    \Text(174,-139)[lb]{\small{\Black{$p_{2}$}}}
    \Text(235,-139)[lb]{\small{\Black{$p_{1}$}}}
    
    \end{picture} }
    
    }

\newcommand{\masterten}{
	\scalebox{0.75}{
		\begin{picture}(65,55) (259,-148)
    \SetWidth{2.7}
    \SetColor{Black}
    \Line(269,-98)(313,-98)
    \SetWidth{1.3}
    \Line[dash,dashsize=5](302,-90)(302,-150)
    \SetWidth{1.0}
    \Line(269,-98)(269,-142)    
    \Line(313,-98)(313,-142)  
    \Line(269,-142)(313,-120)
    \Line(269,-120)(313,-142)
    \Line[arrow,arrowpos=0.5,arrowlength=1.75,arrowwidth=0.7,arrowinset=0.2](256,-98)(269,-98)
    \Line[arrow,arrowpos=0.5,arrowlength=1.75,arrowwidth=0.7,arrowinset=0.2](313,-98)(326,-98)
    \Line[arrow,arrowpos=0.5,arrowlength=1.75,arrowwidth=0.7,arrowinset=0.2](256,-142)(269,-142)
    \Line[arrow,arrowpos=0.5,arrowlength=1.75,arrowwidth=0.7,arrowinset=0.2](313,-142)(326,-142)

    \Text(258,-95)[lb]{\small{\Black{$p_{1}$}}}
    \Text(319,-95)[lb]{\small{\Black{$p_{2}$}}}
    \Text(258,-139)[lb]{\small{\Black{$p_{2}$}}}
    \Text(319,-139)[lb]{\small{\Black{$p_{1}$}}}
    
    \end{picture} }
    }

\newcommand{\mastereleven}{
	\scalebox{0.75}{ 
		\begin{picture}(65,60) (343,-148)
     \SetWidth{2.7}
    \SetColor{Black}
    \Line(353,-98)(397,-98)
    \SetWidth{1.3}
    \Line[dash,dashsize=5](389,-90)(389,-150)
    \SetWidth{1.0}
    \Line(353,-98)(353,-142)    
    \Line(397,-98)(397,-142)
    \Line(369,-98)(397,-142)
    \Line(353,-142)(397,-115)
    \Line[arrow,arrowpos=0.5,arrowlength=1.75,arrowwidth=0.7,arrowinset=0.2](340,-98)(353,-98)   
    \Line[arrow,arrowpos=0.5,arrowlength=1.75,arrowwidth=0.7,arrowinset=0.2](397,-98)(410,-98)
    \Line[arrow,arrowpos=0.5,arrowlength=1.75,arrowwidth=0.7,arrowinset=0.2](340,-142)(353,-142)   
    \Line[arrow,arrowpos=0.5,arrowlength=1.75,arrowwidth=0.7,arrowinset=0.2](397,-142)(410,-142)
    
    \Text(342,-95)[lb]{\small{\Black{$p_{1}$}}}
    \Text(403,-95)[lb]{\small{\Black{$p_{2}$}}}
    \Text(342,-139)[lb]{\small{\Black{$p_{2}$}}}
    \Text(403,-139)[lb]{\small{\Black{$p_{1}$}}}

		\end{picture}  }
}

\newcommand{\mastertwelve}{
	\scalebox{0.75}{
		\begin{picture}(65,55) (259,-81)
		
	\SetWidth{2.7}
	\Line(269,-33)(291,-44)
  	\SetColor{Black}
    \SetWidth{1.3}
    \Line[dash,dashsize=5](280,-30)(280,-80)
    \SetWidth{1.0}
    	\Line(291,-44)(313,-55)
    \Line(269,-77)(313,-55)
    \Line(269,-77)(291,-44)
    \Line(269,-33)(269,-77)    
    \Line[arrow,arrowpos=0.5,arrowlength=1.75,arrowwidth=0.7,arrowinset=0.2](256,-33)(269,-33)
    \Line[arrow,arrowpos=0.5,arrowlength=1.75,arrowwidth=0.7,arrowinset=0.2](256,-77)(269,-77)
    \Line[arrow,arrowpos=0.5,arrowlength=1.75,arrowwidth=0.7,arrowinset=0.2](313,-55)(326,-55)
    \SetWidth{1.0}
    
    \Text(258,-30)[lb]{\small{\Black{$p_{1}$}}}
    \Text(258,-74)[lb]{\small{\Black{$p_{2}$}}}  
    \Text(319,-52)[lb]{\small{\Black{$p$}}} 
		\end{picture} }
}

\newcommand{\masterthirteen}{
	\scalebox{0.75}{
		\begin{picture}(65,55) (259,-81)
		
	\SetWidth{2.7}
	\Line(269,-33)(291,-44)
  	\SetColor{Black}
    \SetWidth{1.3}
    \Line[dash,dashsize=5](280,-30)(280,-80)
    \SetWidth{1.0}
    	\Line(291,-44)(313,-55)
    \Line(269,-77)(313,-55)
    \Line(269,-77)(291,-44)
    \Line(269,-33)(269,-77)    
    \Line[arrow,arrowpos=0.5,arrowlength=1.75,arrowwidth=0.7,arrowinset=0.2](256,-33)(269,-33)
    \Line[arrow,arrowpos=0.5,arrowlength=1.75,arrowwidth=0.7,arrowinset=0.2](256,-77)(269,-77)
    \Line[arrow,arrowpos=0.5,arrowlength=1.75,arrowwidth=0.7,arrowinset=0.2](313,-55)(326,-55)
    \SetWidth{1.0}
    \Vertex(269,-55){2.3}
    
    \Text(258,-30)[lb]{\small{\Black{$p_{1}$}}}
    \Text(258,-74)[lb]{\small{\Black{$p_{2}$}}}  
    \Text(319,-52)[lb]{\small{\Black{$p$}}} 
		\end{picture} }
}

\newcommand{\masterfourteen}{
	\scalebox{0.75}{
		\begin{picture}(65,55) (344,-22)

\SetWidth{2.7}
    \SetColor{Black}
	\Line(353,27)(397,27)
	\Line(397,27)(397,-11.105)
	\SetWidth{1.3}
    \Line[dash,dashsize=5](375,30)(375,-20)
    \SetWidth{1.0}
    \Arc(375,27)(44,240,300)
    \Arc(375,-49.21)(44,60,120)
    \Line(353,27)(353,-11.105)
    \Line[arrow,arrowpos=0.5,arrowlength=1.75,arrowwidth=0.7,arrowinset=0.2](340,27)(353,27)
    \Line[arrow,arrowpos=0.5,arrowlength=1.75,arrowwidth=0.7,arrowinset=0.2](397,27)(410,27)
    \Line[arrow,arrowpos=0.5,arrowlength=1.75,arrowwidth=0.7,arrowinset=0.2](340,-11.105)(353,-11.105)
    \Line[arrow,arrowpos=0.5,arrowlength=1.75,arrowwidth=0.7,arrowinset=0.2](397,-11.105)(410,-11.105)
    
    \Text(342,30)[lb]{\small{\Black{$p_{1}$}}}
    \Text(403,30)[lb]{\small{\Black{$p_{2}$}}}
    \Text(342,-8)[lb]{\small{\Black{$p_{2}$}}}
    \Text(403,-8)[lb]{\small{\Black{$p_{1}$}}}

		\end{picture}	}
}

\newcommand{\masterfifteen}{
	\scalebox{0.75}{
		\begin{picture}(65,60) (343,-81)
    \SetWidth{2.7}
    \Line(353,-55)(397,-55)
    \SetColor{Black}
    \SetWidth{1.3}
    \Line[dash,dashsize=5](364,-30)(363,-80)
    \SetWidth{1.0}
    \Line(353,-77)(397,-55)
    \Line(353,-33)(375,-66)
    \Line(353,-33)(353,-77)    
    \Line[arrow,arrowpos=0.5,arrowlength=1.75,arrowwidth=0.7,arrowinset=0.2](340,-33)(353,-33)
    \Line[arrow,arrowpos=0.5,arrowlength=1.75,arrowwidth=0.7,arrowinset=0.2](340,-77)(353,-77)
    \Line[arrow,arrowpos=0.5,arrowlength=1.75,arrowwidth=0.7,arrowinset=0.2](393,-55)(410,-55)

    \Text(407,-52)[lb]{\small{\Black{$p$}}}
    \Text(342,-30)[lb]{\small{\Black{$p_{1}$}}}
    \Text(342,-74)[lb]{\small{\Black{$p_{2}$}}}

		\end{picture}  }
}

\newcommand{\mastersixteen}{
	\scalebox{0.75}{
		\begin{picture}(65,60) (343,-81)
    \SetWidth{2.7}
    \Line(353,-33)(375,-66)
    \SetColor{Black}
    \SetWidth{1.3}
    \Line[dash,dashsize=5](364,-30)(363,-80)
    \SetWidth{1.0}
    \Line(353,-55)(397,-55)
    \Line(353,-77)(397,-55)
    \Line(353,-33)(353,-77)    
    \Line[arrow,arrowpos=0.5,arrowlength=1.75,arrowwidth=0.7,arrowinset=0.2](340,-33)(353,-33)
    \Line[arrow,arrowpos=0.5,arrowlength=1.75,arrowwidth=0.7,arrowinset=0.2](340,-77)(353,-77)
    \Line[arrow,arrowpos=0.5,arrowlength=1.75,arrowwidth=0.7,arrowinset=0.2](393,-55)(410,-55)

    \Text(407,-52)[lb]{\small{\Black{$p$}}}
    \Text(342,-30)[lb]{\small{\Black{$p_{1}$}}}
    \Text(342,-74)[lb]{\small{\Black{$p_{2}$}}}

		\end{picture}  }
}

\newcommand{\masterseventeen}{
	\scalebox{0.75}{
		\begin{picture}(85,55) (175,-148)
    \SetWidth{2.7}
    \SetColor{Black}
    \Line(185,-98)(219,-98)
    \SetWidth{1.3}
    \Line[dash,dashsize=5](200,-90)(238,-150)
    \SetWidth{1.0}
    \Line(185,-142)(219,-142)
    \Line(185,-98)(185,-142)    
    \Line(219,-98)(219,-142)   
    
    \Line(219,-98)(253,-98)    
    \Line(253,-98)(253,-142)    
    \Line(219,-142)(253,-142)    
 
    \Line[arrow,arrowpos=0.5,arrowlength=1.75,arrowwidth=0.7,arrowinset=0.2](172,-98)(185,-98)
    \Line[arrow,arrowpos=0.5,arrowlength=1.75,arrowwidth=0.7,arrowinset=0.2](253,-98)(266,-98)
    \Line[arrow,arrowpos=0.5,arrowlength=1.75,arrowwidth=0.7,arrowinset=0.2](172,-142)(185,-142)
    \Line[arrow,arrowpos=0.5,arrowlength=1.75,arrowwidth=0.7,arrowinset=0.2](253,-142)(266,-142)

    \Text(174,-95)[lb]{\small{\Black{$p_{1}$}}}
    \Text(259,-95)[lb]{\small{\Black{$p_{2}$}}}
    \Text(174,-139)[lb]{\small{\Black{$p_{2}$}}}
    \Text(259,-139)[lb]{\small{\Black{$p_{1}$}}}
    
    \end{picture} }
    
    }

\newcommand{\mastereighteen}{
	\scalebox{0.75}{
		\begin{picture}(65,55) (259,-148)
    \SetWidth{2.7}
    \SetColor{Black}
    \Line(269,-98)(296,-142)
    \SetWidth{1.3}
    \Line[dash,dashsize=5](283,-90)(283,-150)
    \SetWidth{1.0}
    \Line(269,-98)(269,-142)    
    \Line(313,-98)(313,-142)  
    \Line(269,-142)(313,-142)
    \Line(269,-128)(313,-98)
    \Line[arrow,arrowpos=0.5,arrowlength=1.75,arrowwidth=0.7,arrowinset=0.2](256,-98)(269,-98)
    \Line[arrow,arrowpos=0.5,arrowlength=1.75,arrowwidth=0.7,arrowinset=0.2](313,-98)(326,-98)
    \Line[arrow,arrowpos=0.5,arrowlength=1.75,arrowwidth=0.7,arrowinset=0.2](256,-142)(269,-142)
    \Line[arrow,arrowpos=0.5,arrowlength=1.75,arrowwidth=0.7,arrowinset=0.2](313,-142)(326,-142)

    \Text(258,-95)[lb]{\small{\Black{$p_{1}$}}}
    \Text(319,-95)[lb]{\small{\Black{$p_{1}$}}}
    \Text(258,-139)[lb]{\small{\Black{$p_{2}$}}}
    \Text(319,-139)[lb]{\small{\Black{$p_{2}$}}}
    
    \end{picture} }
    
    }

\newcommand{\masternineteen}{
	\scalebox{0.75}{
		\begin{picture}(85,55) (175,-148)
    \SetWidth{2.7}
    \Line(185,-98)(185,-142)    
    \Line(185,-142)(219,-142)
    \SetColor{Black}
    \SetWidth{1.3}
    \Line[dash,dashsize=5](200,-150)(238,-90)
    \SetWidth{1.0}
    \Line(219,-98)(219,-142)   
    \Line(185,-98)(219,-98)
    \Line(219,-98)(253,-98)    
    \Line(253,-98)(253,-142)    
    \Line(219,-142)(253,-142)    
 
    \Line[arrow,arrowpos=0.5,arrowlength=1.75,arrowwidth=0.7,arrowinset=0.2](172,-98)(185,-98)
    \Line[arrow,arrowpos=0.5,arrowlength=1.75,arrowwidth=0.7,arrowinset=0.2](253,-98)(266,-98)
    \Line[arrow,arrowpos=0.5,arrowlength=1.75,arrowwidth=0.7,arrowinset=0.2](172,-142)(185,-142)
    \Line[arrow,arrowpos=0.5,arrowlength=1.75,arrowwidth=0.7,arrowinset=0.2](253,-142)(266,-142)

    \Text(174,-95)[lb]{\small{\Black{$p_{1}$}}}
    \Text(259,-95)[lb]{\small{\Black{$p_{1}$}}}
    \Text(174,-139)[lb]{\small{\Black{$p_{2}$}}}
    \Text(259,-139)[lb]{\small{\Black{$p_{2}$}}}
    
    \end{picture} }
    
    }

\newcommand{\mastertwenty}{
	\scalebox{0.75}{
		\begin{picture}(65,55) (259,-148)
    \SetWidth{2.7}
    \SetColor{Black}
    \Line(269,-98)(313,-98)
    \Line(313,-98)(313,-120)  
    \SetWidth{1.3}
    \Line[dash,dashsize=5](283,-90)(283,-150)
    \SetWidth{1.0}
    \Line(269,-98)(269,-142)    
    \Line(313,-120)(313,-142)  
    \Line(269,-142)(313,-120)
    \Line(269,-120)(313,-142)
    \Line[arrow,arrowpos=0.5,arrowlength=1.75,arrowwidth=0.7,arrowinset=0.2](256,-98)(269,-98)
    \Line[arrow,arrowpos=0.5,arrowlength=1.75,arrowwidth=0.7,arrowinset=0.2](313,-98)(326,-98)
    \Line[arrow,arrowpos=0.5,arrowlength=1.75,arrowwidth=0.7,arrowinset=0.2](256,-142)(269,-142)
    \Line[arrow,arrowpos=0.5,arrowlength=1.75,arrowwidth=0.7,arrowinset=0.2](313,-142)(326,-142)

    \Text(258,-95)[lb]{\small{\Black{$p_{1}$}}}
    \Text(319,-95)[lb]{\small{\Black{$p_{2}$}}}
    \Text(258,-139)[lb]{\small{\Black{$p_{2}$}}}
    \Text(319,-139)[lb]{\small{\Black{$p_{1}$}}}
    
    \end{picture} }
    }

\newcommand{\mastertwentyone}{
	\scalebox{0.75}{
		\begin{picture}(65,55) (259,-148)
    \SetWidth{2.7}
    \SetColor{Black}
    \Line(269,-98)(313,-120)
    \Line(269,-98)(269,-142)    
    \SetWidth{1.3}
    \Line[dash,dashsize=5](300,-90)(300,-150)
    \SetWidth{1.0}
    \Line(313,-98)(313,-142)  
    \Line(269,-142)(313,-142)
    \Line(283,-142)(313,-98)
    \Line[arrow,arrowpos=0.5,arrowlength=1.75,arrowwidth=0.7,arrowinset=0.2](256,-98)(269,-98)
    \Line[arrow,arrowpos=0.5,arrowlength=1.75,arrowwidth=0.7,arrowinset=0.2](313,-98)(326,-98)
    \Line[arrow,arrowpos=0.5,arrowlength=1.75,arrowwidth=0.7,arrowinset=0.2](256,-142)(269,-142)
    \Line[arrow,arrowpos=0.5,arrowlength=1.75,arrowwidth=0.7,arrowinset=0.2](313,-142)(326,-142)

    \Text(258,-95)[lb]{\small{\Black{$p_{2}$}}}
    \Text(319,-95)[lb]{\small{\Black{$p_{1}$}}}
    \Text(258,-139)[lb]{\small{\Black{$p_{1}$}}}
    \Text(319,-139)[lb]{\small{\Black{$p_{2}$}}}
    
    \end{picture} }
    
    }
    
\newcommand{\mastertwentytwo}{
	\scalebox{0.75}{
		\begin{picture}(63,55) (92,-22)
		
			\SetWidth{2.7}
			\Line(123,27)(123,-17)
			\Arc(123,5)(22,0,90)
    			\SetWidth{1.0}
    			\SetColor{Black}
    			\Arc(123,5)(22,0,180)
    			\Arc(123,5)(22,180,360)
   			\SetWidth{1.3}
   			\Line[dash,dashsize=5](110,30)(135,-20)
   			\SetWidth{1.0}
    			\Line[arrow,arrowpos=0.5,arrowlength=1.75,arrowwidth=0.7,arrowinset=0.2](88,5)(101,5)
   			\Line[arrow,arrowpos=0.5,arrowlength=1.75,arrowwidth=0.7,arrowinset=0.2](145,5)(158,5)
  		    \Text(90,8)[lb]{\small{\Black{$p$}}}
    			\Text(153,8)[lb]{\small{\Black{$p$}}}
		\end{picture}	}
}    

\newcommand{\mastertwentythree}{
	\scalebox{0.75}{
		\begin{picture}(85,55) (175,-148)
    \SetWidth{2.7}
    \Line(219,-98)(219,-142)   
    \Line(185,-98)(219,-98)
    \SetColor{Black}
    \SetWidth{1.3}
    \Line[dash,dashsize=5](200,-150)(238,-90)
    \SetWidth{1.0}
    \Line(185,-98)(185,-142)    
    \Line(185,-142)(219,-142)
    \Line(219,-98)(253,-98)    
    \Line(253,-98)(253,-142)    
    \Line(219,-142)(253,-142)    
 
    \Line[arrow,arrowpos=0.5,arrowlength=1.75,arrowwidth=0.7,arrowinset=0.2](172,-98)(185,-98)
    \Line[arrow,arrowpos=0.5,arrowlength=1.75,arrowwidth=0.7,arrowinset=0.2](253,-98)(266,-98)
    \Line[arrow,arrowpos=0.5,arrowlength=1.75,arrowwidth=0.7,arrowinset=0.2](172,-142)(185,-142)
    \Line[arrow,arrowpos=0.5,arrowlength=1.75,arrowwidth=0.7,arrowinset=0.2](253,-142)(266,-142)

    \Text(174,-95)[lb]{\small{\Black{$p_{1}$}}}
    \Text(259,-95)[lb]{\small{\Black{$p_{1}$}}}
    \Text(174,-139)[lb]{\small{\Black{$p_{2}$}}}
    \Text(259,-139)[lb]{\small{\Black{$p_{2}$}}}
    
    \end{picture} }
    
    }

\newcommand{\mastertwentyfour}{
	\scalebox{0.75}{
		\begin{picture}(63,55) (92,-22)
		
			\SetWidth{2.7}
			\Line(123,27)(123,-17)
    			\SetWidth{1.0}
    			\SetColor{Black}
    			\Arc(123,5)(22,0,180)
    			\Arc(123,5)(22,180,360)
   			\SetWidth{1.3}
   			\Line[dash,dashsize=5](110,30)(135,-20)
   			\SetWidth{1.0}
    			\Line[arrow,arrowpos=0.5,arrowlength=1.75,arrowwidth=0.7,arrowinset=0.2](88,5)(101,5)
   			\Line[arrow,arrowpos=0.5,arrowlength=1.75,arrowwidth=0.7,arrowinset=0.2](145,5)(158,5)
  		    \Text(90,8)[lb]{\small{\Black{$p$}}}
    			\Text(153,8)[lb]{\small{\Black{$p$}}}
		\end{picture}	}
}    

\newcommand{\mastertwentyfive}{
	\scalebox{0.75}{
		\begin{picture}(65,60) (343,-81)
    \SetWidth{2.7}
    \Line(353,-33)(375,-66)
    \Line(353,-33)(353,-55)
    \SetColor{Black}
    \SetWidth{1.3}
    \Line[dash,dashsize=5](364,-30)(363,-80)
    \SetWidth{1.0}
    \Line(353,-55)(397,-55)
    \Line(353,-77)(397,-55)
    \Line(353,-33)(353,-77)    
    \Line[arrow,arrowpos=0.5,arrowlength=1.75,arrowwidth=0.7,arrowinset=0.2](340,-33)(353,-33)
    \Line[arrow,arrowpos=0.5,arrowlength=1.75,arrowwidth=0.7,arrowinset=0.2](340,-77)(353,-77)
    \Line[arrow,arrowpos=0.5,arrowlength=1.75,arrowwidth=0.7,arrowinset=0.2](393,-55)(410,-55)

    \Text(407,-52)[lb]{\small{\Black{$p$}}}
    \Text(342,-30)[lb]{\small{\Black{$p_{1}$}}}
    \Text(342,-74)[lb]{\small{\Black{$p_{2}$}}}

		\end{picture}  }
}

\newcommand{\mastertwentysix}{
	\scalebox{0.75}{
		\begin{picture}(85,55) (175,-148)
    \SetWidth{2.7}
    \Line(219,-98)(219,-142)   
    \SetColor{Black}
    \SetWidth{1.3}
    \Line[dash,dashsize=5](238,-150)(200,-90)
    \SetWidth{1.0}
    \Line(185,-98)(219,-98)
    \Line(185,-98)(185,-142)    
    \Line(185,-142)(219,-142)
    \Line(219,-98)(253,-98)    
    \Line(253,-98)(253,-142)    
    \Line(219,-142)(253,-142)    
 
    \Line[arrow,arrowpos=0.5,arrowlength=1.75,arrowwidth=0.7,arrowinset=0.2](172,-98)(185,-98)
    \Line[arrow,arrowpos=0.5,arrowlength=1.75,arrowwidth=0.7,arrowinset=0.2](253,-98)(266,-98)
    \Line[arrow,arrowpos=0.5,arrowlength=1.75,arrowwidth=0.7,arrowinset=0.2](172,-142)(185,-142)
    \Line[arrow,arrowpos=0.5,arrowlength=1.75,arrowwidth=0.7,arrowinset=0.2](253,-142)(266,-142)

    \Text(174,-95)[lb]{\small{\Black{$p_{1}$}}}
    \Text(259,-95)[lb]{\small{\Black{$p_{1}$}}}
    \Text(174,-139)[lb]{\small{\Black{$p_{2}$}}}
    \Text(259,-139)[lb]{\small{\Black{$p_{2}$}}}
    
    \end{picture} }
    
    }
    
\newcommand{\mastertwentyseven}{
	\scalebox{0.75}{
		\begin{picture}(85,55) (175,-148)
    \SetWidth{2.7}
    \Line(219,-98)(219,-142)   
    \SetColor{Black}
    \SetWidth{1.3}
    \Line[dash,dashsize=5](238,-150)(200,-90)
    \SetWidth{1.0}
    \Line(185,-98)(219,-98)
    \Line(185,-98)(185,-142)    
    \Line(185,-142)(219,-142)
    \Line(219,-98)(253,-98)    
    \Line(253,-98)(253,-142)    
    \Line(219,-142)(253,-142)    
 
    \Line[arrow,arrowpos=0.5,arrowlength=1.75,arrowwidth=0.7,arrowinset=0.2](172,-98)(185,-98)
    \Line[arrow,arrowpos=0.5,arrowlength=1.75,arrowwidth=0.7,arrowinset=0.2](253,-98)(266,-98)
    \Line[arrow,arrowpos=0.5,arrowlength=1.75,arrowwidth=0.7,arrowinset=0.2](172,-142)(185,-142)
    \Line[arrow,arrowpos=0.5,arrowlength=1.75,arrowwidth=0.7,arrowinset=0.2](253,-142)(266,-142)

    \Text(174,-95)[lb]{\small{\Black{$p_{1}$}}}
    \Text(259,-95)[lb]{\small{\Black{$p_{2}$}}}
    \Text(174,-139)[lb]{\small{\Black{$p_{2}$}}}
    \Text(259,-139)[lb]{\small{\Black{$p_{1}$}}}
    
    \end{picture} }
    
    }    
    
\newcommand{\mastertwentyeight}{
	\scalebox{0.75}{ 
		\begin{picture}(65,60) (343,-148)
    \SetWidth{2.7}
    \Line(369,-98)(397,-142)
    \Line(397,-142)(397,-115)
    \SetColor{Black}
    \SetWidth{1.3}
    \Line[dash,dashsize=5](389,-90)(389,-150)
    \SetWidth{1.0}
    \Line(353,-98)(397,-98)
    \Line(353,-98)(353,-142)    
    \Line(397,-98)(397,-115)
    \Line(353,-142)(397,-115)
    \Line[arrow,arrowpos=0.5,arrowlength=1.75,arrowwidth=0.7,arrowinset=0.2](340,-98)(353,-98)   
    \Line[arrow,arrowpos=0.5,arrowlength=1.75,arrowwidth=0.7,arrowinset=0.2](397,-98)(410,-98)
    \Line[arrow,arrowpos=0.5,arrowlength=1.75,arrowwidth=0.7,arrowinset=0.2](340,-142)(353,-142)   
    \Line[arrow,arrowpos=0.5,arrowlength=1.75,arrowwidth=0.7,arrowinset=0.2](397,-142)(410,-142)
    
    \Text(342,-95)[lb]{\small{\Black{$p_{2}$}}}
    \Text(403,-95)[lb]{\small{\Black{$p_{2}$}}}
    \Text(342,-139)[lb]{\small{\Black{$p_{1}$}}}
    \Text(403,-139)[lb]{\small{\Black{$p_{1}$}}}

		\end{picture}  }
}
    
\newcommand{\mastertwentynine}{
	\scalebox{0.75}{ 
		\begin{picture}(65,60) (343,-148)
    \SetWidth{2.7}
    \Line(369,-98)(397,-142)
    \Line(397,-142)(397,-115)
    \SetColor{Black}
    \SetWidth{1.3}
    \Line[dash,dashsize=5](389,-90)(389,-150)
    \SetWidth{1.0}
    \Line(353,-98)(397,-98)
    \Line(353,-98)(353,-142)    
    \Line(397,-98)(397,-115)
    \Line(353,-142)(397,-115)
    \Line[arrow,arrowpos=0.5,arrowlength=1.75,arrowwidth=0.7,arrowinset=0.2](340,-98)(353,-98)   
    \Line[arrow,arrowpos=0.5,arrowlength=1.75,arrowwidth=0.7,arrowinset=0.2](397,-98)(410,-98)
    \Line[arrow,arrowpos=0.5,arrowlength=1.75,arrowwidth=0.7,arrowinset=0.2](340,-142)(353,-142)   
    \Line[arrow,arrowpos=0.5,arrowlength=1.75,arrowwidth=0.7,arrowinset=0.2](397,-142)(410,-142)
    
    \Text(342,-95)[lb]{\small{\Black{$p_{1}$}}}
    \Text(403,-95)[lb]{\small{\Black{$p_{2}$}}}
    \Text(342,-139)[lb]{\small{\Black{$p_{2}$}}}
    \Text(403,-139)[lb]{\small{\Black{$p_{1}$}}}

		\end{picture}  }
}

\newcommand{\masterthirty}{
	\scalebox{0.75}{
		\begin{picture}(65,55) (259,-148)
    \SetWidth{2.7}
    \SetColor{Black}
    \Line(269,-120)(313,-142)
    \Line(313,-120)(313,-142)  
    \SetWidth{1.3}
    \Line[dash,dashsize=5](283,-90)(283,-150)
    \SetWidth{1.0}
    \Line(269,-98)(269,-142)    
    \Line(313,-120)(313,-142)  
    \Line(269,-142)(313,-120)
    \Line(269,-98)(313,-98)
    \Line(313,-98)(313,-120)
    \Line[arrow,arrowpos=0.5,arrowlength=1.75,arrowwidth=0.7,arrowinset=0.2](256,-98)(269,-98)
    \Line[arrow,arrowpos=0.5,arrowlength=1.75,arrowwidth=0.7,arrowinset=0.2](313,-98)(326,-98)
    \Line[arrow,arrowpos=0.5,arrowlength=1.75,arrowwidth=0.7,arrowinset=0.2](256,-142)(269,-142)
    \Line[arrow,arrowpos=0.5,arrowlength=1.75,arrowwidth=0.7,arrowinset=0.2](313,-142)(326,-142)

    \Text(258,-95)[lb]{\small{\Black{$p_{1}$}}}
    \Text(319,-95)[lb]{\small{\Black{$p_{2}$}}}
    \Text(258,-139)[lb]{\small{\Black{$p_{2}$}}}
    \Text(319,-139)[lb]{\small{\Black{$p_{1}$}}}
    
    \end{picture} }
    }

\begin{document}

\title{Double-real corrections at \oaas to single gauge boson production}

\author[a]{Roberto Bonciani,}
\author[b]{Federico Buccioni,}
\author[c]{Roberto Mondini,}
\author[d]{Alessandro Vicini\,}

\affiliation[a]{Dipartimento di Fisica, Universit\`a di Roma ``La Sapienza'' and INFN Sezione di Roma,\\Piazzale Aldo Moro 5, I-00185 Roma, Italy}
\affiliation[b]{Physik-Institut, Universit\"at Z\"urich,\\Winterthurerstrasse 190, CH-8057 Z\"urich, Switzerland}
\affiliation[c]{Department of Physics, University at Buffalo\\ The State University of New York, Buffalo 14260 USA}
\affiliation[d]{Tif lab, Dipartimento di Fisica, Universit\`a di Milano, and INFN, Sezione di Milano, \\Via A. Celoria 16, 20133, Milano, Italy}

\emailAdd{roberto.bonciani@roma1.infn.it}
\emailAdd{buccioni@physik.uzh.ch}
\emailAdd{rmondini@buffalo.edu}
\emailAdd{alessandro.vicini@mi.infn.it}

\abstract{ 
We consider the \oaas corrections to single on-shell gauge boson production at hadron colliders.
We concentrate on the contribution of all the subprocesses where the gauge boson is accompanied
by the emission of two additional real partons and we evaluate the corresponding total cross sections.
The latter are divergent quantities, because of soft and collinear emissions, and are expressed as Laurent
series in the dimensional regularization parameter. The total cross sections are evaluated by means
of reverse unitarity, i.e.~expressing the phase-space integrals in terms of two-loop forward box
integrals with cuts on the final state particles. The results are reduced to a combination of
Master Integrals, which eventually are evaluated in terms of Generalized Polylogarithms. The
presence of internal massive lines in the Feynman diagrams, due to the exchange of electroweak
gauge bosons, causes the appearance of 14 Master Integrals which were not previously known 
in the literature and have been evaluated via Differential Equations.}

\preprint{TIF-UNIMI-2016-9}
\maketitle

\flushbottom


\section{Introduction}
\label{sec:intro}
The electroweak (EW) production of a pair of leptons, each with large transverse momentum, in hadron-hadron
collisions, known as Drell-Yan (DY) process \cite{Drell:1970wh}, is one of the historical testgrounds of
perturbative quantum chromodynamics (QCD). The charged-current (CC) and the neutral-current (NC) processes
are relevant not only to put stringent constraints on the proton parton density functions (PDFs), but also
to perform high-precision measurements of fundamental EW parameters such as the masses and decay widths of
the $W$ and $Z$ bosons or the EW mixing angle. Furthermore, they represent an important background to many
new physics searches (for a recent review see Ref.~\cite{Mangano:2015ejw}). All these studies require
precise calculations of higher-order radiative effects and a corresponding implementation in simulation
tools that can be used to analyze the experimental data (for a discussion on the status of simulation codes
for DY processes see Ref. \cite{Alioli:2016fum}). 

In specific cases like the weak mixing angle or the $W$ boson mass measurements, with a final precision
goal in the $1-2\,\cdot 10^{-4}$ range, all the elements entering the theoretical predictions need to be
scrutinized. For instance, in the $W$ mass case it is necessary to assess the uncertainty due to a still
inaccurate representation of non-perturbative QCD effects parameterized in the proton PDFs
\cite{Bozzi:2011ww,Bozzi:2015hha,Bozzi:2015zja} or in the models present in the QCD Parton Shower, or
stemming from the incomplete knowledge of higher-order perturbative QCD, EW, or mixed QCD$\times$EW
contributions \cite{mwew}. These measurements require an excellent control not only on the absolute
normalization of the observables, but also on their shape. In this respect a major role is played by final
state QED radiation as well as by the interplay of the latter with QCD corrections. A detailed study of
this interplay requires the exact evaluation of the next order of perturbative corrections, namely those of
\oaas, which is not available yet.

A kinematical limit where the EW corrections play an important role is the so-called Sudakov regime, when
the observables are characterized by values of the kinematical invariants (large invariant/transverse
masses or large transverse momenta) much larger than the gauge boson masses, yielding large logarithmic
factors. The EW \oa\, corrections are responsible for the first large correction of this kind
\cite{Dittmaier:2001ay,Baur:2004ig,Baur:2001ze,Chiesa:2013yma, Campbell:2016dks}, but it has been shown
\cite{Kuhn:2005az,Kuhn:2007cv} that also ${\cal O}(\alpha^2)$ terms may still be sizeable. The \oaas
corrections represent the first QCD correction to these large EW factors and their explicit evaluation is
thus needed to get the predictions in the Sudakov regime under control.

The DY cross sections can be expressed as a double perturbative expansion in the strong and
electromagnetic couplings, respectively $\alpha_s$ and $\alpha$, which can be formally written as
follows, with all the phase-space factors understood in the definition of the coefficients $d\sigma$:
\be
d\sigma=d\sigma_0\,+\,
\alpha\, d\sigma_{\alpha}\, +\, 
\alpha^2\, d\sigma_{\alpha^2}\, +\dots +\,
\alpha_s\, d\sigma_{\alpha_s}\, +\, 
\alpha_s^2 d\sigma_{\alpha_s^2}\, +\dots +\,
\alpha \alpha_s\, d\sigma_{\alpha \alpha_s}\, + \dots
\label{eq:xsectot}
\ee
In Eq.~(\ref{eq:xsectot}) we recognize terms purely due to the strong or the EW corrections, and also
terms where the mixed combined effect of the two interactions is present. QCD corrections to the total
cross section have been computed at  next-to-leading-order (NLO) in Ref.~\cite{Altarelli:1979ub} and
at next-to-next-to-leading-order (NNLO) in Refs.~\cite{Hamberg:1990np,vanNeerven:1991gh}. Recently the
next-to-next-to-next-to-leading-order (N3LO) corrections to the Higgs production gluon fusion process
became available \cite{Anastasiou:2014vaa,Anastasiou:2016cez}, allowing in turn the estimate of the
N3LO corrections in the soft approximation also for EW gauge boson production
\cite{Ahmed:2014cla,Catani:2014uta}. The NLO-EW corrections have been computed separately for the
CC-DY in Refs.~\cite{Dittmaier:2001ay,Baur:2004ig} and for the NC-DY in Ref.~\cite{Baur:2001ze}.
Preliminary steps towards the evaluation of the full NNLO-EW corrections have been accomplished with
the discussion of the renormalization of the full two-loop amplitudes
\cite{Degrassi:2003rw,Actis:2006ra,Actis:2006rb,Actis:2006rc}.

The evaluation of the differential distributions of the final-state products is available in the codes
described in Refs. \cite{Catani:2009sm, Gavin:2010az, Gavin:2012sy,Boughezal:2016wmq} and in those of
Refs.~\cite{Dittmaier:2001ay,Baur:2004ig,Baur:2001ze,CarloniCalame:2006zq,CarloniCalame:2007cd,Arbuzov:2005dd,Arbuzov:2007db} 
respectively with NNLO-QCD and NLO-EW accuracy for the cross section. The inclusion of subsets of dominant
higher-order corrections, going beyond the fixed-order description of Eq.~(\ref{eq:xsectot}), 
has been implemented in many codes that match exact 
matrix elements with a Parton Shower (PS). Focusing on the strong interactions,
Refs.~\cite{Alioli:2008gx,Frixione:2002ik} provide the matching with (NLO+PS)-QCD accuracy,
Refs.~\cite{Karlberg:2014qua,Hoeche:2014aia} with (NNLO+PS)-QCD accuracy, and
Ref.~\cite{Alioli:2015toa} performs the matching in the framework of effective theories. 
On the EW side, the consistent matching of fixed- and all-orders effects is performed for instance in Refs.~\cite{CarloniCalame:2006zq,CarloniCalame:2007cd,Placzek:2013moa}. The
resummation to all orders of terms enhanced by logarithms of the lepton-pair transverse momentum is
available with next-to-next-to-leading-logarithm (NNLL) accuracy in the codes of
Refs.~\cite{Balazs:1995nz,Balazs:1997xd,Catani:2015vma}.

The full set of exact \oaas corrections to the total cross section is not available yet due to
 difficulties in the evaluation of the relevant virtual and phase-space integrals and only subsets of corrections are available.
In Ref.~\cite{Kilgore:2011pa} the authors considered the QCD$\times$QED contributions to the production 
of a lepton pair in the $q \bar{q}$ channel. 
The \oaas corrections to the decays of $Z$ and $W$ bosons have been
computed respectively in Refs.~\cite{Czarnecki:1996ei} and \cite{Kara:2013dua}.
In Ref.~\cite{Kotikov:2007vr} the mixed two-loop 
corrections to the form factors for the production of a $Z$ boson have been presented.
Very recently, in Ref.~\cite{Bonciani:2016ypc} 
the authors evaluated all the two-loop virtual master integrals contributing to the \oaas partonic processes of 
production of a $l^-l^+$ or $l^-\overline{\nu}$ pair. Moreover, the Altarelli-Parisi splitting functions
have been computed with \oaas accuracy in Ref.~\cite{deFlorian:2015ujt} thus allowing for a consistent
subtraction of all the initial-state collinearly divergent terms.
NLO-EW corrections to $V$+jet  and NLO-QCD corrections to $V+\gamma$
final states have been computed in Refs. \cite{Denner:2009gj,Denner:2011vu,Denner:2014bna,Denner:2015fca}, including the leptonic decay of the vector boson.
These results are based on the matrix elements describing the production of a gauge boson (and its subsequent decay) accompanied by one  additional hard parton; they therefore include terms of $\oaas$, but are divergent in the limit of vanishing vector boson transverse momentum.

The absence of an exact calculation of the \oaas corrections to the DY processes has been partially
compensated, in the past, by the use of different approximations: the restriction, for the EW
corrections, to the subset of final-state QED corrections allowed the factorized combination of QCD
and QED corrections \cite{Cao:2004yy,Adam:2008pc,Balossini:2009sa}; an additive recipe for the
NNLO-QCD and NLO-EW results has been proposed in Ref.~\cite{Li:2012wna}; the combination of NLO-QCD 
and NLO-EW matrix elements, consistently matched with (QCD+QED)-PS, has been described in
Refs.~\cite{Bernaciak:2012hj,Barze:2012tt,Barze':2013yca}.

A calculation of the \oaas corrections 
to the DY processes near the resonance region has been
performed in Refs.~\cite{Dittmaier:2014qza, Dittmaier:2014koa,Dittmaier:2015rxo}. 
The calculation was done in the pole approximation, 
namely retaining all the leading terms contributing to the $W$ ($Z$) 
boson resonance. 
Among the various contributions that the authors analyze, 
the non-factorizable terms due to soft-photon exchange 
between the production and decay processes result to be negligible 
for current phenomenological purposes. 
The conclusion is, therefore, that the treatment of the process in the
resonance region, 
which effectively decouples the production from the decay processes, 
is sufficient for the level of accuracy needed by current experiments.
In particular, the factorizable contributions due to initial-state QCD with final-state QED corrections (emission of photons from the final state) 
turn out to be the most phenomenologically relevant. 
A comparison is in progress between
these analytical results and the approximation of the mixed QCD$\times$EW effects implemented in the  Shower Monte Carlo of 
Refs.~\cite{Barze:2012tt,Barze':2013yca}. 
However, 
in the analysis of Refs.~\cite{Dittmaier:2014qza,Dittmaier:2014koa,Dittmaier:2015rxo}
the double corrections to the initial state are not calculated; 
they are estimated to be negligible. 

In this paper we face the problem of the exact evaluation of the \oaas corrections to the total cross section for the production of an on-shell weak boson ($W$ or $Z$).
%
The importance of this calculation is two-fold. 
From one side, an exact calculation can give a solid
ground and a quantitative check to the estimation of Refs.~\cite{Dittmaier:2014qza,Dittmaier:2014koa,Dittmaier:2015rxo}. 
From the other side, individual pieces of our calculation can be
important for guiding and checking other ingredients necessary for the treatment of more exclusive observables, such as the gauge boson rapidity distribution, or for the calculation of the mixed QCD$\times$EW infrared subtraction terms.

The evaluation of the \oaas corrections to the production of an on-shell vector boson from $q \bar{q}$ initial-state annihilation 
requires the study of four different subprocesses, with 0, 1, or 2 additional partons (gluon, quark, photon) in the final state.
The respective contributions to the total cross section for on-shell gauge boson production
are obtained by computing the two-loop virtual corrections to the lowest-order amplitude
or by integrating the relevant squared matrix elements over the full phase space of the additional partons.
In the latter cases we adopt a technique called reverse unitarity,
developed for the evaluation of the total cross section for Higgs production  
\cite{Anastasiou:2002yz,Anastasiou:2012kq,Anastasiou:2013srw}. 
The standard phase-space integration is turned into the evaluation of ``cut'' two-loop integrals, namely with the additional condition that the final state particles fulfill the on-shell relation. 
Integrals with up to three internal massive lines appear in the calculation; some of them were not previously available in the literature and required a dedicated study.
The calculation of the total cross section is done 
by reducing the dimensionally regularized scalar integrals coming from the squared amplitude to a set of Master Integrals (MIs) via integration-by-parts (IBP) identities \cite{Tkachov:1981wb,Chetyrkin:1981qh,Studerus:2009ye,
vonManteuffel:2012np,Smirnov:2008iw,Smirnov:2013dia,Smirnov:2014hma,Lee:2012cn}. The MIs are then computed using the differential equations method 
\cite{Kotikov:1990kg,Kotikov:1991pm,Bern:1993kr,Remiddi:1997ny,Gehrmann:1999as,Argeri:2007up,Henn:2014qga,Henn:2013pwa,Gehrmann:2014bfa,Argeri:2014qva,Lee:2014ioa}. 
Their expressions in terms of
Harmonic Polylogarithms (HPLs) \cite{Remiddi:1999ew} and their generalizations \cite{Goncharov:polylog,Goncharov2001,Goncharov2007,Aglietti:2004tq,Bonciani:2010ms} can be found in Refs.~\cite{Aglietti:2003yc,Aglietti:2004ki,Bonciani:2016ypc}. 

In this paper we focus on the evaluation of the double-real contribution to the \oaas\, corrections to the total cross section for on-shell single gauge boson production. 
We consider all possible channels involved at this order in perturbation theory. This includes $q \bar{q}$-initiated process as well as $qg$-, $q \gamma$-, and $\gamma g$-initiated processes. 
Since the $W$ boson is charged, it can emit a photon. 
As a consequence, we need to consider diagrams in which a massive propagator is present along with the massive cut external particle. 
While the diagrams relevant for $Z$ production give rise to MIs
that were already computed in the literature, 
those for $W$ production introduce additional MIs that are presented here, to our knowledge, for the first time. 
The cross sections corresponding to the channels under consideration are expressed as Laurent series of $\eps = (4-d)/2$, where $d$ is the space-time dimension. 
The coefficients of the series are given in terms of generalized polylogarithms up to weight 3.

The paper is organized as follows. 
In section \ref{sec:proc} we present the partonic processes under consideration in more detail and we define their cross sections as linear combinations of a limited number of MIs. 
Moreover, we briefly discuss the prescription of the $\gamma_5$ matrix employed in this computation.
In section \ref{sec:eval} we describe how the MIs are computed. In particular, we focus on the evaluation of the soft limits of the MIs, 
which are used to fix the boundary conditions of the differential equations. 
In section \ref{sec:xsec} we present the analytic expressions of the partonic cross sections of all the relevant processes. 
In section \ref{conclu} we draw our conclusions.
In appendix \ref{sec:appMI} we provide the reader with the analytic expressions of the MIs and with the expressions of the soft limits
with exact dependence on the regulator $\eps$. 
The complete set of cross sections and MIs is also given in an ancillary file that we include in the arXiv submission.


\section{Partonic subprocesses}
\label{sec:proc}
\subsection{Contributions of \oaas to the total cross section}
According to the collinear factorization theorem,
the inclusive total cross section for the production of a single gauge boson in hadron-hadron collisions can be written as
\be
\sigma_{tot}(h_1 h_2\to V+X)=
\sum_{i,j} \int dx_1 dx_2\,
f_{i,1}(x_1,\muf) f_{j,2}(x_2,\muf)\,
\hat\sigma_{tot}(ij\to V+X)\, ,
\ee
where
$V=W^{\pm},Z$,
the sum over $i$ and $j$ runs over all partons present in the proton (quark, gluons, photons),
$f_{i,h}$ are the proton PDFs for a parton $i$ inside hadron $h$,
and
each partonic cross section $\hat\sigma_{tot}(ij\to V+X)$ admits a double perturbative expansion 
as depicted in Eq.~\eqref{eq:xsectot}.
The lowest-order non-vanishing contribution to inclusive single gauge boson production is due to
quark-antiquark annihilation,  with a cross section of ${\cal O}(G_\mu)$ ($G_\mu$ is the Fermi
constant). At higher orders,  for a subprocess initiated by a given pair of partons, one has to
consider the virtual corrections to the lower-order amplitudes as well as the contribution of the
radiative processes  with additional emitted partons in the final state. The cancellation of the soft
infrared divergences occurs after the combination of these different partonic cross sections with the
same initial state. For instance, in the case of \oaas corrections to quark-antiquark annihilation we
have four, separately divergent contributions:
\be
\hat\sigma_\aas(q\bar q \!\to \! V\!+\!X)=
\hat\sigma_\aas^{VV}(q\bar q\!\to \!V)+
\hat\sigma_\aas^{VR}(q\bar q\!\to\! Vg)+
\hat\sigma_\aas^{RV}(q\bar q\!\to\! V\gamma)+
\hat\sigma_\aas^{RR}(q\bar q\!\to\! V\gamma g) ,
\label{eq:contrqqbar}
\ee
with the superscripts $a$ and $b$ in $\sigma_\aas^{ab}$ representing
the correction due to a virtual (V) or real (R) exchange in the EW or in the strong interactions respectively.
In Eq.~\eqref{eq:contrqqbar} 
the sum is free of soft IR divergences and
the inclusion of initial-state collinear subtraction terms makes eventually the result IR finite.
Moreover, at a given higher perturbative order,
more initial states with different combinations of partons $ij$ have to be considered. 
Focusing on the \oaas contributions, we need to include processes initiated by $qg$: 
\be
\hat\sigma_\aas(qg\to V+X)=
\hat\sigma_\aas^{VR}(qg\to q V)+
\hat\sigma_\aas^{RR}(qg\to q V\gamma),
\label{eq:contrqg}
\ee
initiated by $q\gamma$:
\be
\hat\sigma_\aas(q\gamma\to V+X)=
\hat\sigma_\aas^{RV}(q\gamma\to q V)+
\hat\sigma_\aas^{RR}(q\gamma\to q V g),
\label{eq:contrqgamma}
\ee
and by $g\gamma$:
\be
\hat\sigma_\aas(g\gamma\to V+X)=
\hat\sigma_\aas^{RR}(g\gamma\to V q \bar q) \, .
\label{eq:contrggamma}
\ee
In this work we study the partonic subprocesses that contribute at \oaas to the inclusive hadronic cross section
for the production of a gauge boson with two additional partons in the final state (double-real corrections), i.e.~all the processes labeled by
$\hat\sigma^{RR}$ in Eqs.~\eqref{eq:contrqqbar}-\eqref{eq:contrggamma}:
\bea
&&q_i(p_1)\, \bar q_j(p_2) \to W^{\pm}(p_3)\, g(p_4)\, \gamma(p_5) \, ,  \label{proc1}\\
&&q_i(p_1)\, g(p_2) \to W^{\pm}(p_3)\, q_j(p_4)\, \gamma(p_5)\, , \label{proc2}\\
&&q_i(p_1)\, \gamma(p_2) \to W^{\pm}(p_3)\, g(p_4)\, q_j(p_5) \, ,  \label{proc3}\\
&&\gamma(p_1)\, g(p_2) \to W^{\pm}(p_3)\, q_j(p_4)\, \bar q_i(p_5) \, ,  \label{proc4}\\
&&q_i(p_1)\, \bar q_i(p_2) \to Z(p_3)\, g(p_4) \, \gamma(p_5) \, ,  \label{proc5}\\
&&q_i(p_1)\, g(p_2) \to Z(p_3)\, q_i(p_4)\, \gamma(p_5) \, ,  \label{proc6}\\
&&q_i(p_1)\, \gamma(p_2) \to Z(p_3)\, g(p_4)\, q_i(p_5) \, ,  \label{proc7}\\
&&\gamma(p_1)\, g(p_2) \to Z(p_3)\, q_i(p_4)\, \bar q_i(p_5) \label{proc8} \, .
\eea
We note that the squared matrix elements of processes (\ref{proc2})-(\ref{proc4}) and
(\ref{proc6})-(\ref{proc8}) can be obtained by crossing those of processes (\ref{proc1}) and
(\ref{proc5}) respectively. However, in the evaluation of their total cross sections  new MIs, absent
in the first two cases, appear, making a dedicated calculation necessary.

\subsection{Treatment of $\gamma_5$}

The squared matrix element of each subprocess, averaged over initial spin polarizations and color and
summed over final spin polarizations and color, must be computed in an arbitrary number of dimensions
$d=4-2\varepsilon$, in order to include all the finite contributions due to the interplay of the
squared amplitude with the divergent phase-space integration treated in dimensional regularization.

In this respect, to perform our calculation we need to adopt a prescription for the manipulation of the
Dirac matrix $\gamma_5$, as it is not defined in a non-integer number of dimensions. Therefore, in the
present work we consider the proposal of Ref.~\cite{Chanowitz:1979zu}, and take $\gamma_5$
anticommuting with all the other $\gamma^\mu$ matrices in arbitrary $d$ dimensions. Before the
evaluation of the traces, the product of Dirac $\gamma$ matrices is rearranged by shifting all the
$\gamma_5$ to the utmost right position, using the anticommuting property. We do not rely on the
possibility of a cyclic permutation of the matrices inside the trace,  because under the assumption of
anticommuting $\gamma_5$ in all $d$ dimensions the cyclicity property of the trace does not hold.
Moreover, since there are four independent momenta in the process, it is possible to saturate all the
indices of a Levi-Civita tensor resulting from the computation of the traces and thus yielding
non-vanishing factors. These terms containing Levi-Civita tensors are responsible for the two following
problems, after evaluation of the traces: the presence of gauge-dependent terms, when the polarization
sum is done with an arbitrary gauge vector, and the presence of purely imaginary terms out of a squared
matrix element, which should obviously be real-valued. The solution is found,  consistently with the
prescription of Ref.~\cite{Chanowitz:1979zu},  by promoting also the Schouten identity to be valid in
an arbitrary number $d$ of dimensions; all the problematic terms thus exactly cancel.

\subsection{Definition of the total cross section and Reverse Unitarity}

We define the total partonic cross sections of the processes under consideration as:
\bea
\hat\sigma_{12\to V45}(z,d)
&=&
\frac{z}{2 M_{\smallv}^2}
\int \rmd \Phi_{3} \,\overline{\left|{\cal M}\right|^2} \, , \label{eq:xsec} \\
\rmd \Phi_{3} &=& \frac{\rmd^{d}p_{4}\rmd^{d}p_{5}}{(2\pi)^{2d-3}} \, \delta(p^2_{4}) \, \delta(p^2_{5}) \, \delta\pts{\pts{p_{1}+p_{2} - p_{4}-p_{5}}^2-M^2_{\smallv}} \, , 
\eea
where $z=M_{\smallv}^2/\hat s$ is the ratio between the gauge boson mass squared and the partonic
center-of-mass energy squared and we conventionally assign the momentum $p_3$ to the massive gauge
boson. The Reverse Unitarity (RU) technique relies on the remark that, in terms of distributions, the
following replacement (Cutkosky rule) holds
\be
\delta(p^2-m^2) 
\to 
\frac{1}{2\pi i}
\left(
\frac{1}{p^2-m^2+i\eta}
-
\frac{1}{p^2-m^2-i\eta}
\right).
\ee
The phase-space measure of each final state particle can thus be rewritten as the difference of two 
propagators with opposite prescriptions for their imaginary part (with $\eta$ an infinitesimal positive
real number). The integral over the full phase space of the two additional partons,  necessary to
compute the total cross section, is transformed into the evaluation of the imaginary part of two-loop
integrals with the additional constraint that lines corresponding to the final-state particles are cut,
i.e. are on-shell (optical theorem). The calculation of the total cross section of processes
(\ref{proc1})-(\ref{proc8}) can therefore be accomplished by means of the techniques developed for the
study of virtual corrections.

After computing the squared amplitude and applying the Cutkosky rule, the phase-space integral of
Eq.~\eqref{eq:xsec} consists of a very large number of terms. Most of these terms, however, are not
independent. By means of algebraic relations, Lorentz (LI) and IBP identities
(in our case implemented in the codes {\tt Reduze} \cite{Studerus:2009ye,vonManteuffel:2012np} and {\tt
FIRE} \cite{Smirnov:2008iw,Smirnov:2013dia,Smirnov:2014hma}), it is possible to simplify the sum of
these phase-space integrals and express it as a combination of a limited number of irreducible MIs. 
For the processes under consideration, the number of the independent MIs  that eventually have to be
explicitly computed is of ${\cal O}(10)$. The expression of the total cross section of a given process
$X$ can therefore be cast as:
\be
\hat\sigma(X) 
=
\sum_{i}
c_i^{X}(z,d) I_i(z,d) \, ,
\label{eq:structureMI}
\ee
where the coefficients $c^X_i$ are rational functions and are process dependent. The cross section is a
combination of MIs $I_i$, whose precise number and expressions depend on the process and on some
choices applied in the reduction procedure. In our case, the total partonic cross sections of the
processes (\ref{proc1})-(\ref{proc8}) have been expressed in terms of   $(11,13,13,11,7,9,9,7)$ MIs
respectively,  with a total of 30 distinct integrals, of which 16 already known in the literature and
14 new. In Section \ref{sec:eval} we discuss the techniques developed to compute the new MIs and in
Appendix \ref{sec:appMI} we list the explicit expressions of all the new integrals (as well as the
expressions of the others for completeness), written in terms of HPLs.

The total cross sections of the processes (\ref{proc1})-(\ref{proc8}) are IR divergent. In dimensional
regularization the highest-order singularity can be at most an $\eps^{-4}$ pole due to the simultaneous
soft and collinear divergences of both additional partons (e.g. photon and gluon in the $q\bar
q$-initiated process). The rational coefficients and the MIs in Eq.~(\ref{eq:structureMI}) depend in a
non-trivial way on the regularization parameter $\eps$. 
The explicit expressions of the cross sections are obtained by expanding both in powers of $\eps$,
keeping all the terms of the product that are non-vanishing in the limit $\eps\to 0$. The total cross
sections can therefore be written as Laurent series in $\eps$:
\be
\hat\sigma(X) 
=
\sum_{i=-4}^0
\eps^i P_i^{X}(z)
\,+\,{\cal O}(\eps) \, .
\label{eq:structureEPS}
\ee
We remark that in the $q\bar q$-initiated processes, in order to extract the soft singularity $z\to1$
(thus obtaining the $\eps^{-4}$ pole),  the following identity is used:
\begin{align}
(1-z)^{-1-4\eps} = -\frac{\delta(1-z)}{4\eps} + \sum_{n=0}^\infty \frac{(-4\eps)^n}{n!} \Big(\frac{\log^n(1-z)}{1-z}\Big)_+
\end{align}
with the so-called ``plus'' distribution defined as
\begin{align}
\int_0^1 \mathrm{d}x \, \Big(\frac{\log^n(x)}{x}\Big)_{\!+} f(x) = \int_0^1 \mathrm{d}x \, \frac{\log^n(x)}{x} \big[f(x)-f(0)\big] \, .
\end{align}

\begin{figure}[t]
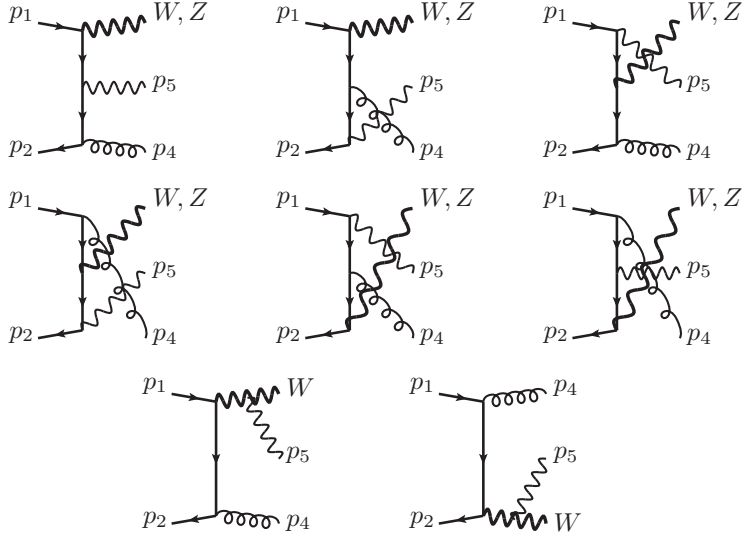

\begin{center}
\amplitudescollection 
\end{center}
\caption{
The Feynman diagrams that contribute to the probability amplitude of the processes $q\bar q\to Z\gamma
g$ and $q\bar q'\to W\gamma g$. Diagrams in the first two rows are common to both processes,  whereas
the two at the bottom are typical of $q\bar q'\to W\gamma g$ because the photon only couples to charged
particles. The Feynman diagrams contained in this article were drawn with \texttt{Jaxodraw}
\cite{Binosi:2003yf}. \label{fig:feynman-diagrams}
}
\end{figure}

Lastly, we note that the assumption that the final-state $W$ boson is on-shell yields additional IR
soft divergences with respect to the off-shell case. The production of a $W$ boson differs with respect
to the case of a $Z$ boson because of its electric charge:  since a photon can be radiated off each
charged leg,  in the case of $W$ production the amplitude receives a contribution from additional
Feynman diagrams. In the case of quark-antiquark annihilation, the additional Feynman diagrams are
those in the last row of Fig.~\ref{fig:feynman-diagrams}. From the point of view of strong
interactions, the amplitude for $W$ production can be thus divided into two gauge-invariant subsets:
the first two rows of Fig.~\ref{fig:feynman-diagrams}, common to $W$ and $Z$ production, and the last
one. The invariance under electromagnetic gauge transformations requires instead the sum of all the
diagrams, and it can be checked by writing explicitly the charges of up-type quarks, down-type quarks,
and of the $W$ boson, respectively $Q_u, Q_d$, and $Q_\smallw$.

\section{Evaluation of the Master Integrals}
\label{sec:eval}

The MIs necessary to compute the total cross sections of processes (\ref{proc1})-(\ref{proc8}) involve at
least one massive line (the EW gauge boson in the final state) and possibly an additional one from those
diagrams where a photon is emitted off a $W$ leg. For the processes under consideration, we found a total of
30 MIs, of which 16 with one massive line, and 14 with two massive lines.
All the integrals with one massive line were already available in the literature after the evaluation of the
NNLO-QCD corrections to the inclusive Higgs boson production in gluon fusion
\cite{Anastasiou:2002yz,Anastasiou:2012kq}. In order to validate the routines developed for the present
calculation, we recomputed them and found complete agreement. The computation of all the necessary MIs has
been performed  using the differential equations method. The system of equations has been written with the
help of the package {\tt Reduze} \cite{Studerus:2009ye,vonManteuffel:2012np}, while the solutions of the
equations have been worked out with dedicated {\tt Mathematica} \cite{math} routines.


By solving the differential equations of the MIs in the dimensionless variable
$z=\frac{M_{\smallv}^2}{\hat{s}}$, the results of the MIs can be naturally expressed in terms of HPLs 
and their generalizations. 
Specifically,  the MIs contributing to processes (\ref{proc1})-(\ref{proc3}) and (\ref{proc5})-(\ref{proc8})
can be written,  as a function of the variable $z$, in terms of HPLs with weights $\{0,\pm1\}$. The process
(\ref{proc4}) requires instead an enlargement of the basis of functions and the use of non-linear weights,
while keeping the variable $z$.  In this case, following Ref.~\cite{Aglietti:2004tq},  we use the set of
weights $\{0,\pm1,-\frac{1}{4},-\frac{r_0}{4}\}$.

In some cases, HPLs with non-linear weights can be transformed into combinations of HPLs with linear weights
at the price of introducing new weights (``letters'') in the set (``alphabet'') (see
e.g.~Ref.~\cite{Bonciani:2010ms}). In our case, by performing the change of variable $\xi=\xi(z)$ defined
through the equations
\begin{equation} \label{eq:xidef1}
z = \frac{\xi}{(1-\xi)^2} \, , 
\quad\quad\quad\quad
\xi = \frac{1+2 z-\sqrt{1+4 z}}{2 z} \, , 
\end{equation}
and by introducing two new linear weights $a_1,a_2$ defined as
\begin{align} \label{eq:defa1a2}
a_1 = \frac{3-\sqrt{5}}{2}, \quad \quad \quad a_2 = \frac{3+\sqrt{5}}{2} \, ,
\end{align}
HPLs of variable $z$ that contain the non-linear weight $-\frac{r_0}{4}$ can be expressed in terms of HPLs of variable $\xi$ and linear weights $\{0,\pm1,a_1,a_2\}$. 
In particular, the additional weights $a_1,a_2$ have to be introduced only for those HPLs of variable $z$ that simultaneously contain the weights $1$ and $-\frac{r_0}{4}$.
We further observe that HPLs with the latter combination of weights exactly cancel in the final results for the partonic cross sections of the process (\ref{proc4}). 
Two explicit examples of the aforementioned transformations are:
\begin{align}
\Hpl{(-\frac{1}{4},-\frac{r_0}{4},0;z)} &= -8 \Hpl{(-1,1,0;\xi)}-16 \Hpl{(-1,1,1;\xi)}-8 \Hpl{(1,1,0;\xi)}-16 \Hpl{(1,1,1;\xi)} \, , \notag \\
\Hpl{(-\frac{r_0}{4},0,1;z)} &= 8 \Hpl{(1,0,1;\xi)} + 4 \Hpl{(1,0,a_{1};\xi)} + 4 \Hpl{(1,0,a_{2};\xi)} + 16 \Hpl{(1,1,1;\xi)} \notag \\
&\quad + 8 \Hpl{(1,1,a_{1};\xi)} + 8 \Hpl{(1,1,a_{2};\xi)}\, .
\end{align}
In the ancillary \texttt{Mathematica} file we list all the transformations needed for the MIs that contribute to process (\ref{proc4}).
The advantage of this type of transformations is that 
the HPLs that appear in the final expressions
can be easily converted into ordinary logarithms and polylogarithms and evaluated numerically.

\subsection{Evaluation of the soft limits}
\label{sec:evalsoft}

We use the soft limit (i.e.~$z\to 1$ limit) of the MIs as boundary conditions to the differential equations.
We compute the soft limit of all the MIs relevant for the present calculation with the method described in
Ref.~\cite{Anastasiou:2013srw}. The main idea of this method is to rescale the momenta of the final-state
partons in the propagators of the MIs by a factor $(1-z)$ and to perform an expansion of the integrals around
the threshold $z=1$. The coefficients of these expansions are integrals with simpler structures, e.g.~eikonal
propagators. By means of the IBP identities it is then possible to reduce these ``soft'' integrals to a
combination of a very small number of ``soft Master Integrals'', which have to be computed explicitly. By
construction, the first term in the threshold expansion of the MIs is the leading behavior as $z\to1$,
i.e.~their soft limit.

For the processes under consideration,  we found that the soft limits of all the necessary MIs  can be
expressed as combinations of three soft MIs,  two of which were already known in the literature  while one,
to our knowledge, was not available yet.  We also observe that  in the case of $I_{21}(z;\eps)$, according to
the indexing of Appendix \ref{sec:appMI}, the integration constants of the differential equation can be fixed
only by computing also the  next-to-leading term in the threshold expansion of the soft limit of the MI.

The first soft MI is the pure phase-space integral. It can be computed using the energy-angles parameterization of Refs.~\cite{Anastasiou:2010pw,Anastasiou:2012kq} and reads
\be
{\cal X}(z;\eps) \equiv
\lim_{z\to 1} \int d\Phi_3 =
{\cal N}(\eps) M_{\smallv}^2 (1-z)^{3-4\eps} \frac{\Gamma(1-\eps)^2}{\Gamma(1+\eps)^2\,\Gamma(4-4\eps)} \, ,
\label{eq:softPS}
\ee
where we defined the normalization factor common to all MIs
\begin{equation}
\Ncal(\eps) = \frac{1}{2}\frac{\Gamma(1+\eps)^2}{\pts{4 \pi}^{3}} \pts{\frac{4 \pi \mu^2}{M^2_{\smallv}}}^{2\eps} \, .
\label{eq:normfac}
\end{equation}
The second soft MI appears in the soft limits of some of the MIs relevant for the $q\gamma$- and $qg$-initiated subprocesses.
Its expression has been discussed in Refs.~\cite{Anastasiou:2012kq,Anastasiou:2013srw} 
and reads\footnote{
The expressions ${\cal Y}(z;\eps)$ and $X_{18}$ in Eq.~(3.4) of Ref.~\cite{Anastasiou:2013srw}
differ by a normalization factor, namely\\
${\cal Y}(z;\eps)= \left[
\left(
\frac{\mu^2}{M_{\smallv}^2}
\right)^{2\eps}
\frac{1}{M_{\smallv}^4}
\right]
X_{18} 
$.
}
\bea
{\cal Y}(z;\eps)&\equiv &
\lim_{z\to 1} 
\displaystyle{\int \rmd \Phi_{3} \frac{1}{ \left(p_{1}-p_{4}\right)^2 \left(p_{2} - p_{5}\right)^2 \left(p_{4} + p_{5} \right)^2}}\label{eq:softY}\\
&=& 
-\frac{\Ncal(\eps)}{M_{\smallv}^4}\,(1-z)^{-1-4 \eps}\,
\frac{4 (1-4 \eps) \, (1-2 \eps)  \,
 \Gamma (1-\eps)^2 }{\eps^3 \, \Gamma (3-4 \eps) \,
\Gamma (1+\eps)^2}\, _3F_2(1,1,-\eps;1-2 \eps,1-\eps;1) \, .\nonumber
\eea
The third soft MI is peculiar of $W$ production. In this case, 
the presence of an additional internal massive line
spoils the factorization of the different integrations over the energy/angles variables discussed in Refs.~\cite{Anastasiou:2010pw,Anastasiou:2012kq}. More specifically,
\bea
{\cal Z}(z;\eps)&\equiv&
\lim_{z\to 1} 
\int
\frac{d\Phi_3}
{(p_1-p_4-p_5)^2 
\left[(p_1+p_2-p_4)^2-M_{\smallw}^2 \right]} \label{eq:softZ}\\
&=&
-\frac{(1-z)^{-2}}{M_{\smallw}^4}  {\cal X}(z;\eps) \frac{\Gamma(4-4\eps)}{\Gamma(1-\eps)^4}
\int_0^1
dx_1\,dx_3\,dx_4\,
\frac{x_1^{1-2\eps} (\bar x_1^2 x_3 \bar x_3 x_4 \bar x_4)^{-\eps}    }
{x_1 \bar x_3 + \bar x_1 \bar x_4 }   \, ,
\eea
where $\bar x_i=1-x_i$. The solution is found by introducing a Mellin-Barnes (MB) representation
for the last denominator, allowing the factorization of the integrals over $x_{1,3,4}$ at the price of an extra integration over the MB transform variable:
\bea
{\cal Z}(z;\eps)
&=&
-\frac{{\cal N}(\eps)}{ M_{\smallw}^2} \frac{1}{\Gamma(1+\eps)^2\Gamma(2-4\eps)}
(1-z)^{1-4\eps}\times\nonumber\\
&\times&
\int_{-i\infty+u_0}^{+i\infty+u_0}
\frac{du}{2\pi i}
\frac{\Gamma(-u)\Gamma(1+u)\Gamma(-\eps-u)\Gamma(1-\eps+u)}{(-2\eps-u)} \, .
\label{eq:MB}
\eea
The integration contour can be chosen such that all the poles of the $\Gamma(a+u)$ are located to the left 
of the vertical line defined by $u_{0}$ and all the poles of the $\Gamma(b-u)$ are located to the right.
The integration can then be solved using the residue theorem by
choosing a finite closed rectangular contour to the left of the vertical line at $u=u_0$ and then taking the limit of an infinitely extended contour. In this limit, the contribution of the additional lines vanishes and 
the result of the integral is thus given by the infinite sum of the residues of the integrand. Explicitly, we find:
\bea
&& \hspace*{-2mm} {\cal Z}(z;\eps)=
\frac{{\cal N}(\eps)}{ M_{\smallw}^2}
(1-z)^{1-4\eps}
\frac{\Gamma(1-\eps)^2}
{\eps^2 \,\Gamma(3-4\eps)\,\Gamma(1+\eps)^2}\times\nonumber\\
&& \hspace*{8mm} \times \!
\left\{2\eps \,
_3F_2(1,1-2\eps,1-\eps;2-2\eps,1+\eps;1)-
\frac{\Gamma(1-3\eps) \Gamma(2-2\eps)\Gamma(1+\eps) \Gamma(1+2\eps)}{\Gamma(1-\eps)^2}\right\} . \nonumber \\
\eea
In Appendix \ref{app:MIsoft} we collect the expressions,
exact in $\eps$, of the soft limits of all the MIs appearing in this calculation.

\section{Total partonic cross sections}
\label{sec:xsec}
We now present the analytic expressions of the total partonic cross sections of the processes under consideration.
For each subprocess we indicate which MIs contribute to the partonic cross section according to the indexing
proposed in Appendix \ref{sec:appMI}. We present the results expressed as Laurent series in the dimensional
regulator $\eps=(4-d)/2$ and as functions of the dimensionless variable $z=M_{\smallv}^2/\hat{s}$ with
$V=W^{\pm},Z$. We remark that all the cross sections are expressed in terms of HPLs up to weight 3, as the
coefficients that contain HPLs of weight 4 in the expansion in $\eps$ of the individual MIs do not contribute to
the cross sections up to $\mathcal{O}(\eps^0)$. The only exceptions to this are the cross sections for the $q\bar
q$-initiated processes, where HPLs of weight 4 coming from integrals $I_1$ and $I_2$ (according to the indexing of
Appendix \ref{sec:appMI}) enter the cross section at $\mathcal{O}(\eps^0)$, but eventually exactly cancel among
each other. 
We remark
that some of the MIs contributing to the process $g \gamma\to W^{\pm} q_i \bar{q}_j$ are represented in terms of
generalized HPLs with non-linear weights,  as it is explicitly shown in the results of Appendix \ref{sec:appMI}. 
Nevertheless, we observe that the generalized HPLs that are eventually part of the cross section can all be
transformed into HPLs with linear weights $\{0,\pm 1\}$ and variable $\xi(z)$ defined in Eq.~\eqref{eq:xidef1}.

Finally, in order to facilitate the numerical evaluation of the results, we convert all the HPLs appearing in the
cross sections into ordinary logarithms and polylogarithms of variables $z$ and $\xi(z)$, and Riemann zeta
functions. We perform this conversion with the \texttt{HPL} package \cite{Maitre:2005uu} and with in-house
\texttt{Mathematica} routines. 

Throughout this section we use the following normalization factor for the cross sections:
\begin{equation}\label{eq:AnormFact}
\mathcal{A}_{\smallv}(\eps) = \Gamma(1+\eps)^2 \pts{\frac{4 \pi \mu^2}{M_{\smallv}^2}}^{2 \eps} \, .
\end{equation}

\subsection{$W$ production}

In the calculation of the different subprocesses that contribute to $W$ boson production we have retained the full
dependence on the quarks and $W$ boson electric charges,  obtaining expressions that are lengthier than those for
$Z$ production. For the sake of brevity, we present here results where the explicit charge values have been
inserted in the formulae, while in the ancillary files we deliver the generic expressions.

\subsubsection{The subprocess $q_i\,\bar{q}_j \to W^{\pm} \,g\,\gamma$}

We present here the fully inclusive partonic cross section for the tree-level processes:
\begin{equation}
q_{i} \; \bar{q}_{j} \rightarrow W^{\pm} \; g \; \gamma \, .
\end{equation}
The cross sections $\hat{\sigma}^{RR}_{q_i\bar{q}_j\to W^{\pm}g\gamma}\pts{z;\eps}$ are obtained by summing the following combination of MIs
\be
\hat{\sigma}^{RR}_{q_i\bar{q}_j\to W^{\pm}g\gamma}\pts{z;\eps} = \sum_{k=1}^{11} c_k^{q_{i} \bar{q}_{j}\to W^{\pm} g \gamma}(z;\eps) I_k(z;\eps) \, ,
\ee
where the explicit expressions of the MIs $I_k\,\,(k=1,\dots,11)$ can be found in Appendix \ref{sec:appMI}.
After expanding in $\eps$ and introducing plus distributions, we recast the results as
\begin{equation}
\hat{\sigma}^{RR}_{q_i\bar{q}_j\to W^{\pm}g\gamma}\pts{z;\eps} =  4 \hat{\sigma}^{0}_{q_i\bar{q}_j\to W^{\pm}}(z) \, C_{F}  \frac{\alpha}{2 \pi}\frac{\alpha_{S}}{2 \pi} \mathcal{A}_{\smallw}(\eps) \sum_{n=-4}^0 \eps^{n} \mathrm{\textbf{P}}^{(n)}_{{\smallw}^{\pm}} \pts{z} \, ,
\end{equation}
where $\mathcal{A}_{\smallw}$ follows from \eqref{eq:AnormFact} with $M_{\smallv}=M_{\smallw}$.
We defined
\begin{equation}\label{eq:SigmaLOW}
\hat{\sigma}^{0}_{q_i\bar{q}_j\to W^{\pm}}(z) = \frac{\pi^2}{N_{c}} \frac{\alpha }{\SWsq}\left| V_{ij}\right|^{2} \frac{z}{M^2_{{\smallw}^{\pm}}}\; ,
\end{equation}
which is the coefficient of $\delta(1-z)$ in the Born cross section,
with $\SWsq$ the squared sinus of the weak mixing angle and $N_{c}$ the number of colors. 
We remark that the total cross sections for
the processes  $q_i\bar{q}_j\to W^{+}g\gamma$ and $q_j\bar{q}_i\to W^{-}g\gamma$ are identical. In terms of
ordinary logarithms and polylogarithms, the functions $\mathrm{\textbf{P}}^{(n)}_{{\smallw}^{\pm}} \pts{z}$ read:
\begingroup
\allowdisplaybreaks
\begin{align}
\mathrm{\textbf{P}}^{(-4)}_{{\smallw}^{\pm}} \pts{z} =& \frac{5}{18} \deltz, \\
\mathrm{\textbf{P}}^{(-3)}_{{\smallw}^{\pm}} \pts{z} =& \frac{2}{9} \deltz - \frac{5}{9} \left(1+z^2\right) \PlusD{\frac{1}{1-z}}, \\
\mathrm{\textbf{P}}^{(-2)}_{{\smallw}^{\pm}} \pts{z} =& \left( \frac{1}{2} - \frac{20}{9} \zed{2} \right) \deltz - \frac{5}{18} \left( 3 (1+z^2) -\frac{14 z}{5} \right) \PlusD{\frac{1}{1-z} } \nonumber \\
+& \frac{20}{9}  \left(1+z^2\right) \PlusD{\frac{\Log{1-z}}{1-z} } -\frac{5}{9} \left(2+3z^2\right) \frac{\Log{z}}{1-z}, \\
\mathrm{\textbf{P}}^{(-1)}_{{\smallw}^{\pm}} \pts{z} =& 
\frac{50}{9} \Log{1-z} \Log{z} \frac{\left(1+z^2 \right)}{1-z}
+\left(\frac{25}{36} - \frac{17}{8} \left(1+z^2\right)\right) \frac{\PLog{z}{2}}{1-z} \nonumber \\
+& \frac{10}{9} (1+z) \Li{2}{z}
-\left(2 -\frac{37}{18} z + \frac{41}{18} z^2 \right) \frac{\Log{z}}{1-z} \nonumber \\
+& \left(1-\frac{16}{9}\zed{2} -\frac{50}{9} \zed{3} \right) \deltz 
+\left(\frac{10}{3} \left(1+z^2 \right)-\frac{28 z}{9}\right) \PlusD{\frac{\Log{1-z}}{1-z}} \nonumber \\
+& \left(\frac{113}{18} z + \frac{10}{9} \left(3 + 5 z^2 \right) \zed{2} - \frac{149}{36} \left(1 + z^2 \right)\right) \PlusD{\frac{1}{1-z}} \nonumber \\
-& \frac{40}{9} \left(1+z^2\right) \PlusD{\frac{\PLog{1-z}{2}}{1-z}} ,
\\
\mathrm{\textbf{P}}^{(0)}_{{\smallw}^{\pm}} \pts{z} =& 
4 -\frac{64 \zed{2}}{9} -(1-z) \left(\frac{1981}{108}-\frac{z}{4}\right)
+(1-z) \left(\frac{71}{12} -z -\frac{z^2}{4}\right) \zed{2} \nonumber \\
+& (1+z) \left(-\frac{38 \Li{3}{1-z}}{9}-\frac{40}{9} \Li{2}{z} \Log{1-z}+\frac{38}{9} \zed{2} \Log{1-z}\right) \nonumber \\
- & \left(\frac{10}{3} - 2 z -\frac{10 z^2}{9} \right) \frac{\Li{3}{z}-\zed{3}}{1-z}
+\left(\frac{25}{9} - z-\frac{8 z^2}{3} \right) \Li{2}{z} \frac{\Log{z}}{1-z} \nonumber \\
+& \left(31 + 37 z - 27 z^2 - 9 z^3 \right) \frac{\Li{2}{z}-\zed{2}}{36}
+\left(\frac{85}{9} -2 z + \frac{143 z^2}{9}\right) \zed{2} \frac{\Log{z}}{1-z} \nonumber \\
-&  \left(\frac{89}{72} + \frac{z}{6} + \frac{341 z^2}{216} \right)  \frac{\PLog{z}{3}}{1-z}
+\left(\frac{123 + 115 z^2}{18} \right) \Log{1-z}  \frac{\PLog{z}{2}}{1-z} \nonumber \\
-& \frac{40}{9} \left(3 + 2 z^2\right) \PLog{1-z}{2} \frac{\Log{z}}{1-z}
-\left(\frac{103}{9} - \frac{209}{12} z + \frac{101}{9} z^2 - \frac{z^3}{4} \right) \frac{\Log{z}}{1-z} \nonumber \\
-& \left(\frac{197}{72} - \frac{8 z}{3} + \frac{3 z^2}{2} + \frac{z^3}{4} +\frac{z^4}{8} \right) \frac{\PLog{z}{2}}{1-z} +\big(319 -290 z + 264 z^2 + 18 z^3 \nonumber \\
+& 9 z^4 \big) \Log{1-z} \frac{\Log{z}}{36 (1-z)}
+ 2  \left(1 -2 \zed{2}-\frac{20}{9} \zed{3} +\frac{20}{9} \zed{4} \right)\deltz \nonumber \\
+&  \left(\frac{100}{9} \zed{3} \left(1+z^2 \right) + \frac{64}{9} \zed{2} -4\right)\PlusD{\frac{1}{1-z}} 
+\frac{160}{27} \left(1+z^2 \right) \PlusD{\frac{\PLog{1-z}{3}}{1-z}} \nonumber \\ 
+& \left( \frac{4 \zed{2}}{9} -\frac{226 z}{9} +  \left(\frac{149}{9}-18 \zed{2}\right) \left(1+z^2\right) \right) \PlusD{\frac{\Log{1-z}}{1-z}} \nonumber \\
-& \left(\frac{20}{3} \left(1+z^2 \right)-\frac{56 z}{9}\right) \PlusD{\frac{\PLog{1-z}{2}}{1-z}} \, .
\end{align}
\endgroup

\subsubsection{The subprocess $q_i \,g \to W^{\pm} \,q_j \,\gamma$}

We present here the fully inclusive partonic cross section for the tree-level processes:
\begin{equation}
q_{i} \; g \rightarrow W^{\pm} \; q_{j} \; \gamma .
\end{equation}
The cross sections $\hat{\sigma}^{RR}_{q_i g \to W^{\pm} q_j \gamma}\pts{z;\eps}$ are obtained by summing the following combination of MIs
\be
\hat{\sigma}^{RR}_{q_i g \to W^{\pm} q_j \gamma}\pts{z;\eps} = \sum_{k=1}^{23} c_k^{q_i g\to W^{\pm} q_j \gamma}(z;\eps) I_k(z;\eps) \, ,
\ee
with the explicit expressions of the MIs $I_k\,\,(k=1,\dots,23)$ collected in Appendix \ref{sec:appMI}.
We observe that $c_k^{q_i g\to W^{\pm} q_j \gamma}=0$ for $k=3,7\!-\!11,14,19\!-\!21$. 
The cross section expressed as a Laurent series in the dimensional regulator $\eps$ has the form:
\begin{equation}
\hat{\sigma}^{RR}_{q_i g \to W^{\pm} q_j \gamma}\pts{z;\eps} =  2 \hat{\sigma}^{0}_{q_i\bar{q}_j\to W^{\pm}}(z)  \frac{\alpha}{2 \pi}\frac{\alpha_{S}}{2 \pi} \mathcal{A}_{\smallw}(\eps) \sum_{n=-3}^0 \eps^{n} \mathrm{\textbf{Q}}^{(n)}_{q_i g,{\smallw}^{\pm}} \pts{z} \, ,
\end{equation}
where $\mathcal{A}_{\smallw}$ has been defined in Eq.~\eqref{eq:AnormFact} and $\hat{\sigma}^{0}_{q_i\bar{q}_j\to W^{\pm}}(z)$ in Eq.~\eqref{eq:SigmaLOW}.
We remark that the cross sections of the subprocesses initiated by a gluon and an up- or a down-type quark differ because of the different electric charge flow probed by the final state photon.
For the specific process $u \,g \to W^+ d \,\gamma$, the functions $\mathrm{\textbf{Q}}^{(n)}_{ug,{\smallw}^{+}} \pts{z}$ read\footnote{
The results for the subprocess $d\,g \to W^- u\,\gamma$ can be easily obtained
with the expressions present in the ancillary files, written with generic electric charges.
}:
\begingroup
\allowdisplaybreaks
\begin{align}
\mathrm{\textbf{Q}}^{(-3)}_{u g,{\smallw}^{+}} \pts{z} =& -\frac{5}{18} \ptsq{\frac{(1-z)^2 + z^2}{2}} , \\
\mathrm{\textbf{Q}}^{(-2)}_{u g,{\smallw}^{+}} \pts{z} =& - \frac{1}{4} z^2 \Log{z} +\frac{5}{18} \ptsq{\frac{(1-z)^2 + z^2}{2}} \left(4 \Log{1-z}-\frac{21}{10} \Log{z} \right) \nonumber \\
-& \frac{5}{16} +\frac{47 z}{36}-\frac{9 z^2}{8} , \\
\mathrm{\textbf{Q}}^{(-1)}_{u g,{\smallw}^{+}} \pts{z} =& -\frac{29}{72} +\frac{641 z}{144} -\frac{623 z^2}{144} +\left(\frac{5}{4} -\frac{47 z}{9} + \frac{9 z^2}{2}\right) \Log{1-z}
-z^2 (\Li{2}{z}-\zed{2}) \nonumber \\
+& \frac{5}{18} \ptsq{ \frac{(1-z)^2+z^2}{2} } \left(\frac{13 \Li{2}{z}}{5}-8 \PLog{1-z}{2} + 11 \Log{z} \Log{1-z} \right. \nonumber \\
-& \left. \frac{61}{20}\PLog{z}{2} +\frac{22 \zed{2}}{5}\right)
-\frac{19}{72} z^2 \PLog{z}{2}
+\left(-\frac{17}{144} +\frac{119 z}{36} -\frac{27 z^2}{8} \right) \Log{z} , \\
\mathrm{\textbf{Q}}^{(0)}_{u g, {\smallw}^{+}} \pts{z} =&  -\frac{5}{36} \left(\frac{389}{40} -\frac{586 z}{5} + \frac{901 z^2}{8}\right)
+ \frac{5}{18} \ptsq{ \frac{(1-z)^2 + z^2}{2} } \left(
-\frac{44}{5} \Li{3}{1-z} \right. \nonumber \\
-& \frac{16}{5} \Li{3}{-z} 
- 13 \Li{3}{z}
-\frac{52}{5} \Li{2}{z} \Log{1-z}
+\frac{8}{5} \Li{2}{-z} \Log{z} \nonumber \\
+&  \frac{58}{5} \Li{2}{z} \Log{z}
-\frac{88}{5} \zed{2} \Log{1-z}
+\frac{77}{5} \zed{2} \Log{z}
+\frac{32}{3} \PLog{1-z}{3} \nonumber \\
-& \left. \frac{37}{12} \PLog{z}{3}
-\frac{136}{5} \Log{z} \PLog{1-z}{2}
+\frac{173}{10} \PLog{z}{2} \Log{1-z}
+\frac{122 \zed{3}}{5}\right) \nonumber \\
-& \frac{1}{18} z^2 \Log{z} (49 \Li{2}{z}-29 \zed{2})
+4 z^2 \Log{1-z} (\Li{2}{z}-\zed{2}) \nonumber \\
+& \frac{1}{72} \left(213 -100 z -78 z^2 \right) \Li{2}{z}
+\frac{2}{9} z^2 (17 \Li{3}{1-z}+14 \Li{3}{z}) \nonumber \\
-&\frac{2}{9} \left((1-z)^2-4 z^2\right) (\Li{2}{-z}+\Log{z} \Log{1+z} )
-\frac{28 z^2 \zed{3}}{9} -\frac{13}{72} z^2 \PLog{z}{3} \nonumber \\
+& \frac{5}{18} \left(\frac{69 z^2}{2}-\frac{139 z}{5}-\frac{51}{20}\right) \zed{2}
-\frac{1}{18} \left(45 -188 z + 162 z^2\right) \PLog{1-z}{2} \nonumber \\
+& \frac{1}{72} \left(247-1052 z+894 z^2\right) \Log{1-z} \Log{z}
+\left(-\frac{23}{4} +\frac{257 z}{2} -\frac{1377 z^2}{4} \right. \nonumber \\
+& \left. 334 z^3 -\frac{233 z^4}{2} \right) \frac{ \PLog{z}{2}}{24 (1-z)^2}
+\frac{z^2}{9} \Log{1-z} \Log{z}(18 \Log{1-z}-\Log{z})  \nonumber \\
+& \left(58 - 641 z + 623 z^2\right) \frac{\Log{1-z}}{36}
+\Big(35 +1552 z -3504 z^2 \nonumber \\
+&  1869 z^3 \Big) \frac{\Log{z}}{144 (1-z)} \, .
\end{align}
\endgroup

\subsubsection{The subprocess $q_i \,\gamma \to W^{\pm} \,q_j\, g$}

We now focus on the partonic cross section for the tree-level processes:
\begin{equation}
q_{i} \; \gamma \rightarrow W^{\pm} \; q_{j} \; g .
\end{equation}
We obtain the cross section as
\be
\hat{\sigma}^{RR}_{q_i \gamma \to W^{\pm} q_j g}\pts{z;\eps} = \sum_{k=1}^{21} c_k^{q_i \gamma \to W^{\pm} q_j g}(z;\eps) I_k(z;\eps) \, ,
\ee
where $c_k^{q_i \gamma \to W^{\pm} q_j g}=0$ for $k=3,5\!-\!11$. As a Laurent series, the cross section can be rewritten as
\begin{equation}
\hat{\sigma}^{RR}_{q_i \gamma \to W^{\pm} q_j g}\pts{z;\eps} =  4 \hat{\sigma}^{0}_{q_i\bar{q}_j\to W^{\pm}}(z) \, C_{A} C_{F} \frac{\alpha}{2 \pi}\frac{\alpha_{S}}{2 \pi} \mathcal{A}_{\smallw}(\eps) \sum_{n=-3}^0 \eps^{n} \mathrm{\textbf{G}}^{(n)}_{q_i\gamma,{\smallw}^{\pm}} \pts{z} \, ,
\end{equation}
with $\mathcal{A}_{\smallw}$ and $\hat{\sigma}^{0}_{q_i\bar{q}_j\to W^{\pm}}(z)$ as earlier defined.
We remark that the cross sections of the subprocesses initiated by a photon and an up- or a down-type quark differ because of the different electric charge flow probed by the photon.
For the specific process $u \,\gamma \to W^+ d \,g$, the functions $\mathrm{\textbf{G}}^{(n)}_{u\gamma,{\smallw}^{+}} \pts{z}$ read\footnote{
The results for the subprocess $d\,\gamma \to W^- u\,g$ can be easily obtained
with the expressions present in the ancillary files, written with generic electric charges.
}:
\begingroup
\allowdisplaybreaks
\begin{align}
\mathrm{\textbf{G}}^{(-3)}_{u \gamma,{\smallw}^{+}} \pts{z} =& -\frac{1}{9} \ptsq{\frac{(1-z)^2 + z^2}{2} } , \\
\mathrm{\textbf{G}}^{(-2)}_{u \gamma,{\smallw}^{+}} \pts{z} =& \frac{35}{48} +\frac{11 z}{36} - \frac{3 z^2}{2}+\frac{4}{9} \ptsq{\frac{(1-z)^2 + z^2}{2} } \Log{1-z}
+\frac{1}{2} \ptsq{\frac{1+(1-z)^2}{z}} \nonumber \\
-& \left(\frac{1}{6}- 5 (1+z) \right) \frac{\Log{z}}{4} , \\
\mathrm{\textbf{G}}^{(-1)}_{u \gamma,{\smallw}^{+}} \pts{z} =&  \frac{557}{144} -\frac{133 z}{144} - \frac{67 z^2}{18}
+\left(38 z^2+142 z+193\right) \frac{\Li{2}{z}}{36} \nonumber \\
-& \left(44 (1+z) + z^2\right) \frac{\zed{2}}{9} + \left(\frac{221}{4}+\frac{109 z}{2} + 5 z^2\right) \frac{\PLog{z}{2}}{36} \nonumber \\
+& \frac{1}{9} \ptsq{ \frac{ (1-z)^2 + z^2 }{2} } \left(\frac{19}{2} \Log{1-z} \Log{z}-8 \PLog{1-z}{2}\right) \nonumber \\
-& \left(\frac{35}{12} + \frac{11 z}{9} - 6 z^2 \right) \Log{1-z}+
\left(\frac{13}{144}+\frac{11 z}{9} - 5 z^2\right) \Log{z} \nonumber \\
+& \ptsq{ \frac{ 1 + (1-z)^2 }{z} } \left(\frac{3}{4}-2 \Log{1-z}\right) , \\
\mathrm{\textbf{G}}^{(0)}_{u \gamma, {\smallw}^{+}} \pts{z} =& 
\frac{1759}{144} -\frac{11 z}{8} -\frac{473 z^2}{36}
-\frac{1}{9} \left(39 z^2-\frac{23 z}{2}+\frac{501}{8}\right) \Li{2}{z} \nonumber \\
-&\frac{1}{9} \left(36 z^2+146 z+191\right) \Li{3}{1-z}
-\frac{1}{6} \left(25 z^2+\frac{157 z}{3} +\frac{527}{6}\right) \Li{3}{z} \nonumber \\
+&\frac{2}{9} \ptsq{\frac{(1-z)^2 + z^2}{2} } \left(-4 \Li{3}{-z}+2 \Li{2}{-z} \Log{z}+\frac{16}{3} \PLog{1-z}{3}\right) \nonumber \\
+& \frac{1}{9} \left(\frac{87 z^2}{2} + 73 z + \frac{263}{2}\right) \Li{2}{z} \Log{z}
+ \left(-14 z^2+305 z+\frac{445}{2}\right) \frac{\PLog{z}{3}}{216} \nonumber \\
-& \frac{1}{9} \left(38 z^2+142 z+193\right) \Li{2}{z} \Log{1-z}
-\frac{\zed{2}}{9} \left(4 z^2+\frac{161 z}{2}+\frac{343}{4}\right) \Log{z} \nonumber \\
-&\frac{2}{9} \left((1-z)^2-4 z^2\right) (\Li{2}{-z}+\Log{z} \Log{1+z}) \nonumber \\
-& \frac{1}{9} \left(38 z^2+142 z+193\right) \Li{2}{z} \Log{1-z} + \left(147 z^2-\frac{61 z}{2}+\frac{87}{8}\right) \frac{\zed{2}}{9} \nonumber \\
-&\ptsq{\frac{1+(1-z)^2}{z}} \Big(5 \Li{2}{z}-4 \PLog{1-z}{2}+5 \Log{z} \Log{1-z} -\zed{2}  \nonumber \\ 
+& 3 \Log{1-z}-2 \Big) +\left(\frac{921}{16} -76 z -\frac{1433 z^2}{16} + \frac{207 z^3}{2}\right) \frac{\Log{z}}{9 (1-z)}  \nonumber \\
-& \frac{2}{9} \left(53 +26 z + 19 z^2 \right) \PLog{1-z}{2} \Log{z} + \left(\frac{35}{6} +\frac{22 z}{9} -12 z^2\right) \PLog{1-z}{2} \nonumber \\
+& \frac{4}{9} \left(44 (1+z) + z^2\right) \zed{2} \Log{1-z}+\frac{1}{9} \left(54 z^2+62 z+140\right) \zed{3} \nonumber \\
+& \frac{1}{9} \left(\frac{79 z^2}{4} -\frac{83 z}{4}+\frac{83}{8}\right) \Log{1-z} \PLog{z}{2}
- \left(\frac{207}{32}-\frac{627 z}{16}+\frac{3591 z^2}{32} \right. \nonumber \\
-& \left. \frac{549 z^3}{4} +60 z^4\right) \frac{\PLog{z}{2}}{9 (1-z)^2} 
+ \left(\frac{134 z^2}{9}+\frac{133 z}{36}-\frac{557}{36}\right) \Log{1-z} \nonumber \\
-&\frac{1}{9} \left(\frac{527}{8} +\frac{65 z}{2} -141 z^2\right) \Log{1-z} \Log{z} \, .
\end{align}
\endgroup

\subsubsection{The subprocess $g\,\gamma \to W^{\pm} \,q_i\, \bar{q}_j$}

Finally, we consider the tree-level processes:
\begin{equation}
g \; \gamma \rightarrow W^{\pm} \; q_{i} \; \bar{q}_{j} \, .
\end{equation}
The partonic cross sections are written as
\be
\hat{\sigma}^{RR}_{g \gamma\to W^{\pm} q_i \bar{q}_j}\pts{z;\eps} = \sum_{i=k}^{30} c_k^{g \gamma\to W^{\pm} q_i \bar{q}_j}(z;\eps) I_k(z;\eps) \, ,
\ee
with $c_k^{g \gamma\to W^{\pm} q_i \bar{q}_j}=0$ for $k=3,5\!-\!8,10\!-\!23$. Expressed as Laurent series, the cross sections have the form:
\begin{equation}
\hat{\sigma}^{RR}_{g \gamma\to W^{\pm} q_i \bar{q}_j}\pts{z;\eps} =  \hat{\sigma}^{0}_{q_i\bar{q}_j\to W^{\pm}}(z) \, C_{A} \frac{\alpha}{2 \pi}\frac{\alpha_{S}}{2 \pi} \mathcal{A}_{\smallw}(\eps) \sum_{n=-2}^0 \eps^{n} \mathrm{\textbf{T}}^{(n)}_{{\smallw}^{\pm}} \pts{z} \, .
\end{equation}
We remark that the total cross sections of the processes 
$g\,\gamma \to W^+ \,d \,\bar{u}$ and $g\,\gamma \to W^- \,u \,\bar{d}$
are identical.
The functions $\mathrm{\textbf{T}}^{(n)}_{{\smallw}^{\pm}} \pts{z}$ read:
\begingroup
\allowdisplaybreaks
\begin{align}
\mathrm{\textbf{T}}^{(-2)}_{{\smallw}^{\pm}} \pts{z} =& -\frac{5}{18} \ptsq{ 2(1-z)(1+3 z) + (1+2 z)^2 \Log{z} } , \\
\mathrm{\textbf{T}}^{(-1)}_{{\smallw}^{\pm}} \pts{z} =&  \frac{17}{4} - \frac{194}{9} z + \frac{647}{36} z^2 + \frac{1}{9} (1 + 2 z)^2 \left(10 \zed{2} -10 \Li{2}{z}+\frac{13}{2} \PLog{z}{2} \right) \nonumber \\
+& \left(1 -\frac{46}{9}z -\frac{16 z^2}{9}\right) \Log{z} 
-\frac{2}{3} \pts{\frac{1+(1-z)^2}{z}} - (6+4 z) z \PLog{z}{2} \nonumber \\
+& \frac{20}{9} (1-z) (1+3 z) \Log{1-z} , \\
\mathrm{\textbf{T}}^{(0)}_{{\smallw}^{\pm}} \pts{z} =& \frac{1337}{72} -\frac{203 z}{3} +\frac{3511 z^2}{72}
+10 \left( (1+z)^2 + z^2\right) \left(\frac{1}{8} \Hpl{\big(-\frac{r_{0}}{4},-\frac{r_{0}}{4},0;1\big)} \right. \nonumber \\
+&  2 \ _4F_3\left(\frac{1}{2},\frac{1}{2},\frac{1}{2},\frac{1}{2};\frac{3}{2},\frac{3}{2},\frac{3}{2};-\frac{1}{4}\right)
-2 \Li{2}{\xi} \Log{1-\xi}-\Log{z} \Li{2}{\xi} \nonumber \\
+& \Li{2}{\xi} \Log{\xi}
+\frac{4}{3} \PLog{1-\xi}{3}
+\Log{z} \PLog{1-\xi}{2}
-2 \Log{\xi} \PLog{1-\xi}{2} \nonumber \\
+& \frac{1}{2} \PLog{\xi}{2} \Log{1-\xi}
-\Log{z} \Log{\xi} \Log{1-\xi}
-\frac{2}{5} \left(\sqrt{5}-7\right) \Log{\frac{1+\sqrt{5}}{2}} \zed{2} \nonumber \\
-& \left. \zed{2} \left(2-\frac{2}{\sqrt{5}}\right)  \text{csch}^{-1}(2)\right)
-10 z^2 \Li{3}{\xi}
-10 \sqrt{1+4 z} \bigg(\Log{1-\xi} \Log{\xi} \nonumber \\
-& \left. \PLog{1-\xi}{2} + \Li{2}{\xi}
+\frac{\PLog{z}{2}}{4}-\frac{2 \zed{2}}{5}\right)
+\frac{2}{3} (1+z)^2 \left(\frac{4}{3}\Li{3}{-z} \right. \nonumber \\
-& 15 \Li{3}{\xi}
-\frac{8}{3} \Li{3}{\frac{1}{1+z}}
-\frac{4}{3} \Li{2}{-z} \Log{z}-\frac{4}{3} \zed{2} \Log{1+z}
+\frac{4}{9} \PLog{z}{3} \nonumber \\ 
-& 2 \PLog{z}{2} \Log{1+z}\bigg)
+\frac{2}{9} \left(43 - 42 z -107 z^2 \right) \Li{2}{z}
-\frac{16}{9} z^2 \Li{3}{-z} \nonumber \\
+& \frac{2}{9} \left(55 - 20 z + 64 z^2\right) \Li{3}{z}
-\left(\frac{73}{9}+\frac{244 z}{9}+\frac{268 z^2}{9}\right) \Li{2}{z} \Log{z} \nonumber \\
+& \frac{40}{9} (1+2 z)^2 \bigg( \Li{3}{1-z}+\Li{2}{z} \Log{1-z}+\Li{2}{z} \Log{z}-\zed{2} \Log{1-z} \nonumber \\
+& \left. \frac{1}{2} \Log{z} \PLog{1-z}{2}\right)
+8 \zed{2} \left((1+z)^2 + z^2\right) \Log{1-\xi} \nonumber \\
-& \frac{2}{9} \left(25+80 z + 47 z^2\right) \zed{2}
-\frac{8}{9} (1+2 z) \Li{2}{-z} \Log{z} \nonumber \\
-& \frac{8}{9} (1+z) (\Li{2}{-z}+\Log{z} \Log{1+z})
-\frac{2}{9} \left(\frac{25}{2} -134 z -62 z^2\right) \zed{2} \Log{z} \nonumber \\
-& \frac{20}{3} \left(3 +\frac{5 z}{3} + 5 z^2\right) \zed{3}
+\frac{1}{9} \left(\frac{25}{6} -37 z -13 z^2 \right) \PLog{z}{3} \nonumber \\
+& \left(\frac{65}{4} -40 z - 67 z^2 \right) \frac{\PLog{z}{2}}{9}
+\frac{2}{9} \left(\frac{565}{8} -188 z +\frac{511 z^2}{2} \right) \Log{z} \nonumber \\
+& \frac{50}{9} \left((1+z)^2-4 z^2\right) \left(\Log{1-z} \Log{z}-\frac{4}{5} \PLog{1-z}{2}\right) +\frac{4}{9} z (47 z+80) \zed{2} \nonumber \\
-& \frac{2}{3} \left(\frac{51}{2} -\frac{388 z}{3} + \frac{647 z^2}{6} \right) \Log{1-z}
-\frac{4}{9} (1-z)^2 \Log{1-z} \PLog{z}{2} \nonumber \\
+& \frac{2}{3} \left[ \frac{1+ (1-z)^2}{z} \right] \left(\frac{1}{2}+4 \Log{1-z}\right) \, ,
\end{align}
where $\Hpl{\big(-\frac{r_{0}}{4},-\frac{r_{0}}{4},0;1\big)}$ explicitly reads
\begin{align}
\Hpl{\pts{-\frac{r_{0}}{4},-\frac{r_{0}}{4},0;1}} &= \, 16 \, \pts{ \zed{2} \Log{\frac{\sqrt{5}-1}{2}} + \zed{3} - \frac{1}{3}\PLog{\frac{\sqrt{5}-1}{2}}{3} - \Li{3}{\frac{\sqrt{5}-1}{2}} } .
\end{align}
\endgroup

\subsection{$Z$ production}

\subsubsection{The subprocess $q_i\,\bar{q}_i \to Z \,g\,\gamma$}

We present here the partonic cross section for the tree-level process:
\begin{equation}
q_{i} \, \bar{q}_{i} \rightarrow Z \, g \, \gamma \, .
\end{equation}
The cross section $\hat{\sigma}^{RR}_{q_{i} \bar{q}_{i}\to Z g \gamma}\pts{z;\eps}$ is obtained by summing the following combination of MIs
\be
\hat{\sigma}^{RR}_{q_{i} \bar{q}_{i}\to Z g \gamma}\pts{z;\eps} = \sum_{k=1}^{10} c_k^{q_{i} \bar{q}_{i}\to Z g \gamma}(z;\eps) I_k(z;\eps) \, ,
\ee
where $c_k^{q_{i} \bar{q}_{i}\to Z g \gamma}=0$ for $k=5\!-\!7$. We can then rewrite the cross section as
\begin{equation}
\hat{\sigma}^{RR}_{q_{i} \bar{q}_{i}\to Z g \gamma}\pts{z;\eps} =  4 \hat{\sigma}^{0}_{q_{i} \bar{q}_{i}\to Z}(z) \, C_{F} \, Q^2_{i} \frac{\alpha}{2 \pi}\frac{\alpha_{S}}{2 \pi} \mathcal{A}_{\smallz}(\eps) \sum_{n=-4}^0 \eps^{n} \mathrm{\textbf{P}}^{(n)}_{\smallz} \pts{z} \, ,
\end{equation}
where $\mathcal{A}_{\smallz}$ is obtained from Eq.~\eqref{eq:AnormFact}.
We defined
\begin{equation}\label{eq:SigmaLOZ}
\hat{\sigma}^{0}_{q_{i} \bar{q}_{i}\to Z}(z) =  \frac{4 \pi^2}{N_{c}} \frac{\alpha}{\SWsq} \frac{C^2_{v,i} + C^2_{a,i}}{\CWsq} \frac{z}{M^2_{\smallz}} \;,
\end{equation}
which is the coefficient of $\delta(1-z)$ in the Born cross section,
with $C_{v,i},\,C_{a,i}$ the coefficients of the vector and axial-vector couplings of the $Z$ boson to a fermion $i$.

The functions $\mathrm{\textbf{P}}^{(n)}_{\smallz} \pts{z}$ read:
\begingroup
\allowdisplaybreaks
\begin{align}
\mathrm{\textbf{P}}^{(-4)}_{\smallz} \pts{z} =& \,\deltz , \\
\mathrm{\textbf{P}}^{(-3)}_{\smallz} \pts{z} =& -\deltz - 2 \pts{1+z^2} \PlusD{\frac{1}{1-z}} , \\
\mathrm{\textbf{P}}^{(-2)}_{\smallz} \pts{z} =& - 2 \pts{2+3z^2}\frac{\Log{z}}{1-z} - 8 \zed{2} \deltz +  4z\PlusD{\frac{1}{1-z}}  \nonumber \\
+& 8 \pts{1+z^2} \PlusD{\frac{\Log{1-z}}{1-z}} , \\
\mathrm{\textbf{P}}^{(-1)}_{\smallz} \pts{z} =&  4 (1+z) \Li{2}{z}
- \left(\frac{17 + 27 z^2}{4} \right)  \frac{\PLog{z}{2}}{1-z}
+20 \left(1+z^2\right) \Log{1-z} \frac{ \Log{z}}{1-z} \nonumber \\
+& z (11-z) \frac{\Log{z}}{1-z}
+(8 \zed{2}-20 \zed{3}) \deltz
-16 \left(1+z^2 \right) \PlusD{\frac{\PLog{1-z}{2}}{1-z}} \nonumber \\
-&16 z \PlusD{\frac{\Log{1-z}}{1-z}}
+ \left(12 \zed{2}-\frac{15}{2}\right) \left(1+z^2\right) \PlusD{\frac{1}{1-z}} \nonumber \\
+ & z \left(15 + 8 \zed{2} z \right) \PlusD{\frac{1}{1-z}} , \\
\mathrm{\textbf{P}}^{(0)}_{\smallz} \pts{z} =& -17(1-z) + (31 - 4 z) \zed{2} -(19-26 z) \Log{z}
+(1-41 z) \Log{1-z}\frac{\Log{z}}{1-z} \nonumber \\
+& \left(6 + 22 z^2 \right) \frac{(\Li{3}{z}-\zed{3})}{1-z}
+\left(1-15 z^2\right) \Li{2}{z} \frac{\Log{z}}{1-z} +(1+4 z) \Li{2}{z} \nonumber \\
-& (1+z) (16 \Li{2}{z}-17 \zed{2}) \Log{1-z}
-17 (1+z) \Li{3}{1-z} \nonumber \\
- & \left(\frac{1}{4} - 12 z + \frac{3 z^2}{4}-\left(21 + 23 z^2 \right) \Log{1-z}  \right)\frac{\PLog{z}{2}}{1-z}
+ \left(25 + 41 z^2 \right) \zed{2} \frac{\Log{z}}{1-z} \nonumber \\
-& \left(\frac{13}{4} + \frac{61 z^2}{12} \right) \frac{\PLog{z}{3}}{1-z}
-16 \left(3 + 2 z^2\right) \PLog{1-z}{2} \frac{\Log{z}}{1-z} \nonumber \\
+& 20 (\zed{3} + \frac{4}{5} \zed{4}) \deltz 
+\left(40 \left(1 + z^2\right) \zed{3} - 32 \zed{2}\right) \PlusD{\frac{1}{1-z}} \nonumber \\
+& \ptsq{(30-63 \zed{2}) \left(1+z^2 \right) -2 (30 z +\zed{2} ) } \PlusD{\frac{\Log{1-z}}{1-z}} \nonumber \\
+& 32 z \PlusD{\frac{\PLog{1-z}{2}}{1-z}}
+ \frac{64}{3} \left(1+z^2\right) \PlusD{\frac{\PLog{1-z}{3}}{1-z}} \, .
\end{align}
\endgroup

\subsubsection{The subprocess $q_i\,g \to Z \,q_i\,\gamma$}

We consider here the tree-level process:
\begin{equation}
q_{i} \, g \rightarrow Z \, q_{i} \, \gamma \, .
\end{equation}
The partonic cross section is obtained by summing the following combination of MIs
\be
\hat{\sigma}^{RR}_{q_{i} g\to Z q_i \gamma}\pts{z;\eps} = \sum_{k=1}^{18} c_k^{q_{i} g\to Z q_i \gamma}(z;\eps) I_k(z;\eps) \, ,
\ee
with $c_k^{q_{i} g\to Z q_i \gamma}=0$ for $k=3,5\!-\!11,14$. The cross section
expressed as a Laurent series in the dimensional regulator $\eps$ has the form:
\begin{equation}
\hat{\sigma}^{RR}_{q_{i} g\to Z q_i \gamma}\pts{z;\eps} =  2 \hat{\sigma}^{0}_{q_{i} \bar{q}_{i}\to Z}(z) \, Q^2_{i} \frac{\alpha}{2 \pi}\frac{\alpha_{S}}{2 \pi} \mathcal{A}_{\smallz}(\eps) \sum_{n=-3}^0 \eps^{n} \mathrm{\textbf{Q}}^{(n)}_{\smallz} \pts{z} \, ,
\end{equation}
with $\mathcal{A}_{\smallz}$ and $\hat{\sigma}^{0}_{q_{i} \bar{q}_{i}\to Z}(z)$ as earlier defined.
The functions $\mathrm{\textbf{Q}}^{(n)}_{\smallz} \pts{z}$ read:
\begingroup
\allowdisplaybreaks
\begin{align}
\mathrm{\textbf{Q}}^{(-3)}_{\smallz} \pts{z} =& - \frac{(1-z)^2 + z^2}{2} , \\
\mathrm{\textbf{Q}}^{(-2)}_{\smallz} \pts{z} =&  \frac{5}{16} + \frac{1-z}{4} \pts{ 7 - 9 (1-z) } -\frac{3}{4} z^2 \Log{z} \nonumber \\
+& \ptsq{\frac{(1-z)^2 + z^2}{2}} \pts{4 \Log{1-z} - \frac{9}{4} \Log{z}} , \\
\mathrm{\textbf{Q}}^{(-1)}_{\smallz} \pts{z} =& \frac{5}{16} +\frac{113 z}{16} -\frac{61 z^2}{8} +\left(9 (1-z)^2-7 (1-z)-\frac{5}{4}\right) \Log{1-z} \nonumber \\ 
-& 3 z^2 (\Li{2}{z}-\zed{2}) -\frac{7}{8} z^2 \PLog{z}{2} + \left(\frac{7}{16} + \frac{13}{2} z -\frac{27}{4} z^2  \right) \Log{z} \nonumber \\ 
+& \ptsq{\frac{(1-z)^2 + z^2}{2}} \left( \frac{19}{2}\Log{z}\Log{1-z} + \frac{1}{2} \Li{2}{z} - 8 \PLog{1-z}{2} \right. \nonumber \\ 
-& \left. \frac{19}{8} \PLog{z}{2} + 8 \zed{2} \right) , \\
\mathrm{\textbf{Q}}^{(0)}_{\smallz} \pts{z} =& \frac{13}{4}+\frac{165 z}{8}-\frac{389 z^2}{16}
+ \ptsq{ \frac{(1-z)^2 + z^2}{2} } 
\left(\frac{13}{2}\Li{3}{z} +4 \Li{3}{-z} \right. \nonumber \\
-& 4 \Li{3}{1-z} -2 \Li{2}{z} \Log{1-z}-2 \Li{2}{-z} \Log{z}
-4 \Li{2}{z} \Log{z} \nonumber \\
-& 32 \zed{2} \Log{1-z}+\frac{35}{2} \zed{2} \Log{z}+\frac{32}{3} \PLog{1-z}{3}-\frac{47 \PLog{z}{3}}{24} +19 \zed{3} \nonumber \\
-& 20 \Log{z} \PLog{1-z}{2}+\frac{35}{4} \PLog{z}{2} \Log{1-z}\bigg)
+7 z^2 \pts{\Li{3}{z}-\zed{3}} \nonumber \\
+&13 z^2 \Li{3}{1-z}
+12 z^2 \Log{1-z} \left(\Li{2}{z}+\frac{1}{2} \Log{1-z} \Log{z}-\zed{2}\right) \nonumber \\
-&\frac{13}{2} z^2 \Li{2}{z} \Log{z}+\left((1-z)^2-4 z^2\right) \pts{\Li{2}{-z}+\Log{z} \Log{1+z}} \nonumber \\
+& \left(\frac{21}{8} +\frac{5 z}{2}-\frac{15 z^2}{2} \right) \Li{2}{z}
-\left(\frac{3}{8} +\frac{55}{2}z -\frac{51}{2} z^2 \right) \zed{2}
+\frac{11}{2} \zed{2}  z^2 \Log{z} \nonumber \\
-&\frac{5}{8} z^2 \PLog{z}{3}
+\left(\frac{15}{32} +6 z -\frac{39 z^2}{8} \right) \PLog{z}{2}
+\frac{1}{2} z^2 \Log{1-z} \PLog{z}{2} \nonumber \\
-& \left(\frac{3}{2} -22 z +18 z^2 \right) \PLog{1-z}{2}
+\left(\frac{3}{16} +\frac{251 z}{16} -\frac{183 z^2}{8} \right) \Log{z} \nonumber \\
+& \left(\frac{7}{8} -\frac{47 z}{2} + \frac{39 z^2}{2}\right) \Log{1-z} \Log{z}
-\left(\frac{5}{4}+\frac{113 z}{4} - \frac{61 z^2}{2}\right) \Log{1-z} \, .
\end{align}
\endgroup

 \subsubsection{The subprocess $q_i\,\gamma \to Z \,q_i\, g$}

We present here the cross section for the tree-level process:
\begin{equation}
q_{i} \, \gamma \rightarrow Z \, q_{i} \, g \, .
\end{equation}
The result is obtained by summing the following combination of MIs
\be
\hat{\sigma}^{RR}_{q_i \gamma\to Z q_i g}\pts{z;\eps} = \sum_{k=1}^{18} c_k^{q_i \gamma\to Z q_i g}(z;\eps) I_k(z;\eps) \, ,
\ee
with $c_k^{q_i \gamma\to Z q_i g}=0$ for $k=3,5\!-\!11,14$. The cross section can be rewritten as
\begin{equation}
\hat{\sigma}^{RR}_{q_i \gamma\to Z q_i g}\pts{z;\eps} =  4 \hat{\sigma}^{0}_{q_{i} \bar{q}_{i}\to Z}(z) \, C_{A} \, C_{F} \, Q^2_{i} \frac{\alpha}{2 \pi}\frac{\alpha_{S}}{2 \pi} \mathcal{A}_{\smallz}(\eps) \sum_{n=-3}^0 \eps^{n} \mathrm{\textbf{G}}^{(n)}_{\smallz} \pts{z} \, .
\end{equation}
For the functions $\mathrm{\textbf{G}}^{(n)}_{\smallz} \pts{z}$ we have $\mathrm{\textbf{G}}^{(n)}_{\smallz} \pts{z} = \mathrm{\textbf{Q}}^{(n)}_{\smallz} \pts{z}$. Since the $Z$ boson does not couple to the photon, the two subprocesses $q_i\,\gamma \to Z \,q_i\, g$ and $q_i\,g \to Z \,q_i\,\gamma$ have the same Feynman diagrams upon exchanging the photon with the gluon. Therefore, the two cross sections are identical apart from a color factor due to the sum over final state color in one case or average over initial state color configurations in the other case.

\subsubsection{The subprocess $g\,\gamma \to Z \,q_i\, \bar{q}_i$}

Finally, we present the cross section of the tree-level process:
\begin{equation}
g \, \gamma \rightarrow Z \, q_{i} \, \bar{q}_{i} \, .
\end{equation}
The result is obtained by summing the following combination of MIs
\be
\hat{\sigma}^{RR}_{g \gamma\to Z q_i \bar{q}_i}\pts{z;\eps} = \sum_{k=1}^{27} c_k^{g \gamma\to Z q_i \bar{q}_i}(z;\eps) I_k(z;\eps) \, ,
\ee
with $c_k^{g \gamma\to Z q_i \bar{q}_i}=0$ for $k=3,5\!-\!8,10\!-\!23,25$. We rewrite the cross section as
\begin{equation}
\hat{\sigma}^{RR}_{g \gamma\to Z q_i \bar{q}_i}\pts{z;\eps} = \hat{\sigma}^{0}_{q_{i} \bar{q}_{i}\to Z}(z) \, C_{A} \, Q^2_{i} \frac{\alpha}{2 \pi}\frac{\alpha_{S}}{2 \pi} \mathcal{A}_{\smallz}(\eps) \sum_{n=-2}^0 \eps^{n} \mathrm{\textbf{T}}^{(n)}_{\smallz} \pts{z} \, .
\end{equation}
The functions $\mathrm{\textbf{T}}^{(n)}_{\smallz} \pts{z}$ read:
\begingroup
\allowdisplaybreaks
\begin{align}
\mathrm{\textbf{T}}^{(-2)}_{\smallz} \pts{z} =& - 2(1-z)(1+3 z) - (1+2 z)^2 \Log{z} , \\
\mathrm{\textbf{T}}^{(-1)}_{\smallz} \pts{z} =&  -\frac{3}{2} - 26 z + \frac{55}{2}z^2 +  4 (1+2 z)^2 \left(\zed{2}-\Li{2}{z}-\frac{\PLog{z}{2}}{4}\right) \nonumber \\
-& 4 z (1-2 z) \Log{z} + 8 (1-z) (1+3 z) \Log{1-z} , \\
\mathrm{\textbf{T}}^{(0)}_{\smallz} \pts{z} = & -\frac{17}{4} -90 z+\frac{377 z^2}{4}
+4 \left(5 + 6 z -7 z^2 \right) \Li{2}{z}+2 (1-z)^2 \Log{1-z} \PLog{z}{2} \nonumber \\
-&4 \left((1+z)^2-2 z^2\right) \Li{3}{-z}
+16 (1+2 z)^2 \bigg(\Li{3}{1-z}+\Li{2}{z} \Log{1-z} \nonumber \\
-& \zed{2} \Log{1-z}+\frac{1}{2} \Log{z} \PLog{1-z}{2}\bigg)
+4 \left(2 (1+z)^2-z^2\right) \Li{2}{-z} \Log{z} \nonumber \\
+& \left(8 + 56 z + 44 z^2\right) \Li{3}{z}
-6 \left(1+8 z+6 z^2\right) \Li{2}{z} \Log{z} \nonumber \\
+& (1+z)^2 \left(8 \Li{3}{\frac{1}{1+z}}+4 \zed{2} \Log{1+z}-\frac{4}{3} \PLog{1+z}{3}+6 \PLog{z}{2} \Log{1+z}\right) \nonumber \\
+& 4 (1+z) (\Li{2}{-z}+\Log{z} \Log{1+z})
-2 \left(1 - 5z + 10 z^2\right) \zed{2} \nonumber \\
+& 4 \zed{2} \left(2 + 7z + 7 z^2 \right) \Log{z}
-2 \left(9 + 38 z + 24 z^2 \right) \zed{3}
-\frac{4}{3} \left(1 + 3 z + 3 z^2 \right) \PLog{z}{3} \nonumber \\
-& \left(\frac{5}{2} +7 z - 2 z^2\right) \PLog{z}{2}
+20 (1-z) (1+ 3 z) \Log{1-z} \left(\Log{z}-\frac{4}{5} \Log{1-z}\right)\nonumber \\
+& \left(6 + 104 z -110 z^2 \right) \Log{1-z} -\left(\frac{13}{2} + 46 z - 70 z^2 \right) \Log{z} \, .
\end{align}
\endgroup

\section{Conclusions}
\label{conclu}
In this work we presented the analytical calculation of the total cross sections
of all the partonic subprocesses that contribute at \oaas to inclusive
single on-shell gauge boson production, with two additional partons in the final state (double-real corrections).
The results are expressed as Laurent series in the dimensional regularization parameter, contain HPLs up to weight 3, and can be cast in terms of logarithms and ordinary Euler polylogarithmic functions.
This calculation required the evaluation of 14 new two-loop cut MIs with two internal massive lines, whose explicit expressions are presented in the Appendices.
These results are part of the \oaas corrections to the total cross section for inclusive on-shell single gauge boson production.
The complete evaluation of the latter requires the calculation of the two-loop virtual corrections to the lowest-order process for gauge boson production (double-virtual corrections)
and of the virtual corrections to the processes with a single emission of an additional real parton (real-virtual corrections).

\section{Acknowledgments}
\label{acknow}
We would like to thank Kirill Melnikov for interesting comments on the evaluation of the Master Integrals.
We would like to thank the Galileo Galilei Institute for Theoretical Physics, where part of this work was carried out, for the hospitality.
AV would like to thank 
the Department of Physics of the SUNY University at Buffalo for the warm hospitality and 
the Kavli Institute for Theoretical Physics at Santa Barbara for hospitality and support during the workshop ``LHC Run II and the Precision Frontier'',
where part of this work was carried out.
RB was partly supported by European Community Seventh Framework Programme FP7/2007-2013, under grant agreement N.302997. 
FB's research was supported in part by the Swiss National Science Foundation (SNF) under contract BSCGI0\_157722. RM is supported by the National Science Foundation through awards number PHY-1417317 and PHY-1619877. 
AV is supported by the European Commission through the HiggsTools Initial Training Network PITN-GA2012-316704.

\appendix

\section{Analytical expressions of the Master Integrals}
\label{sec:appMI}
In this Appendix we present the MIs relevant for the evaluation of the total cross sections of processes
(\ref{proc1})-(\ref{proc8}).  We recall that these are phase-space integrals with phase-space measure
\begin{equation}
\rmd \Phi_{3} = \frac{\rmd^{d}p_{4} \, \rmd^{d}p_{5}}{(2\pi)^{2d-3}} \,\delta(p^2_{4})\,\delta(p^2_{5})\,\delta\pts{\pts{p_{1}+p_{2} - p_{4}-p_{5}}^2-M^2_{\smallv}} , \\
\end{equation}
where the last term is the on-shellness delta function of the vector boson. 

In the following, we separate the result of each MI $I_{k}(z;\eps)$ into soft and hard part (borrowing the notation from Ref.~\cite{Anastasiou:2012kq}):
\begin{equation}
I_{k}(z;\eps) = I^S_{k}(z;\eps) + I^H_{k}(z;\eps),
\end{equation}
where the soft part comes first and both terms are expanded in $\eps$. For each MI we present the two expansions
truncated at the last order in $\eps$ that is relevant for the cross sections (we note that for some MIs the last
relevant order is different between soft and hard part). In addition, in the ancillary \texttt{Mathematica} file
we present the expansion of each MI truncated at the order that contains at most HPLs of weight 4. The soft part
of all the MIs is also available exact to all orders in $\eps$ in
Eqs.~\eqref{eq:softfirst}-\eqref{eq:softlast}\footnote{\,The soft limits collected in Appendix \ref{app:MIsoft}
differ from the soft part of the MIs presented here by the overall $z^{a+b\eps}$ factors ($a,b\in\mathbb{Z}$),
which in the soft limit are exactly 1.}.

We remark that all MIs are written in terms of HPLs of argument $z$, linear weights $\{0,\pm1,-\frac{1}{4}\}$, and
the non-linear weight $-\frac{r_0}{4}$. As discussed in Section \ref{sec:eval}, these HPLs can be converted into
HPLs of argument $\xi(z)$ and linear weights $\{0,\pm1,a_1,a_2\}$, where $a_1$ and $a_2$ are defined in
Eq.~\eqref{eq:defa1a2}. In the ancillary file we provide the explicit transformations for all the HPLs involved in
this calculation and containing the non-linear weight $-\frac{r_0}{4}$. 

Lastly, in all the expressions below we extract an overall normalization factor ${\cal N}(\eps)$ defined in Eq.~(\ref{eq:normfac}).

\subsection{Definitions and results expanded in $\eps$}
\label{app:MIfull}

\begingroup
\allowdisplaybreaks
\begin{align}
I_{1}(z;\eps) = & \vcenter{\hbox{\scalebox{1.2}{\masterone}}} = \displaystyle{\int \rmd \Phi_{3} } \nonumber \\
= & \,\Ncal(\eps) M^2_{\smallv} \, z^{-1+ 2\eps }\pts{1-z}^{3-4\eps} \bigg\lbrace \frac{1}{6} +\frac{11}{9}\eps + 
\Bigl( \frac{170}{27} - \frac{4 \zed{2}}{3} \Bigr) \eps^2 + \Bigl(\frac{2300}{81}-\frac{88 \zed{2}}{9} \nn\\ 
- & \frac{10 \zed{3}}{3} \Bigr)\eps^3 + \Bigl( \frac{29288}{243} -\frac{1360 \zed{2}}{27}-\frac{220 \zed{3}}{9} 
+ \frac{8 \zed{4}}{3} \Bigr) \eps^4 + \Ocal(\eps^5) \bigg\rbrace \nonumber \\
+ & \,\Ncal(\eps) M^2_{\smallv} \, z^{-1+2\eps} \pts{1-z}^{3-4\eps} \bigg\lbrace  \frac{z \Hpl{(0;z)}}{(1-z)^3} 
+ \frac{ 2 + 5 z - z^2}{6 (1-z)^2}  + \frac{\eps}{(1-z)^3} \Bigl( \frac{73}{36} +\frac{11 z^3}{9} \nn\\
- & \frac{83 z^2}{12}-4 z \zed{2}+\frac{11 z}{3} -\frac{z^2}{2} \Hpl{(0;z)} 
+ 3 z \Hpl{(0;z)} - z \Hpl{(0,0;z)}-4 z \Hpl{(1,0;z)} \Bigr) \nn\\
+ & \frac{\eps^2}{(1-z)^3} \biggl[ \frac{1745}{216} +\frac{170 z}{9} 
- \frac{2395 z^2}{72} +\frac{170 z^3}{27}  - \zed{2}\Bigl( \frac{5}{3} + 16 z - 9 z^2 +\frac{4}{3} z^3 \Bigr) \nn\\
- & \Bigl( 3 - \frac{z}{2} \Bigr) z \Hpl{(0,0;z)} + \Bigl( 6 - \frac{13}{4} z - 2 \zed{2}  \Bigr) z \Hpl{(0;z)} 
+ (1 - 12 z +  z^2 ) \Hpl{(1,0;z)} \nn\\
+ & 16 z \zed{2} \Hpl{(1;z)} + z\Hpl{(0,0,0;z)}+6 z \Hpl{(0,1,0;z)} + 4 z \Hpl{(1,0,0;z)} + 16 z \Hpl{(1,1,0;z)} \nonumber \\
- & 8 z \zed{3} \biggr] + \frac{\eps^3}{(1-z)^3} \biggl[ \frac{33265}{1296} -\frac{175 \zed{2}}{18}-\frac{5 \zed{3}}{3} + z \Bigl( \frac{2300}{27} - \frac{160 \zed{2}}{3} - 34 \zed{3} +20 \zed{4} \Bigr) \nn\\
+ & z^2 \Bigl( -\frac{60155}{432} + \frac{371 \zed{2}}{6}+19 \zed{3} \Bigr) + z^3 \Bigl(\frac{2300}{81} -\frac{88 \zed{2}}{9}
- \frac{10 \zed{3}}{3} \Bigr) + \Bigl( \frac{15}{2} -6 \zed{2} -2 \zed{3} \nn\\
- & \frac{115}{8} z + \zed{2} z  \Bigr) z \Hpl{(0;z)} - 4 \Bigl( \zed{2} ( 1 - 12 z + z^2 ) - 8 z \zed{3} \Bigr) \Hpl{(1;z)} + \Bigl( \frac{13}{4} z + 2 \zed{2} \nn\\
- & 6 \Bigr) z \Hpl{(0,0;z)} + \Bigl( \frac{13}{2} +8 z \zed{2} -24 z + \frac{13}{2} z^2 \Bigr) \Hpl{(1,0;z)} +\Bigl( 3 - \frac{z}{2} \Bigr) z \Hpl{(0,0,0;z)} \nonumber \\ 
- & ( 1 -12 z + z^2 ) ( 4 \Hpl{(1,1,0;z)} + \Hpl{(1,0,0;z)} ) + 3 (6 - z ) z \Hpl{(0,1,0;z)} \nn\\
- & 24 z \zed{2} \Hpl{(0,1;z)} - 64 z \zed{2} \Hpl{(1,1;z)} - z \Hpl{(0,0,0,0;z)}  - 6 z \Hpl{(0,0,1,0;z)} \nn\\
- & 6 z \Hpl{(0,1,0,0;z)} - 24 z \Hpl{(0,1,1,0;z)} -4 z \Hpl{(1,0,0,0;z)} - 24 z \Hpl{(1,0,1,0;z)} \nonumber \\
-& 16 z \Hpl{(1,1,0,0;z)} -64 z \Hpl{(1,1,1,0;z)} \bigg] + \Ocal(\eps^4) \bigg\rbrace \, . 
\label{eq:MasterIntegralOneT} \\
& \nonumber \\
I_{2}(z;\eps) = & \vcenter{\hbox{\scalebox{1.2}{\mastertwo}}} = \displaystyle{\int \rmd \Phi_{3} \pts{p_{1}-p_{4}}^2 } \nonumber \\
=& \,\Ncal(\eps) M^4_{\smallv} \, z^{-2+2\eps}\pts{1-z}^{4-4 \eps} \bigg\lbrace -\frac{1}{24} - \frac{11}{36} \eps 
+ \Bigl( \frac{\zed{2}}{3} -\frac{85}{54} \Bigr) \eps^2 
+ \Bigl( \frac{22}{9} \zed{2} + \frac{5}{6} \zed{3} \nn\\
- & \frac{575}{81} \Bigr) \eps^3 + \Ocal(\eps^4)\bigg\rbrace \nonumber \\
+ & \,\Ncal(\eps) M^4_{\smallv} \, z^{-2+2\eps}\pts{1-z}^{4-4 \eps}\bigg\lbrace \frac{1}{(1-z)^3} \left(-\frac{1}{8} -\frac{13 z}{24} +\frac{5 z^2}{24} -\frac{z^3}{24} \right) - \frac{z}{2} \frac{\Hpl{(0;z)}}{(1-z)^4} \nn\\
+ & \frac{\eps}{(1-z)^4} \biggl[ (1-z) \Bigl( -\frac{7}{9} -\frac{83 z}{24} +\frac{35 z^2}{24} -\frac{11 z^3}{36} \Bigr) 
+ 2 \zed{2} z - \frac{z}{12} ( 18 - 6 z + z^2 ) \Hpl{(0;z)} \nn\\
+ & \frac{z}{2} \Hpl{(0,0;z)} + 2 z \Hpl{(1,0;z)} \biggr] + \frac{\eps^2}{(1-z)^4} \biggl[ (1-z) \Bigl( -\frac{695}{216} 
-\frac{2225 z}{144} +\frac{1025 z^2}{144} \nn\\
- & \frac{85 z^3}{54} \Bigr) + \Bigl( \frac{2}{3} + \frac{53}{6} z -7 z^2 + \frac{13}{6} z^3 - \frac{z^4}{3} \Bigr) \zed{2} + \frac{z}{12} ( 18 - 6 z + z^2 ) \Hpl{(0,0;z)} \nonumber \\
+ & \Bigl( \zed{2} -\frac{19}{6} +\frac{35}{12}  z - \frac{13}{24} z^2 \Bigr) z \Hpl{(0;z)} + 4 \zed{3} z 
- 8 \zed{2} z \Hpl{(1;z)} + \frac{1}{6} (-2 + 33 z - 6 z^2 \nn\\
+ & z^3 )\Hpl{(1,0;z)} - \frac{z}{2} \Hpl{(0,0,0;z)} - 3z \Hpl{(0,1,0;z)} -2z \Hpl{(1,0,0;z)} -8z \Hpl{(1,1,0;z)} \bigg] 
\nonumber \\
+ &  \frac{\eps^3}{(1-z)^4} \biggl[ \frac{73}{18}\zed{2} +\frac{5}{6}\zed{3} -\frac{14155}{1296} + \frac{1123}{36} \zed{2} z 
-10 \zed{4} z + \frac{107 \zed{3} z}{6} -\frac{128005 z}{2592} \nn\\
- & \frac{269}{6} \zed{2} z^2 -14 \zed{3} z^2 +\frac{39145 z^2}{432} +\frac{547}{36} \zed{2} z^3 
+ \frac{29}{6}\zed{3} z^3 -\frac{96955 z^3}{2592}
- \frac{22}{9} \zed{2} z^4 - \frac{5}{6}\zed{3} z^4 \nn\\
+ & \frac{575 z^4}{81} + \Bigl( 3 \zed{2} -5 +\zed{3} + \frac{95}{8} z -\zed{2} z-\frac{115}{48} z^2
+\zed{2} \frac{z^2}{6} \Bigr) z \Hpl{(0;z)} - 16 \zed{3} z \Hpl{(1;z)}  \nn\\
+ & \Bigl( \frac{4}{3} -22 z  +4 z^2  -\frac{2}{3} z^3 \Bigr) \zed{2} \Hpl{(1;z)} 
+ \left(\frac{19}{6} -\zed{2} -\frac{35}{12} z +\frac{13}{24} z^2 \right) z \Hpl{(0,0;z)} \nonumber \\
- & \Bigl( \frac{13}{6} - \frac{39}{4}  z + 4  \zed{2} z + \frac{11}{2}  z^2 - \frac{13}{12}  z^3 \Bigr) \Hpl{(1,0;z)} -\frac{z}{12} ( 18 - 6 z + z^2) ( \Hpl{(0,0,0;z)} \nn\\
+ & 6 \Hpl{(0,1,0;z)} ) + \frac{1}{6} ( 2 - 33 z + 6 z^2 - z^3) ( \Hpl{(1,0,0;z)} 
+ 4 \Hpl{(1,1,0;z)} ) \nonumber \\
+ & \frac{z}{2} \Hpl{(0,0,0,0;z)} + 3 z \Hpl{(0,0,1,0;z)} +3 z \Hpl{(0,1,0,0;z)} + 12 z \Hpl{(0,1,1,0;z)} \nn\\
+ & 2 z \Hpl{(1,0,0,0;z)} +12 z \Hpl{(1,0,1,0;z)} + 8  z \Hpl{(1,1,0,0;z)} + 32 z\Hpl{(1,1,1,0;z)} \nn\\
+ & 12 \zed{2} z \Hpl{(0,1;z)} + 32 \zed{2} z \Hpl{(1,1;z)} \bigg] + \Ocal(\eps^4)\bigg\rbrace
\label{eq:MasterIntegralTwoT} \, . \\
& \nonumber \\
I_{3}(z;\eps) = & \vcenter{\hbox{\scalebox{1.2}{\masterthree}}} = \displaystyle{\int \rmd \Phi_{3} \frac{1}{\pts{p_{1}-p_{4}-p_{5}}^2 \pts{p_{2}-p_{4}-p_{5}}^2} } \nonumber \\
= & \,\Ncal(\eps) \frac{1}{M^{2}_{\smallv}} z^{1+2\eps} \pts{1-z}^{1-4 \eps}\bigg\lbrace 1+8\eps+ \Ocal(\eps^2)\bigg\rbrace \nonumber \\
+ & \,\Ncal(\eps) \frac{1}{M^{2}_{\smallv}} z^{1+2\eps} \pts{1-z}^{1-4 \eps}\bigg\lbrace  \frac{\Hpl{(0,0;z)}}{1-z} + \frac{\Hpl{(1,0;z)} + \zed{2}}{1-z} - 1 \nn\\
+ & \frac{\eps}{1-z} \bigl[ -8(1-z) + 2 \zed{2}-\zed{3} -5 \zed{2} \Hpl{(0;z)} -6 \zed{2} \Hpl{(1;z)} 
+ 2 \Hpl{(0,0;z)} + 2 \Hpl{(1,0;z)} \nn\\
- & 2 \Hpl{(0,0,0;z)}-5 \Hpl{(0,1,0;z)} - 3 \Hpl{(1,0,0;z)} - 6 \Hpl{(1,1,0;z)} \bigr] 
+ \Ocal(\eps^2) \bigg\rbrace \label{eq:MasterIntegralThreeT} \, . \\
& \nonumber \\
I_{4}(z;\eps) = & \vcenter{\hbox{\scalebox{1.2}{\masterfour}}} = \displaystyle{\int \rmd \Phi_{3} \frac{1}{\pts{p_{1}-p_{4}}^2 \pts{p_{2}-p_{5}}^2} } \nonumber \\
=& \,\Ncal(\eps) \frac{1}{M^{2}_{\smallv}} z^{1+2\eps}\pts{1-z}^{1-4 \eps}\bigg\lbrace \frac{1}{\eps^2} + \frac{4}{\eps} +16 
- 8 \zed{2} + \Ocal(\eps)\bigg\rbrace \nonumber \\
+& \,\Ncal(\eps) \frac{1}{M^{2}_{\smallv}} z^{1+2\eps} \pts{1-z}^{1-4 \eps} \bigg\lbrace  
- \frac{1}{\eps^2} \Bigl( 1+\frac{\Hpl{(0;z)}}{1-z} \Bigr) 
+ \frac{1}{\eps} \Bigl( 4 \, \frac{(\Hpl{(1,0;z)}+\zed{2}) -4  }{1-z} \nn\\
+ & 2 \, \frac{\Hpl{(0,0;z)}}{1-z} \Bigr) 
-\frac{8}{1-z} \Bigl( \frac{\zed{2}}{4} \Hpl{(0;z)} + (4 - 2 \zed{2})(1-z) + 2 \zed{2} \Hpl{(1;z)}+\frac{1}{2} \Hpl{(0,0,0;z)}\nn\\
+ & \frac{5}{4} \Hpl{(0,1,0;z)} - \frac{\zed{3}}{2} +  \Hpl{(1,0,0;z)} + 2 \Hpl{(1,1,0;z)} \Bigr) 
+ \Ocal(\eps)\bigg\rbrace \label{eq:MasterIntegralFourT} \, . \\
& \nonumber \\
I_{5}(z;\eps) = & \vcenter{\hbox{\scalebox{1.2}{\masterfive}}} = \displaystyle{\int \rmd \Phi_{3} \frac{1}{\pts{p_{1}-p_{4}-p_{5}}^2 \left[\pts{p_{1}+p_{2}-p_{4}}^2-M^2_{\smallw}\right] }} \nonumber \\
=&  \,\Ncal(\eps) \frac{1}{M^{2}_{\smallw}} z^{1+2\eps} \pts{1-z}^{1-4 \eps}\bigg\lbrace -\zed{2} +\Ocal(\eps)\bigg\rbrace \nonumber \\
+& \,\Ncal(\eps) \frac{1}{M^{2}_{\smallw}} z^{1+2\eps} \pts{1-z}^{1-4 \eps}\bigg\lbrace \zed{2} + 2 \zed{2} \frac{\Hpl{(0;z)} }{1-z} + \frac{\Hpl{(0,0,0;z)} }{1-z} \nonumber \\
+& \frac{\Hpl{(0,1,0;z)} + 2 \zed{3}}{1-z} 
+ \Ocal(\eps) \bigg\rbrace \label{eq:MasterIntegralFiveT} \, . \\
& \nonumber \\
I_{6}(z;\eps) = & \vcenter{\hbox{\scalebox{1.2}{\mastersix}}} = \displaystyle{\int \rmd \Phi_{3} \frac{1}{\pts{p_{1}-p_{4}-p_{5}}^4 \left[\pts{p_{1}+p_{2}-p_{4}}^2-M^2_{\smallw}\right] }} \nonumber \\
= & \,\Ncal(\eps) \frac{1}{M^{4}_{\smallw}} z^{1+2\eps} \pts{1-z}^{-4 \eps}\bigg\lbrace \frac{1}{6\eps^2}- 2 \zed{2} - 8 \zed{3} \eps - 4 \zed{4} \eps^{2} +\Ocal(\eps^3)\bigg\rbrace \nonumber \\
+ &  \,\Ncal(\eps) \frac{1}{M^{4}_{\smallw}} z^{1+2\eps} \pts{1-z}^{-4 \eps}\bigg\lbrace \frac{1}{\eps^2} (1-z) + \frac{1}{\eps} \frac{\Hpl{(0;z)}}{6 } - \frac{5}{6} \Hpl{(0,0;z)} - \Hpl{(1,0;z)} \nn\\
- & (3-2 z) \zed{2} + \eps \, \Bigl( 5 \zed{2} \Hpl{(0;z)}+4 \zed{2} \Hpl{(1;z)} + \frac{13}{6} \Hpl{(0,0,0;z)}+5 \Hpl{(0,1,0;z)} \nn\\
+ &\Hpl{(1,0,0;z)}+4 \Hpl{(1,1,0;z)} + (8 z -3) \zed{3} \Bigr) 
+ \Ocal(\eps^2) \bigg\rbrace  \label{eq:MasterIntegralSixT} \, . \\
& \nonumber \\
I_{7}(z;\eps) = & \vcenter{\hbox{\scalebox{1.2}{\masterseven}}} = \displaystyle{\int \rmd \Phi_{3} \frac{1}{\pts{p_{1}-p_{4}-p_{5}}^2 \pts{p_{1}-p_{5}}^2 \left[\pts{p_{1}+p_{2}-p_{4}}^2-M^2_{\smallw}\right] } } \nonumber \\
=& \,\Ncal(\eps) \frac{1}{M^{4}_{\smallw}} z^{2+2\eps} \pts{1-z}^{-4 \eps}\bigg\lbrace - \frac{1}{3\eps^3} + \frac{3 \zed{2}}{\eps}  +9 \zed{3} + \Ocal(\eps)\bigg\rbrace \nonumber \\
+& \,\Ncal(\eps) \frac{1}{M^{4}_{\smallw}} z^{2+2\eps} \pts{1-z}^{-4 \eps}\bigg\lbrace  -\frac{1}{\eps^2} \frac{4 \Hpl{(0;z)}}{3}
- \frac{1}{\eps} \Bigl( \frac{\Hpl{(0,0;z)}}{3}  -  5 \Hpl{(1,0;z)} \nn\\
- & 5 \zed{2} \Bigr) + 12 \zed{2} \Hpl{(0;z)}-20 \zed{2} \Hpl{(1;z)}+\frac{5}{3} \Hpl{(0,0,0;z)} + 3 \Hpl{(1,0,0;z)} 
- 20 \Hpl{(1,1,0;z)} \nn\\
+ & 17 \zed{3} 
+ \Ocal(\eps)\bigg\rbrace \label{eq:MasterIntegralSevenT} \, . \\
& \nonumber \\
I_{8}(z;\eps) = & \vcenter{\hbox{\scalebox{1.2}{\mastereight}}} = \displaystyle{\int \rmd \Phi_{3}\frac{1}{\pts{p_{1}-p_{4}}^2 \pts{p_{2}-p_{4}}^2 \pts{p_{1}-p_{4}-p_{5}}^2 \pts{p_{2}-p_{4}-p_{5}}^2} } \nonumber \\ 
= & \,\Ncal(\eps) \frac{1}{M^{6}_{\smallv}} z^{3+2\eps} \pts{1-z}^{-1-4 \eps} \bigg\lbrace -\frac{1}{\eps^3} + \frac{8\zed{2}}{\eps} + 20 \zed{3} - 16 \zed{4} \eps + \Ocal(\eps^2) \bigg\rbrace \nonumber \\
+ & \,\Ncal(\eps) \frac{1}{M^{6}_{\smallv}} z^{3+2\eps} \pts{1-z}^{-1-4 \eps}\bigg\lbrace  -\frac{2}{\eps^2} \Hpl{(0;z)}
+\frac{4}{\eps} \left( \Hpl{(1,0;z)} + \zed{2} \right) \! + \! 14 \zed{2} \Hpl{(0;z)} \nn\\
- & 10 \zed{2} \Hpl{(1;z)} +2\Hpl{(0,0,0;z)} -2 \Hpl{(0,1,0;z)} + 2 \Hpl{(1,0,0;z)} - 10 \Hpl{(1,1,0;z)} \nn\\
+ & 4 \zed{3} 
+ \Ocal(\eps)\bigg\rbrace \label{eq:MasterIntegralEightT} \, . \\
& \nonumber \\ 
I_{9}(z;\eps) = & \vcenter{\hbox{\scalebox{1.2}{\masternine}}} = \displaystyle{\int \rmd \Phi_{3}\frac{1}{\pts{p_{1}-p_{4}}^2 \pts{p_{1}-p_{5}}^2 \pts{p_{2}-p_{4}}^2 \pts{p_{2}-p_{5}}^2} } \nonumber \\ 
=& \,\Ncal(\eps) \frac{1}{M^{6}_{\smallv}} z^{3+2\eps} \pts{1-z}^{-1-4 \eps} \bigg\lbrace -\frac{4}{\eps^3} + \frac{32}{\eps} \zed{2} + 80 \zed{3} - 64 \zed{4} \eps + \Ocal(\eps^2)\bigg\rbrace \nonumber \\
+& \,\Ncal(\eps) \frac{1}{M^{6}_{\smallv}} z^{3+2\eps} \pts{1-z}^{-1-4 \eps} \bigg\lbrace -\frac{2}{\eps^2} \Hpl{(0;z)}
+\frac{4}{\eps} \Hpl{(0,0;z)} - 4 \zed{2} \Hpl{(0;z)} \nn\\
- & 8 \Hpl{(0,0,0;z)}-20 \Hpl{(0,1,0;z)}-4 \Hpl{(1,0,0;z)} - 36 \zed{3} 
+ \Ocal(\eps)\bigg\rbrace \label{eq:MasterIntegralNineT} \, . \\
& \nonumber \\
I_{10}(z;\eps) = & \vcenter{\hbox{\scalebox{1.2}{\masterten}}} = \displaystyle{\int \rmd \Phi_{3} \frac{1}{\pts{p_{1}-p_{4}}^2 \pts{p_{1}-p_{4}-p_{5}}^2 \pts{p_{2}-p_{5}}^2 \pts{p_{2}-p_{4}-p_{5}}^2 } } \nonumber \\  
=& \,\Ncal(\eps) \frac{1}{M^{6}_{\smallv}} z^{2+2 \eps} \pts{1-z}^{-1-4 \eps} \bigg\lbrace -\frac{1}{\eps^3} + \frac{8\zed{2}}{\eps} + 20 \zed{3} - 16 \zed{4} \eps + \Ocal(\eps^2)\bigg\rbrace \nonumber \\
+& \,\Ncal(\eps) \frac{1}{M^{6}_{\smallv}} z^{2+2 \eps} \pts{1-z}^{-1-4 \eps} \bigg\lbrace \frac{1}{\eps^2} \Hpl{(0;z)}
-\frac{4}{\eps} \left( \Hpl{(1,0;z)} + \zed{2} \right) - 10 \zed{2} \Hpl{(0;z)} \nn\\
+ & 10 \zed{2} \Hpl{(1;z)} \! - \!4 \Hpl{(0,0,0;z)} \!-\! 2 \Hpl{(0,1,0;z)} \!- \! 14 \zed{3} \!+ \! 10 \Hpl{(1,1,0;z)}
+ \Ocal(\eps) \bigg\rbrace \label{eq:MasterIntegralTenT} . \\
%
%
I_{11}(z;\eps) = & \vcenter{\hbox{\scalebox{1.2}{\mastereleven}}} \nn\\
= & \displaystyle{\int \rmd \Phi_{3}\frac{1}{\pts{p_{1}-p_{4}-p_{5}}^2 \pts{p_{2}-p_{4}}^2 \pts{p_{1}-p_{5}}^2 \left[\pts{p_{1}+p_{2}-p_{4}}^2-M^2_{\smallw}\right] }}  \nonumber \\ 
= & \,\Ncal(\eps) \frac{1}{M^{6}_{\smallw}} z^{2+ 2\eps} \pts{1-z}^{-1-4 \eps}\bigg\lbrace \frac{4}{3\eps^3} -\frac{12 \zed{2}}{\eps} -36 \zed{3} + 8 \zed{4} \eps + \Ocal({\eps^2}) \bigg\rbrace \nonumber \\
+& \,\Ncal(\eps) \frac{1}{M^{6}_{\smallw}} z^{2+ 2\eps} \pts{1-z}^{-1-4 \eps} \bigg\lbrace 
- \frac{1-z}{3 \eps^3}
+\frac{1}{\eps^2} \frac{(4 z-3)}{3} \Hpl{(0;z)} \nonumber \\
+ & \frac{1}{\eps} \Bigl( \frac{1}{3} z \Hpl{(0,0;z)} + 5(1 - z) \Hpl{(1,0;z)} + 8 (1-z) \zeta_2 \Bigr) 
- 20 \zed{2} (1-z) \Hpl{(1;z)} \nn\\
+ & 2 \zed{2} (5 - 6z) \Hpl{(0;z)} - 20 (1-z) \Hpl{(1,1,0;z)} + \frac{12-5 z}{3} \Hpl{(0,0,0;z)} \nn\\
- & (1+3 z) \Hpl{(1,0,0;z)} + 2 \zed{3} (15 - 13 z) 
+ \Ocal(\eps) \bigg\rbrace \label{eq:MasterIntegralElevenT} \, . \\
& \nonumber \\
I_{12}(z;\eps) = & \vcenter{\hbox{\scalebox{1.2}{\mastertwelve}}} = \displaystyle{\int \rmd \Phi_{3} } \frac{1}{\left(p_{1}+p_{2}-p_{4}\right)^2 \left(p_{2} - p_{4} - p_{5} \right)^2}  \nonumber \\ 
=& \,\Ncal(\eps) \frac{1}{M^{2}_{\smallv}} z^{1+ 2\eps} \pts{1-z}^{2-4 \eps}\bigg\lbrace -\frac{1}{2} + \Ocal({\eps}) \bigg\rbrace \nonumber \\
+& \,\Ncal(\eps) \frac{1}{M^{2}_{\smallv}} z^{1+ 2\eps} \pts{1-z}^{2-4 \eps} \bigg\lbrace 
\frac{1}{2} \!+\! \frac{\Hpl{(0,0,0;z)}}{(1-z)^2} \! + \frac{\zed{2} \Hpl{(0;z)}\!+\!\Hpl{(0,1,0;z)}\! +\!2 \zed{3}}{(1-z)^2} \nn\\
+&  \Ocal(\eps) \bigg\rbrace \label{eq:MasterIntegralTwelveT} \, . \\
& \nonumber \\
I_{13}(z;\eps) = & \vcenter{\hbox{\scalebox{1.2}{\masterthirteen}}} = \displaystyle{\int \rmd \Phi_{3} \frac{1}{\left(p_{1}+p_{2}-p_{4}\right)^2 \left(p_{2} - p_{4} - p_{5} \right)^4}} \nonumber \\
= & \,\Ncal(\eps) \frac{1}{M^{4}_{\smallv}} z^{1+ 2\eps} \pts{1-z}^{1-4 \eps} \bigg\lbrace 
-\frac{1}{2 \eps} - 3 + (4 \zed{2}-14) \eps + \Ocal({\eps^2}) \bigg\rbrace \nonumber \\
+ & \,\Ncal(\eps) \frac{1}{M^{4}_{\smallv}} z^{1+ 2\eps} \pts{1-z}^{1-4 \eps} \bigg\lbrace \frac{1}{\eps} \left(\frac{1}{2} 
+ \frac{1}{2} \frac{\Hpl{(0;z)}}{1-z} \right) + 3 - 3 \frac{\left( \Hpl{(1,0;z)} + \zed{2} \right)}{1-z} \nn\\
- & \frac{3}{2} \frac{\Hpl{(0,0;z)}}{1-z} + \frac{\eps}{1-z} \Bigl( 2(7- 2 \zed{2})(1-z) +  5 \zed{2} \Hpl{(0;z)}+14 \zed{2} \Hpl{(1;z)} + \frac{7}{2} \Hpl{(0,0,0;z)}\nn\\
+ & 9 \Hpl{(0,1,0;z)} + 5 \Hpl{(1,0,0;z)} +14 \Hpl{(1,1,0;z)}-\zed{3} \Bigr) 
+ \Ocal(\eps^2) \bigg\rbrace \label{eq:MasterIntegralThirteenT} \, . \\
& \nonumber \\
I_{14}(z;\eps) = & \vcenter{\hbox{\scalebox{1.2}{\masterfourteen}}} = \displaystyle{\int \rmd \Phi_{3} \frac{1}{ \left(p_{2} - p_{4} - p_{5} \right)^2 \left[ \left(p_{1}-p_{4}-p_{5}\right)^2 - M^2_{\smallw}\right]}} \nonumber \\
=& \,\Ncal(\eps) \frac{1}{M^{2}_{\smallw}} z^{1+ 2\eps} \pts{1-z}^{2-4 \eps}\bigg\lbrace \frac{1}{2}+ 4 \eps + \Ocal({\eps^2}) \bigg\rbrace \nonumber \\
+& \,\Ncal(\eps) \frac{1}{M^{2}_{\smallw}} z^{1+ 2\eps} \pts{1-z}^{2-4 \eps}\bigg\lbrace 
-\frac{1}{2} + \frac{\Hpl{(0,0;z)}}{(1-z)^2} + \frac{\eps}{(1-z)^2} \big( -4(1-z)^2 \nn\\
- & 4 \zed{2} \Hpl{(0;z)} + 2 \Hpl{(0,0;z)}-2 \Hpl{(0,0,0;z)} - 4 \Hpl{(0,1,0;z)}-4 \Hpl{(1,0,0;z)}-4 \zed{3} \bigr) \nn\\
+ & \Ocal(\eps^2) \bigg\rbrace \label{eq:MasterIntegralFourteenT} \, . \\
& \nonumber \\
I_{15}(z;\eps) = & \vcenter{\hbox{\scalebox{1.2}{\masterfifteen}}} = \displaystyle{\int \rmd \Phi_{3} \frac{1}{ \left(p_{1}-p_{4}\right)^2 \left(p_{2} - p_{5}\right)^2 \left(p_{4} + p_{5} \right)^2}} \nonumber \\
=& \,\Ncal(\eps) \frac{1}{M^{4}_{\smallv}} z^{2+ 2\eps} \pts{1-z}^{-1-4 \eps}\bigg\lbrace -\frac{3}{\eps^3}+\frac{26 \zed{2}}{\eps} +70 \zed{3} + \Ocal({\eps}) \bigg\rbrace \nonumber \\
+& \,\Ncal(\eps) \frac{1}{M^{4}_{\smallv}} z^{2+ 2\eps} \pts{1-z}^{-1-4 \eps}\bigg\lbrace -\frac{1}{\eps^2} \Hpl{(0;z)}
+\frac{2}{\eps} \Hpl{(0,0;z)} - 2 \zed{2} \Hpl{(0;z)} \nn\\
- & 4 \Hpl{(0,0,0;z)}-10 \Hpl{(0,1,0;z)}-20 \zed{3}
+ \Ocal(\eps) \bigg\rbrace \label{eq:MasterIntegralFifteenT} \, . \\
& \nonumber \\
I_{16}(z;\eps) = & \vcenter{\hbox{\scalebox{1.2}{\mastersixteen}}} = \displaystyle{\int \rmd \Phi_{3} \frac{1}{ \left(p_{1}+p_{2}-p_{4}\right)^2 \left(p_{2} - p_{5}\right)^2 \left(p_{2} - p_{4} - p_{5} \right)^2}} \nonumber \\
= & \,\Ncal(\eps) \frac{1}{M^{4}_{\smallv}} z^{2+ 2\eps} \pts{1-z}^{1-4 \eps}\bigg\lbrace  \frac{1}{2 \eps^2} +\frac{2}{\eps}-4 \zed{2}
+ 8 + \Ocal({\eps}) \bigg\rbrace \nonumber \\
+ & \,\Ncal(\eps) \frac{1}{M^{4}_{\smallv}} z^{2+ 2\eps} \pts{1-z}^{1-4 \eps}\bigg\lbrace \frac{1}{\eps^2} \Bigl( -\frac{1}{2} 
- \frac{\Hpl{(0;z)} }{1-z^2} \Bigr) - \frac{2}{\eps} \Bigl( 1 + \frac{\Hpl{(0,0;z)}}{1-z^2} \nn\\
- & \frac{2 \Hpl{(-1,0;z)} +2 \Hpl{(1,0;z)} +3 \zed{2}}{1-z^2} \Bigr) + \frac{1}{1-z^2} 
\big( (4 \zed{2}-8)(1-z^2) -24 \zed{2} \Hpl{(-1;z)}\nn\\
+ & 16 \zed{2} \Hpl{(0;z)} - 24 \zed{2} \Hpl{(1;z)} - 16 \Hpl{(-1,-1,0;z)}+8 \Hpl{(-1,0,0;z)} +30 \zed{3} \nonumber \\
- & 16 \Hpl{(-1,1,0;z)} + 6 \Hpl{(0,0,0;z)}-16 \Hpl{(1,-1,0;z)} + 8 \Hpl{(1,0,0;z)} +  8 \Hpl{(0,1,0;z)} \nn\\
- & 16 \Hpl{(1,1,0;z)}  \bigr) 
+ \Ocal(\eps) \bigg\rbrace \label{eq:MasterIntegralSixteenT} \, . \\
& \nonumber \\
I_{17}(z;\eps) = & \vcenter{\hbox{\scalebox{1.1}{\masterseventeen}}} = \displaystyle{\int \rmd \Phi_{3} \frac{1}{ \left(p_{1}-p_{4}\right)^2 \left(p_{1}+p_{2}-p_{4}\right)^2 \left(p_{2} - p_{4} - p_{5} \right)^2 \left(p_{4} + p_{5}\right)^2}} \nonumber \\
=& \,\Ncal(\eps) \frac{1}{M^{6}_{\smallv}} z^{3+ 2\eps} \pts{1-z}^{-1-4 \eps}\bigg\lbrace  -\frac{1}{\eps^3}+\frac{8 \zed{2}}{\eps} +20 \zed{3} + \Ocal({\eps}) \bigg\rbrace \nonumber \\
+& \,\Ncal(\eps) \frac{1}{M^{6}_{\smallv}} z^{3+ 2\eps} \pts{1-z}^{-1-4 \eps}\bigg\lbrace 
-\frac{2 }{\eps^2} \Hpl{(0;z)} + \frac{4}{\eps} \left( \Hpl{(1,0;z)}+\zed{2}\right) \nn\\
+ & 14 \zed{2} \Hpl{(0;z)}-10 \zed{2} \Hpl{(1;z)} + 2 \Hpl{(0,0,0;z)} - 2 \Hpl{(0,1,0;z)}
+ 2 \Hpl{(1,0,0;z)} \nn\\
- & 10 \Hpl{(1,1,0;z)}+4 \zed{3} 
+ \Ocal(\eps) \bigg\rbrace \label{eq:MasterIntegralSeventeenT} \, . \\
& \nonumber \\
I_{18}(z;\eps) = & \vcenter{\hbox{\scalebox{1.2}{\mastereighteen}}} = \displaystyle{\int \rmd \Phi_{3} \frac{1}{ \left(p_{1}-p_{4}\right)^2 \left(p_{1}+p_{2}-p_{4}\right)^2 \left(p_{2} - p_{4} - p_{5} \right)^2 \left(p_{2} - p_{5}\right)^2}} \nonumber \\
= & \,\Ncal(\eps) \frac{1}{M^{6}_{\smallv}} z^{2+ 2\eps} \pts{1-z}^{-4 \eps}\bigg\lbrace \frac{1}{2 \eps^3}-\frac{4 \zed{2}}{\eps}-10 \zed{3} + \Ocal({\eps}) \bigg\rbrace \nonumber \\
+ & \,\Ncal(\eps) \frac{1}{M^{6}_{\smallv}} z^{2+ 2\eps} \pts{1-z}^{-4 \eps}\bigg\lbrace  
\frac{1}{\eps^2} \frac{z}{1+z} \Hpl{(0;z)} + \frac{2}{\eps} \Bigl[ \frac{1-z}{1+z} \Hpl{(-1,0;z)} \nn\\
- & \frac{\Hpl{(0,0;z)}}{1+z} - 2 \frac{z}{1+z} \Hpl{(1,0;z)} + \frac{(1-5 z)}{1+z} \frac{\zed{2}}{2} \Bigr] 
- \frac{1-z}{1+z} \bigl[ 2 \Hpl{(1,-1,0;z)} + 3 \zed{2} \Hpl{(-1;z)} \nn\\
- & 2 \zed{2} \Hpl{(0;z)} + 2 \Hpl{(-1,-1,0;z)} - \Hpl{(-1,0,0;z)} +2 \Hpl{(-1,1,0;z)} \bigr] 
- 4 \Hpl{(0,-1,0;z)} \nn\\
+ & \frac{8+2z}{1+z} \Hpl{(0,0,0;z)} +\frac{10+2z}{1+z} \Hpl{(0,1,0;z)} + \frac{6-2z}{1+z} \Hpl{(1,0,0;z)} \nn\\
+ & \frac{2+18 z}{1+z} \Hpl{(1,1,0;z)} +\frac{22 z-2}{1+z} \zed{2} \Hpl{(1;z)} + \frac{9-21 z}{1+z} \zed{3} 
+ \Ocal(\eps) \bigg\rbrace \label{eq:MasterIntegralEighteenT} \, . \\
& \nonumber \\
I_{19}(z;\eps) = & \vcenter{\hbox{\scalebox{1.1}{\masternineteen}}} \nn\\
= & \displaystyle{\int \rmd \Phi_{3} \frac{1}{ \left(p_{1}-p_{4}\right)^2 \left(p_{1}+p_{2}-p_{4}\right)^2  \left(p_{4} + p_{5}\right)^2
\left[\left(p_{1} - p_{4} - p_{5} \right)^2 - M^2_{\smallw} \right]}} \nonumber \\
=& \,\Ncal(\eps) \frac{1}{M^{6}_{\smallw}} z^{2+ 2\eps} \pts{1-z}^{-4 \eps}\bigg\lbrace -\frac{1}{2 \eps^3} +\frac{4 \zed{2}}{\eps} +10 \zed{3} + \Ocal({\eps}) \bigg\rbrace \nonumber \\
+& \,\Ncal(\eps) \frac{1}{M^{6}_{\smallw}} z^{2+ 2\eps} \pts{1-z}^{-4 \eps}\bigg\lbrace \frac{1}{\eps} \left(2 \Hpl{(0,0;z)}-\Hpl{(1,0;z)}-\zed{2} \right) - 8 \zed{2} \Hpl{(0;z)} \nn\\
+ & 4 \zed{2} \Hpl{(1;z)}-8 \Hpl{(0,0,0;z)}-8 \Hpl{(0,1,0;z)} - 11 \Hpl{(1,0,0;z)}+4 \Hpl{(1,1,0;z)} \nn\\
- & 9 \zed{3}
+ \Ocal(\eps) \bigg\rbrace \label{eq:MasterIntegralNineteenT} \, . \\
& \nonumber \\
I_{20}(z;\eps) = & \vcenter{\hbox{\scalebox{1.1}{\mastertwenty}}} \nn\\
= & \displaystyle{\int \rmd \Phi_{3} \frac{1}{ \left(p_{1}-p_{4}\right)^2 \left(p_{2}-p_{5}\right)^2 \left(p_{2} - p_{4} - p_{5} \right)^2 \left[\left(p_{1} - p_{4} - p_{5} \right)^2 - M^2_{\smallw} \right]}} \nonumber \\
= & \, I_{19}(z;\eps) \label{eq:MasterIntegralTwentyT} \, . \\
& \nonumber \\
I_{21}(z;\eps) = & \vcenter{\hbox{\scalebox{1.2}{\mastertwentyone}}} = \displaystyle{\int \rmd \Phi_{3} \frac{1}{ \left(p_{1}-p_{4}\right)^2 \left(p_{2}-p_{5}\right)^2 \left(p_{4} + p_{5} \right)^2 \left[ \left(p_{1} - p_{4} - p_{5} \right)^2 - M^2_{\smallw} \right]}} \nonumber \\
=& \,\Ncal(\eps) \frac{1}{M^{6}_{\smallw}} z^{2+ 2\eps} \pts{1-z}^{-1-4 \eps}\bigg\lbrace  \frac{3}{\eps^3}-\frac{26 \zed{2}}{\eps} -70 \zed{3} + \Ocal({\eps}) \bigg\rbrace \nonumber \\
+& \,\Ncal(\eps) \frac{1}{M^{6}_{\smallw}} z^{2+ 2\eps} \pts{1-z}^{-1-4 \eps} \bigg\lbrace -\frac{3}{\eps^3} \frac{(1-z)}{2} 
+ \frac{1}{\eps^2} (3 z-2) \Hpl{(0;z)} \nn\\
+ & \frac{1}{\eps} \bigl[ 2(1-2 z) \Hpl{(0,0;z)} + ( 23 \zed{2} + 10 \Hpl{(1,0;z)} )(1-z) \bigr] 
- 4 ( 10  \zed{2} \Hpl{(1;z)} \nn\\
+ & 3  \Hpl{(1,0,0;z)} + 10  \Hpl{(1,1,0;z)} ) (1-z)  +(67-47 z) \zed{3} + 2 (4-3 z) \zed{2} \Hpl{(0;z)} \nonumber \\
+& 4 z \Hpl{(0,0,0;z)} -10(1-2z) \Hpl{(0,1,0;z)} 
+ \Ocal(\eps) \bigg\rbrace \label{eq:MasterIntegralTwentyoneT} \, . \\
& \nonumber \\
I_{22}(z;\eps) = & \vcenter{\hbox{\scalebox{1.2}{\mastertwentytwo}}} = \displaystyle{\int \rmd \Phi_{3} \frac{1}{ \left(p_{1}+p_{2}-p_{5}\right)^2 \left[\left(p_{1} + p_{2} - p_{4} \right)^2 - M^2_{\smallw} \right]}} \nonumber \\
=& \, I_{14}(z;\eps) \label{eq:MasterIntegralTwentytwoT} \, . \\
& \nonumber \\
I_{23}(z;\eps) = & \vcenter{\hbox{\scalebox{1.2}{\mastertwentythree}}} \nn\\
= & \displaystyle{\int \rmd \Phi_{3} \frac{1}{ \left(p_{1}-p_{5}\right)^2 \left(p_{1}+p_{2}-p_{5}\right)^2 \left(p_{2}-p_{4}\right)^2\left[\left(p_{1} + p_{2} - p_{4} \right)^2 - M^2_{\smallw} \right]}} \nonumber \\
= & \, I_{19}(z;\eps) \label{eq:MasterIntegralTwentythreeT} \, . \\
& \nonumber \\
I_{24}(z;\eps) = & \vcenter{\hbox{\scalebox{1.2}{\mastertwentyfour}}} =  \displaystyle{\int \rmd \Phi_{3} \frac{1}{ \left(p_{1}+p_{2}-p_{5}\right)^2 \left(p_{1} + p_{2} - p_{4} \right)^2 }} \nonumber \\
=& \,\Ncal(\eps) \frac{1}{M^{2}_{\smallv}} z^{1+ 2\eps} \pts{1-z}^{3-4 \eps}\bigg\lbrace \frac{1}{6} +\frac{11}{9} \eps + \Ocal({\eps^2}) \bigg\rbrace \nonumber \\
+& \,\Ncal(\eps) \frac{1}{M^{2}_{\smallv}} z^{1+ 2\eps} \pts{1-z}^{3-4 \eps}\bigg\lbrace 
\frac{\Hpl{(0,0;z)} -2 \Hpl{(-1,0;z)} - \zed{2}}{(1-z)^3} - \frac{1}{6} \nonumber \\
+ & \frac{\eps}{(1-z)^3} \bigl[ ( 9 \Hpl{(-1;z)} -3  \Hpl{(0;z)} +4 \Hpl{(1;z)} ) \zed{2} - 2 \zed{2} 
- 7 \zed{3} -4 \Hpl{(-1,0;z)} \nn\\
+ & 2 \Hpl{(0,0;z)} +2 \Hpl{(-1,-1,0;z)} + \Hpl{(-1,0,0;z)} +8 \Hpl{(-1,1,0;z)} +2 \Hpl{(0,-1,0;z)} \nn\\
- & 2 \Hpl{(0,0,0;z)} - 4 \Hpl{(0,1,0;z)} +8 \Hpl{(1,-1,0;z)} -4 \Hpl{(1,0,0;z)}  -\frac{11}{9}(1-z)^3 \bigr] \nonumber \\
+ & \Ocal(\eps^2) \bigg\rbrace \label{eq:MasterIntegralTwentyfourT} \, . \\
& \nonumber \\
I_{25}(z;\eps) = & \vcenter{\hbox{\scalebox{1.2}{\mastertwentyfive}}} = \displaystyle{\int \rmd \Phi_{3} \frac{1}{ \left(p_{1}+p_{2}-p_{5}\right)^2 \left(p_{2} - p_{4} \right)^2 \left[ \left(p_{2} - p_{4} - p_{5}\right)^2 - M^2_{\smallw}\right] } } \nonumber \\
=& \,\Ncal(\eps) \frac{1}{M^{4}_{\smallw}} z^{1+ 2\eps} \pts{1-z}^{2-4 \eps}\bigg\lbrace -\frac{1}{2 \eps} -3 + \Ocal({\eps}) \bigg\rbrace \nonumber \\
+& \,\Ncal(\eps) \frac{1}{M^{4}_{\smallw}} z^{1+ 2\eps} \pts{1-z}^{2-4 \eps}\bigg\lbrace  
\frac{1}{\eps} \Bigl[ \frac{1}{2} - \frac{5 z }{\sqrt{1+4 z} \,(1-z)^2} \Bigl( \Hpl{(0,0;z)} - \frac{4}{5} \zed{2} \nn\\
+ & \frac{1}{2} \Hpl{(-\frac{r_0}{4},0;z)} \Bigr) \Bigr] + 3 + \frac{z}{\sqrt{1+4 z}\, (1-z)^2} \Bigl[ 
13  \Hpl{(0,0,0;z)}+20 \Hpl{(0,1,0;z)} \nonumber \\
- & 5  \Big(\Hpl{(-\frac{1}{4},-\frac{r_{0}}{4},0;1)} - \Hpl{(-\frac{1}{4},-\frac{r_{0}}{4},0;z)} \Big)
+ 10  \Hpl{(-\frac{1}{4},0,0;z)} + 20  \Hpl{(1,0,0;z)} \nn\\
+ & 10  \Hpl{(1,-\frac{r_{0}}{4},0;z)}
+\frac{13}{2}  \Hpl{(-\frac{r_{0}}{4},0,0;z)} + 20 \zed{2} \Hpl{(0;z)} \nn\\
+ & 10  \Hpl{(-\frac{r_{0}}{4},0,1;1)}
+10  \Hpl{(-\frac{r_{0}}{4},1,0;z)}
-8 \zed{2}  \Hpl{(-\frac{1}{4};z)} - 16 \zed{2}  \Hpl{(1;z)} \nonumber \\
+ & 10 \zed{2}  \Hpl{(-\frac{\text{r0}}{4};z)} +40 \Log{\frac{1+\sqrt{5}}{2}} \zed{2} 
+ 52 \ _4F_3\left(\frac{1}{2},\frac{1}{2},\frac{1}{2},\frac{1}{2};\frac{3}{2},\frac{3}{2},\frac{3}{2};-\frac{1}{4}\right) 
\nn\\
+ & 8 \zed{2} \Log{5} -6 \zed{3} + 10 \text{Li}_3(-4) \Bigr] + \Ocal(\eps) \bigg\rbrace \label{eq:MasterIntegralTwentyfiveT} \, .
\\
& \nonumber \\
I_{26}(z;\eps) = & \vcenter{\hbox{\scalebox{1.1}{\mastertwentysix}}} = \displaystyle{\int \rmd \Phi_{3} \frac{1}{\left(p_{1} - p_{5} \right)^2 \left(p_{1}+p_{2}-p_{5}\right)^2 \left(p_{2} - p_{4} \right)^2 \left(p_{1}+p_{2}-p_{4}\right)^2 } } \nonumber \\
=& \,\Ncal(\eps) \frac{1}{M^{6}_{\smallv}} z^{2+ 2\eps} \pts{1-z}^{1-4 \eps}\bigg\lbrace   \frac{1}{\eps^2}+\frac{4}{\eps} -8 \zed{2} + 16 +\Ocal({\eps}) \bigg\rbrace \nonumber \\
+ & \,\Ncal(\eps) \frac{1}{M^{6}_{\smallv}} z^{2+ 2\eps} \pts{1-z}^{1-4 \eps}\bigg\lbrace 
- \frac{1}{\eps^2} \Bigl(\frac{\Hpl{(0;z)}}{1-z}+1 \Bigr) - \frac{4}{\eps} \Bigl( 1 - \frac{\Hpl{(0,0;z)}}{1-z} \nn\\
+ & \frac{\Hpl{(1,0;z)}+\Hpl{(-1,0;z)}}{1-z} - \frac{\zed{2}}{2(1-z)}  \Bigr) + \frac{1}{1-z} \Big[ (8\zed{2} - 16)(1-z) \nn\\
+ & 4 \Hpl{(-1,-1,0;z)} +2 \Hpl{(-1,0,0;z)} + 16 \Hpl{(-1,1,0;z)}  +2 \zed{2} \bigl[ 9 \Hpl{(-1;z)}-3 \Hpl{(0;z)}\nn\\
- & 4 \Hpl{(1;z)} \bigr] - 4 \zed{3} + 8 \Hpl{(0,-1,0;z)} -12\Hpl{(0,0,0;z)}
-18 \Hpl{(0,1,0;z)} \nonumber \\
+& 16 \Hpl{(1,-1,0;z)} - 16 \Hpl{(1,0,0;z)} -16 \Hpl{(1,1,0;z)} \Big] 
+ \Ocal(\eps) \bigg\rbrace \label{eq:MasterIntegralTwentysixT} \, . \\
& \nonumber \\
I_{27}(z;\eps) = & \vcenter{\hbox{\scalebox{1.1}{\mastertwentyseven}}} = \displaystyle{\int \rmd \Phi_{3} \frac{1}{\left(p_{1} - p_{5} \right)^2 \left(p_{1}+p_{2}-p_{5}\right)^2 \left(p_{1} - p_{4} \right)^2 \left(p_{1}+p_{2}-p_{4}\right)^2 } } \nonumber \\
=& \,\Ncal(\eps) \frac{1}{M^{6}_{\smallv}} z^{3+ 2\eps} \pts{1-z}^{1-4 \eps}\bigg\lbrace  \frac{1}{\eps^2}+\frac{4}{\eps} -8 \zed{2}+16 + \Ocal({\eps}) \bigg\rbrace \nonumber \\
+ & \,\Ncal(\eps) \frac{1}{M^{6}_{\smallv}} z^{3+ 2\eps} \pts{1-z}^{1-4 \eps}\bigg\lbrace 
- \frac{1}{\eps^2}\left(1+ 2 \frac{\Hpl{(0;z)}}{1-z^2} \right) - \frac{4}{\eps} \Bigl( 1 + \frac{1}{2} \frac{\Hpl{(0,0;z)}}{1-z^2} \nn\\
- & \frac{\Hpl{(-1,0;z)} +2 \Hpl{(1,0;z)}}{1-z^2} - \frac{5}{2} \frac{\zed{2}}{1-z^2} \Bigr) 
+ \frac{1}{1-z^2} \Big[ (8 \zed{2} - 16)(1- z^2) +34 \zed{3} \nn\\
+ & 2 \zed{2} \big[ 9 \Hpl{(0;z)} - 10 \Hpl{(-1;z)} - 20 \Hpl{(1;z)} \big]
-8 \Hpl{(-1,-1,0;z)} +4 \Hpl{(-1,0,0;z)} \nonumber \\ 
- & 16 \Hpl{(-1,1,0;z)} - 4 \Hpl{(0,-1,0;z)}+8 \Hpl{(0,0,0;z)} +4 \Hpl{(0,1,0;z)} - 16 \Hpl{(1,-1,0;z)} \nn\\
+ & 8 \Hpl{(1,0,0;z)}-32 \Hpl{(1,1,0;z)} \Big] 
+ \Ocal(\eps) \bigg\rbrace \label{eq:MasterIntegralTwentysevenT} \, . \\
& \nonumber \\
I_{28}(z;\eps) = & \vcenter{\hbox{\scalebox{1.2}{\mastertwentyeight}}} \nonumber \\ 
= & \displaystyle{\int \rmd \Phi_{3} \frac{1}{\left(p_{1} - p_{5} \right)^2 \left(p_{1}+p_{2}-p_{5}\right)^2 \left(p_{2} - p_{4} \right)^2 \left[ \left(p_{2}-p_{4}-p_{5}\right)^2 - M^2_{\smallw} \right] } } \nonumber \\
=& \,\Ncal(\eps) \frac{1}{M^{6}_{\smallw}} z^{2+ 2\eps} \pts{1-z}^{1-4 \eps}\bigg\lbrace  -\frac{1}{\eps^2} -\frac{4}{\eps} -16 + 8 \zed{2}  
+ \Ocal({\eps}) \bigg\rbrace \nonumber \\
+ & \,\Ncal(\eps) \frac{1}{M^{6}_{\smallw}} \frac{z^{2+ 2\eps} \pts{1-z}^{-4 \eps}}{\sqrt{1+4z}} \bigg\lbrace  
\frac{1}{\eps^2} [ (1-z) \sqrt{1+4z} +  \sqrt{1+4z} \,\Hpl{(0;z)} ] \nonumber\\
+ & \frac{1}{\eps} \biggl[ \biggl( - \frac{5}{2} - \frac{3}{2} \sqrt{1+4z} \biggr) \Hpl{(0, 0;z)} 
-  4 \sqrt{1+4z} \,\Hpl{(1, 0;z)} - \frac{5}{4} \Hpl{(-\frac{r_0}{4}, 0;z)} \nonumber\\
+ & \sqrt{1+4z} (4 - 4 z - 4 \zeta_2) + 2 \zeta_2 \biggr] 
+ 5 \Hpl{(-\frac{1}{4}, 0, 0;z)} - \frac{5}{2} \bigg(\Hpl{(-\frac{1}{4}, -\frac{r_0}{4}, 0; 1)} \nonumber \\
- & \Hpl{(-\frac{1}{4}, -\frac{r_0}{4}, 0;z)} \bigg) + (\frac{13}{2} + \frac{11}{2} \sqrt{1+4z}) 
     \Hpl{(0, 0, 0;z)} + (10 + 8 \sqrt{1+4z}) \Hpl{(0, 1, 0;z)} \nonumber \\ 
+ & \frac{5}{2} \sqrt{1+4z} 
  \Hpl{(0, -\frac{r_0}{4}, 0;z)} + (10 + 6 \sqrt{1+4z}) 
     \Hpl{(1, 0, 0;z)} + 16 \sqrt{1+4z} \,\Hpl{(1, 1, 0;z)} \nonumber \\
+ & 5 \Hpl{(1, -\frac{r_0}{4}, 0;z)} + (\frac{13}{4} + \frac{5}{2} \sqrt{1+4z}) 
     \Hpl{(-\frac{r_0}{4}, 0, 0;z)} + 5 \Hpl{(-\frac{r_0}{4}, 0, 1; 1)} \nonumber \\
+ & 5 \Hpl{(-\frac{r_0}{4}, 1, 0;z)} - 
   \frac{5}{4} \sqrt{1+4z} \left( \Hpl{(-\frac{r_0}{4}, -\frac{r_0}{4}, 0; 1)} - 
   \Hpl{(-\frac{r_0}{4}, -\frac{r_0}{4}, 0;z)} \right) \nonumber \\
+ & (26-20\sqrt{1+4z}) \ _4F_3 \left( {\frac{1}{2}, \frac{1}{2}, \frac{1}{2}, \frac{1}{2}}, {\frac{3}{2}, \frac{3}{2}, 
\frac{3}{2}}, -\frac{1}{4} \right) + 5 \text{Li}_3(-4) - 4 \zeta_2 \Hpl{(-\frac{1}{4};z)} \nn\\
- & 20 \ln(2) \zeta_2 + 4 \ln(5) \zeta_2 + 20 \ln(1 + \sqrt{5}) \zeta_2 +  \Hpl{(0;z)} (10 \zeta_2 - 4 \sqrt{1+4z} \zeta_2) \nn\\
+ & \Hpl{(-\frac{r_0}{4};z)} (5 \zeta_2 - 2 \sqrt{1+4z} \zeta_2) + \Hpl{(1;z)} (-8 \zeta_2 + 16 \sqrt{1+4z} \zeta_2) - 3 \zeta_3 \nn\\
+ & \sqrt{1+4z} (16 - 16 z - 8 \zeta_2 + 8 z \zeta_2 + 4( 5 - \sqrt{5} ) \text{csch}^{-1}(2) \zeta_2 
+ 28 \ln(2) \zeta_2 \nn\\
- & 4 \sqrt{5} \ln(2) \zeta_2 - 28 \ln(1 + \sqrt{5}) \zeta_2 + 4 \sqrt{5} \ln(1 + \sqrt{5}) \zeta_2 
+ 4 \zeta_3)
+ \Ocal(\eps) 
\bigg\rbrace \label{eq:MasterIntegralTwentyeightT} \, . \\
& \nonumber \\
I_{29}(z;\eps) = & \vcenter{\hbox{\scalebox{1.2}{\mastertwentynine}}} \nn\\
= & \displaystyle{\int \rmd \Phi_{3} \frac{1}{\left(p_{2} - p_{5} \right)^2 \left(p_{1}+p_{2}-p_{5}\right)^2 \left(p_{2} - p_{4} \right)^2 \left[ \left(p_{2}-p_{4}-p_{5}\right)^2 - M^2_{\smallw} \right] } } \nonumber \\
=& \, I_{28}(z;\eps) \label{eq:MasterIntegralTwentynineT} \, . \\
& \nonumber \\
I_{30}(z;\eps) = & \vcenter{\hbox{\scalebox{1.2}{\masterthirty}}} = \displaystyle{\int \rmd \Phi_{3} \frac{1}{\left(p_{1} - p_{4} \right)^2 \left(p_{2}-p_{5}\right)^2 \left(p_{2} - p_{4} \right)^2 \left[ \left(p_{2}-p_{4}-p_{5}\right)^2 - M^2_{\smallw} \right] } } \nonumber \\
=& \,\Ncal(\eps) \frac{1}{M^{6}_{\smallw}} z^{2+ 2\eps} \pts{1-z}^{-4 \eps}\bigg\lbrace  -\frac{1}{\eps^3}+\frac{8 \zed{2}}{\eps} +20 \zed{3} + \Ocal({\eps}) \bigg\rbrace \nonumber \\
+& \,\Ncal(\eps) \frac{1}{M^{6}_{\smallw}} z^{2+ 2\eps} \pts{1-z}^{-4 \eps}\bigg\lbrace  \frac{1}{\eps^2} \Hpl{(0;z)}
-\frac{6}{\eps}\left( \Hpl{(1,0;z)}+\zed{2} \right) - 6 \zed{2} \Hpl{(0;z)} \nn\\
+ & 24 \zed{2} \Hpl{(1;z)}-4 \Hpl{(0,0,0;z)}+2 \Hpl{(0,1,0;z)} + 2 \Hpl{(1,0,0;z)} + 24 \Hpl{(1,1,0;z)}\nn\\
- & 22 \zed{3} 
+ \Ocal(\eps) \bigg\rbrace \label{eq:MasterIntegralThirtyT} \, .
\end{align}%
\endgroup

\subsection{Soft limits with exact dependence on $\eps$}

We present the explicit expressions of the soft limits of the Master Integrals $I_k(z;\eps)$ with $k=1,\dots,30$. 
For convenience we repeat the expressions of the soft Master Integrals ${\cal X}(z;\eps)$, ${\cal Y}(z;\eps)$, and ${\cal Z}(z;\eps)$ introduced in Eqs.~(\ref{eq:softPS}), (\ref{eq:softY}), (\ref{eq:softZ}) and write all the results as combinations of these functions:
\label{app:MIsoft}
\bea
{\cal X}(z;\eps)&=&
\Ncal(\eps)\, M_{\smallv}^2 \, (1-z)^{3-4 \eps}\,
\frac{  \Gamma \
(1-\eps)^2}{\Gamma (4-4 \eps) \Gamma (1+\eps)^2} \, ,  \nonumber \\
{\cal Y}(z;\eps)&=&
-\frac{\Ncal(\eps)}{M_{\smallv}^4}\,(1-z)^{-1-4 \eps}\,
\frac{4 (1-4 \eps) (1-2 \eps) }{\eps^3 }\,
\frac{ \Gamma (1-\eps)^2 }{\Gamma (3-4 \eps) \Gamma (1+\eps)^2} \times \nonumber \\
&&\times
\, _3F_2(1,1,-\eps;1-2\eps,1-\eps;1) \, , \nonumber \\
{\cal Z}(z;\eps)&=&
 \frac{\Ncal(\eps)}{M_{\smallw}^2} \,(1-z)^{1-4 \eps}\,
\frac{ \Gamma (1-\eps)^2 }{\eps^2 \, \Gamma (3-4 \eps) \
\Gamma (1+\eps)^2} \biggl( 2 \,
\eps \ _3F_2(1,1-2 \eps,1-\eps;2-2 \eps,1+\eps;1) \nn\\
& & - \, \frac{\Gamma (1-3 \eps) \, \Gamma (2-2 \eps) \, 
\Gamma (1+\eps) \, \Gamma (1+2 \eps)}{\Gamma (1-\eps)^2}
\biggr) \, .
\eea
The soft limits $I^{soft}_{k}(z;\eps)$ with $k=1,\dots,30$ read:
\begin{align}
I^{soft}_{1}(z;\eps)&=
{\cal X}(z;\eps) \, , \label{eq:softfirst}\\
I^{soft}_{2}(z;\eps)&=
-\frac{1}{4} M_{\smallv}^2 (1-z) {\cal X}(z;\eps) \, , \\
I^{soft}_{3}(z;\eps)&=
\frac{2 (3-4 \eps) }{(1-2 \eps) M_{\smallv}^4 (1-z)^2} \,{\cal X}(z;\eps) \, , \\
I^{soft}_{4}(z;\eps)&=
\frac{2 (1-2 \eps) (3-4 \eps)}{\eps^2 M_{\smallv}^4 (1-z)^2} \,{\cal X}(z;\eps) \, , \\
I^{soft}_{5}(z;\eps)&=
{\cal Z}(z;\eps) \, , \\
I^{soft}_{6}(z;\eps)&=
\frac{(1-2 \eps) (3-4 \eps) (1-4 \eps) \
}{3 \eps^2 M_{\smallw}^6 (1-z)^3}\,{\cal X}(z;\eps)
+\frac{2 (1-4 \,\eps) }{3 M_{\smallw}^2 (1-z)}\,{\cal Z}(z;\eps) \, , \\
I^{soft}_{7}(z;\eps)&=
-\frac{2 (1-2 \eps) (3-4 \eps) (1-4 \eps) \
}{3 \eps^3 M_{\smallw}^6 (1-z)^3}\,{\cal X}(z;\eps)
-\frac{(1-4 \eps) }{3 \eps M_{\smallw}^2 \
(1-z)}\,{\cal Z}(z;\eps) \, , \\
I^{soft}_{8}(z;\eps)&=
-\frac{2 (1-2 \eps) (3-4 \eps) (1-4 \eps) \
}{\eps^3 M_{\smallv}^8 (1-z)^4}\,{\cal X}(z;\eps) \, , \\
I^{soft}_{9}(z;\eps)&=
-\frac{8 (1-2 \eps) (3-4 \eps) (1-4 \eps) \
}{\eps^3 M_{\smallv}^8 (1-z)^4}\,{\cal X}(z;\eps) \, , \\
I^{soft}_{10}(z;\eps)&=
-\frac{2 (1-2 \eps) (3-4 \eps) (1-4 \eps) \
}{\eps^3 M_{\smallv}^8 (1-z)^4}\,{\cal X}(z;\eps) \, , \\
I^{soft}_{11}(z;\eps)&=
\frac{8 (1-2 \eps) (3-4 \eps) (1-4 \eps) \
}{3 \eps^3 M_{\smallw}^8 (1-z)^4}\,{\cal X}(z;\eps)
+\frac{4 (1-4 \,
\eps) }{3 \eps M_{\smallw}^4 \
(1-z)^2}\,{\cal Z}(z;\eps) \, , \\
I^{soft}_{12}(z;\eps)&=
-\frac{(3-4 \eps) }{(1-2 \eps) M_{\smallv}^4 \
(1-z)}\,{\cal X}(z;\eps) \, , \\
I^{soft}_{13}(z;\eps)&=
-\frac{(3-4 \eps) }{\eps M_{\smallv}^6 \
(1-z)^2}\,{\cal X}(z;\eps) \, , \\
I^{soft}_{14}(z;\eps)&=
\frac{(3-4 \eps) }{(1-2 \eps) M_{\smallw}^4 \
(1-z)}\,{\cal X}(z;\eps) \, , \\
I^{soft}_{15}(z;\eps)&=
{\cal Y}(z;\eps) \, , \\
I^{soft}_{16}(z;\eps)&=
\frac{(1-2 \eps) (3-4 \eps) }{\eps^2 \
M_{\smallv}^6 (1-z)^2}\,{\cal X}(z;\eps) \, , \\
I^{soft}_{17}(z;\eps)&=
-\frac{2 (1-2 \eps) (3-4 \eps) (1-4 \eps) \
}{\eps^3 M_{\smallv}^8 (1-z)^4}\,{\cal X}(z;\eps) \, , \\
I^{soft}_{18}(z;\eps)&=
\frac{(1-2 \eps) (3-4 \eps) (1-4 \eps) \
}{\eps^3 M_{\smallv}^8 (1-z)^3}\,{\cal X}(z;\eps) \, , \\
I^{soft}_{19}(z;\eps)&=
-\frac{(1-2 \eps) (3-4 \eps) (1-4 \eps) \
}{\eps^3 M_{\smallw}^8 (1-z)^3}\,{\cal X}(z;\eps) \, , \\
I^{soft}_{20}(z;\eps)&=
-\frac{(1-2 \eps) (3-4 \eps) (1-4 \eps) \
}{\eps^3 M_{\smallw}^8 (1-z)^3}\,{\cal X}(z;\eps) \, , \\
I^{soft}_{21}(z;\eps)&=
-\frac{1}{\,M_{\smallw}^2}\,{\cal Y}(z;\eps)
-\frac{2 (1-2 \eps) (3-4 \eps) (1-4 \eps) \
}{\eps^2 M_{\smallw}^8 \
(1-z)^3}\,{\cal X}(z;\eps) \, , 
\\
I^{soft}_{22}(z;\eps)&=
\frac{(3-4 \eps) }{(1-2 \eps) M_{\smallw}^4 \
(1-z)}\,{\cal X}(z;\eps) \, , \\
I^{soft}_{23}(z;\eps)&=
-\frac{(1-2 \eps) (3-4 \eps) (1-4 \eps) \
}{\eps^3 M_{\smallw}^8 (1-z)^3}\,{\cal X}(z;\eps) \, , \\
I^{soft}_{24}(z;\eps)&=
\frac{1}{M_{\smallv}^4}\,{\cal X}(z;\eps) \, , \\
I^{soft}_{25}(z;\eps)&=
-\frac{(3-4 \eps) }{\eps M_{\smallw}^6 \
(1-z)}\,{\cal X}(z;\eps) \, , \\
I^{soft}_{26}(z;\eps)&=
\frac{2 (1-2 \eps) (3-4 \eps) \
}{\eps^2 M_{\smallv}^8 (1-z)^2}\,{\cal X}(z;\eps) \, , \\
I^{soft}_{27}(z;\eps)&=
\frac{2 (1-2 \eps) (3-4 \eps) \
}{\eps^2 M_{\smallv}^8 (1-z)^2}\,{\cal X}(z;\eps) \, , \\
I^{soft}_{28}(z;\eps)&=
-\frac{2 (1-2 \eps) (3-4 \eps) \
}{\eps^2 M_{\smallw}^8 (1-z)^2}\,{\cal X}(z;\eps) \, , \\
I^{soft}_{29}(z;\eps)&=
-\frac{2 (1-2 \eps) (3-4 \eps) \
}{\eps^2 M_{\smallw}^8 (1-z)^2}\,{\cal X}(z;\eps) \, , \\
I^{soft}_{30}(z;\eps)&=
-\frac{2 (1-2 \eps) (3-4 \eps) (1-4 \eps) \
}{\eps^3 M_{\smallw}^8 (1-z)^3}\,{\cal X}(z;\eps) \, . \label{eq:softlast}
\end{align}
\clearpage

\bibliographystyle{JHEP}
\bibliography{bbmv}

\end{document}